\documentclass{aastex62}

\usepackage[clockwise,figuresright]{rotating}

\newcommand\kmps{\mbox{$\rm km\,s^{-1}$}}
\newcommand\Msun{\mbox{$M_\sun$}}
\newcommand\Mspy{\mbox{$\Msun\rm\,yr^{-1}$}}
\newcommand\dunit{\mbox{g\,cm$^{-3}$}}
\newcommand\columndensity{\mbox{g\,cm$^{-2}$\,(km/s)$^{-1}$}}
\newcommand{\vct}[1]{\vec{\mathbf{#1}}}
\newcommand{\W}{\mbox{$\mathcal{W}$}}

\shorttitle{Templates of binary-induced spiral-shell patterns}
\shortauthors{KIM et al.}

\begin{document}

\title{Templates of binary-induced spiral-shell patterns around mass-losing post-main sequence stars}

\author{Hyosun Kim}
\affiliation{Korea Astronomy and Space Science Institute, 776, Daedeokdae-ro, Yuseong-gu, Daejeon 34055, Republic of Korea}
\affiliation{Academia Sinica Institute of Astronomy and Astrophysics - TIARA, P.O. Box 23-141, Taipei 10617, Taiwan}

\author{Sheng-Yuan Liu}
\affiliation{Academia Sinica Institute of Astronomy and Astrophysics - TIARA, P.O. Box 23-141, Taipei 10617, Taiwan}

\author{Ronald E. Taam}
\affiliation{Academia Sinica Institute of Astronomy and Astrophysics - TIARA, P.O. Box 23-141, Taipei 10617, Taiwan}
\affiliation{Center for Interdisciplinary Exploration and Research in Astrophysics (CIERA) and Department of Physics and Astronomy, Northwestern University, 2145 Sheridan Road, Evanston, IL 60208, USA}

\correspondingauthor{Hyosun Kim}
\email{hkim@kasi.re.kr}

\begin{abstract}
The morphological properties of the outflowing circumstellar envelopes surrounding mass-losing stars in eccentric binary systems are presented from a set of three-dimensional hydrodynamical model simulations. Based on four template models of the envelope viewed for a range of inclination angles of the systems, we implement interpretative tools for observations at high spectral/angular resolutions (as illustrated via velocity channel maps as well as position-velocity, radius-velocity, and angle-radius diagrams). Within this framework, the image and kinematical structures can be used to place constraints on the orbital parameters of the system. Specifically, three unique characteristic patterns in the envelopes are found that distinguish these systems from those in binary systems in circular orbits. Bifurcation of the spiral pattern, asymmetry in the interarm density depression, and a concurrent spiral/ring appearance all point to a binary system with an eccentric orbit. The methodology presented in this paper is illustrated in an analysis of recent radio observations of several asymptotic giant branch stars.

\end{abstract}


\section{INTRODUCTION}\label{sec:int}
The circumstellar features surrounding asymptotic giant branch (AGB) stars and their preplanetary nebulae (pPNe) and planetary nebulae (PNe) descendants have been discovered for more than two decades using, e.g., the Hubble Space Telescope in detecting dust scattered light \citep[e.g.,][]{sah98,mau00,lat00,gon03}. Due to the anisotropic illumination either by central starlight (through the polar holes in the cases of pPNe and PNe) or by diffuse interstellar radiation (in the cases of AGB circumstellar envelopes totally obscuring the central stars), many of the optical images only partially reveal recurrent features in the form of patched arcs inscribed in the extended halos of circumstellar media. The observed recurrent rings and arcs surrounding 10 evolved stars were first collected and summarized by \citet{su04}. The collection has been updated based on searches for fainter patterns in the literature and the archival imaging data from Hubble/Spitzer Space Telescopes, increasing the number of evolved circumstellar envelopes with detected ring/arc patterns to $\sim60$ \citep{cor04,ram16}.

Albeit the optical/infrared imaging of dust scattered light led to the discovery of recurrent patterns around a number of AGB and (p)PN sources, the golden era has arrived when the spectroimaging of molecular line emission can be used to characterize the kinematical description of such patterns. Given that the photometric optical/infrared imaging observations offer limited information on the three-dimensional (3-D) pattern that is projected onto the sky plane, molecular line observations provide velocity information, thereby, facilitating the 3-D reconstruction of the pattern. Such a diagnostic probe allows for the investigation of the varying mass-loss history (i.e., temporal variation of the wind density and velocity). The exquisite capability of (sub)millimeter interferometers has revealed the kinematics of circumstellar patterns at high angular/spectral resolutions, as demonstrated by many works on AGB circumstellar patterns: CIT 6 \citep{cla11,kim13,kim15}, AFGL 3068 \citep{kim17}, and IRC+10216 \citep[e.g.,][]{gue18}. Furthermore, the interferometeric facility, Atacama Large Millimeter/submillimeter Array (ALMA), since its operation in 2011, has provided new discoveries of yet fainter and/or smaller recurrent circumstellar patterns thanks to its unprecedented high sensitivity, resolution, and fidelity: R Scl \citep{mae12}, Mira \citep{ram14}, W Aql \citep{ram17}, EP Aqr \citep{hom18}, II Lup \citep{lyk18}, and OH 26.5+0.6/OH 30.1-0.7 \citep{dec19}.

As shown in some of the above papers, the observed recurrent patterns are well reproduced by binary models. Here, the orbital motions of mass-losing stars result in anisotropic winds which lead to the formation of spiral-shell patterns surrounding the binary stars. Theoretical investigations and numerical simulations of the hydrodynamics of a stellar wind in a wide binary system show that recurrent spiral/shell/arc patterns can be explained by single spiral-shell patterns viewed at different inclination angles \citep[e.g,][]{mas99,kim12b,kim13}. It has been shown by the above papers, though yet to be fully recognized that a spiral-shell pattern can appear on the sky not only as a perfect spiral when the binary is viewed face-on but also as bicentric half-circles or circular rings when the binary is viewed edge-on. The variation of the edge-on shape from bicentric half-circles to concentric rings theoretically depends on the magnitude of the orbital velocity of the mass-losing star compared to its radial wind velocity; high for the former and low for the latter \citep{kim12b}. 

The generalization of the model to an eccentric orbit \citep{kim15,kim17} is particularly pertinent for the long-period binary systems under consideration as the tidal interaction and energy dissipation in the envelopes of the stellar components are unimportant for circularization of the orbit in these systems \citep{zah77,zah89}. If the pericenters of binary stars are sufficiently close, the consequential periodic variation of the mass outflow rates \citep{har97,cer15} further diversifies the binary-induced spiral-shell patterns in explaining greater structural complexity, but at the same time maintaining the repetitive circumstellar patterns observed in many AGB, pPNe, and PNe. 

For the case of a substellar-mass companion (or planet), its gravity is insufficient to introduce a sufficient reflex motion in the orbit of the mass-losing star to result in such patterns.  However, the companion's gravitational drag of the background material (supplied from the mass-losing star) produces a trailing tail wake attached to the companion that propagates in a spiral shape. Such a gravitational wake of the companion has a smaller scale height from the orbital plane and a significantly smaller density contrast between the spiral arm region and interarm region, compared to those of the pattern induced by the orbital motion of the mass-losing star \citep{kim12a,kim12b}.

In this paper, we focus on the unique structural characteristics of the circumstellar envelope in our theoretical model that enable the inference of an eccentric orbit for the assumed binary system. Towards this end, we provide an interpretative framework for the increasing number of discoveries of recurrent patterns inscribed in the circumstellar envelopes of AGB stars and (p)PNe. Specifically, we present four template models illustrated at four different viewing angles to distinguish the effects of orbital eccentricity and companion mass on its structural appearance. In Section\,\ref{sec:sim}, we describe the numerical method, underlying assumptions, and input parameters for the 3-D hydrodynamical simulations. In Section\,\ref{sec:res}, we present the characteristics of individual models that can be used for interpreting current and future observations revealing recurrent patterns around mass-losing stars. Within this section, the midplane density distribution and the radial profiles of physical quantities (density, temperature, and velocity) are first inspected to illustrate the overall 3-D structural characteristics of the spiral-shell pattern. The velocity channel maps, the position-velocity diagrams along different position angles, the radius-velocity diagrams (i.e., position-velocity diagrams averaged over a 360\arcdeg-sector), and the angle-radius diagrams for representative channels are also presented for eventual comparison to observations. The implementation of these tools in thoroughly analyzing the observed spectroimaging data can be used to place constraints on the binary orbital parameters. We show that the bifurcation of the spiral pattern, the one-sided interarm density depression, and the spiral/ring combination are all unique features of the eccentric binary model. In Section\,\ref{sec:dis}, illustrative applications to the observations are discussed. Finally, we summarize the main results in Section\,\ref{sec:sum}.

\section{NUMERICAL METHOD}\label{sec:sim}

To simulate the 3-D hydrodynamical interaction of the stellar wind of a mass-losing star in a binary system, we use the Eulerian code FLASH version 4.3 \citep{fry00} with adaptive mesh refinement in Cartesian coordinates.  It is assumed that the flow is adiabatic with the ratio of specific heats, $\gamma$, equaling 1.4 for a diatomic gas, appropriate for the low temperatures (see below) under consideration and that the equation of state is ideal. The flow is described by the equation of mass conservation and equation of motion given as 
\begin{equation}\label{equ:con}
  \frac{\partial\rho}{\partial t}+\vct{\nabla}\cdot(\rho\vct{V})=0,
\end{equation}
and
\begin{equation}\label{equ:mom}
  \frac{\partial\vct{V}}{\partial t}+\vct{V}\cdot\vct{\nabla}\vct{V}
  = -\frac{1}{\rho}\vct{\nabla}P-\vct{\nabla}\Phi_1^{\rm eff}
  -\vct{\nabla}\Phi_2.
\end{equation}
Here, the effective gravitational forces attributed to the mass-losing star, denoted by subscript 1, and its companion, denoted by subscript 2, are given by
\begin{equation}\label{equ:f1}
  -\vct{\nabla}\Phi_1^{\rm eff}
  = -\frac{GM_1}{|\vct{r}-\vct{r_1}|^2+\epsilon_1^2} \times (1-f),
\end{equation}
and
\begin{equation}\label{equ:f2}
  -\vct{\nabla}\Phi_2 = -\frac{GM_2}{|\vct{r}-\vct{r_2}|^2+\epsilon_2^2} \times\W,
\end{equation}
respectively, where f is an acceleration factor representing the force due to the radiation pressure onto dust grains in the dust forming zone. Here, $\vct{r}$ is a specific point where the gravitational force is calculated in the coordinates with the origin at the center of mass of the binary, $\vct{r_1}$ is the position of the mass-losing star, and $\vct{r_2}$ is the position of the companion star. The pressure and density of the gas is denoted as $P$ and $\rho$ respectively, and the parameter \W\ is defined as 0 in Models 1 and 3 and as 1 in Models 2 and 4 (see below). The stellar masses are set to be $M_1=0.8\,\Msun$ and $M_2=2.2\,\Msun$. We adopt the sum of semi-major axes of the two stars as $a_1+a_2=68$\,AU. The separation for a circular-orbit model is defined as $a_1+a_2$, and the separation for an eccentric-orbit model with an eccentricity of $e$ ranges from $(a_1+a_2)(1-e)$ to $(a_1+a_2)(1+e)$. The gravitational softening radius of the mass-losing star, $\epsilon_1$, is set to 2 AU, close to the photospheric radius of AGB stars \citep[e.g.,][for IRC+10216]{men12}. By setting a large value of 10\,AU for the softening radius of the companion, $\epsilon_2$, we intentionally ignore the 
possible formation of an accretion disk around the companion. The presence of an accretion disk does not change the overall features of the circumstellar spiral-shell pattern distributed over a much larger extent \citep[e.g.,][]{the93,che17,sal18}. Although the gravitational softening length of the companion star is large, we note that it is less than its Bondi-Hoyle capture radius by a factor of 2. Our choice leads to the capture of less mass in the wake region in comparison to the case where no softening is used. However, the hydrodynamics of the gas flow in the wind region beyond several gravitational capture radii is unaffected.

The mass-losing star is assumed to lose mass at a rate given by  $\dot{M}_1=3.2\times10^{-6}\,\Mspy$. We assume that the internal wind quantities are set at a radius of 10\,AU from the mass-losing star, at which the gas temperature, $T_{\rm gas}$, equals 470 K and the wind velocity, $V_w$, equals 5.7\,\kmps\ are refreshed at each time step. With an empirically adopted value for the acceleration factor $f=2$, the resulting terminal wind velocity of an isolated mass-losing star is $\sim$ 14.5\, \kmps, of which 94\% is achieved within 68\,AU from the mass-losing star. 

The computational domain is defined as a 3-D cube in Cartesian coordinates, $x^\prime y^\prime z^\prime$, with the orbital axis taken as $z^\prime$. In eccentric orbit models, the mass-losing star has the apocenter of its orbit at $-x^\prime$ and the pericenter at $+x^\prime$. The companion's apocenter and pericenter are located on the opposite sides of those of the mass-losing star. On the other hand, the sky coordinates, $xyz$, are defined so that the observer resides at an infinite distance in the direction of the $+z$-axis. The skyplane ($xy$-plane) is labelled with the horizontal axis as $x$ and the vertical axis as $y$ in all figures presented in this paper.

The computational domain size corresponds to $6.4\times6.4\times3.2$ kAU, assuming mirror-symmetry about the orbital plane. With the maximum refinement level of 7, the resulting resolution of the adaptive simulation mesh ranges from 1.5625 AU to 25 AU. At the finest resolution applied near the stars, the wind reset region (diameter of 20\,AU) is well resolved, ensuring that the wind is intrinsically isotropic. The least refined numerical resolution is sufficient to resolve the resulting circumstellar arm patterns that are characterized by a width $>$ 100 AU (see Section\,\ref{sec:dtv}).

To elucidate the characteristics of the structure in the circumstellar envelope and to facilitate its physical understanding, we consider four distinct model simulations as follows.  
\begin{itemize}
    \item Model 1: We adopt a circular orbit for the binary system as a reference model to which models (see below) with a binary eccentricity can be compared. In addition, it is assumed that the companion has no direct gravitational influence on the wind (\W=0). In this way, we isolate the features produced in the circumstellar envelope due to the effects associated with the anisotropic distribution of the wind caused by the orbital motion of the mass-losing star. 
    \item Model 2: A circular orbit is adopted as in Model 1, but the gravitational interaction of the companion with the stellar wind is taken into account (\W=1).
    \item Model 3: The eccentricity of the binary is taken to be equal to 0.8, however, the gravitational effects exerted by the companion on the stellar wind are neglected (\W=0).  The numerical results from this simulation can be directly compared to the results obtained in Model 1. 
    \item Model 4: As in Model 3 we adopt a binary eccentricity of 0.8, but take account of the gravitational interaction of the companion with the stellar wind (\W=1) for direct comparison with the results of Models 2 and 3.
\end{itemize}

As noted in \citet{kim17}, the appearance of a spiral-shell pattern is governed by eight key parameters: individual masses of the stellar components of the binary system, sum of the semi-major axes, orbital inclination angle (i.e., viewing angle), orbital eccentricity, position angle of the current location of the mass-losing star (i.e., orbital phase of the mass-losing star), position angle of the pericenter, and position angle of the line of nodes of the orbit. The latter three position angle parameters lead to additional modification of the circumstellar structures for the case of non-circular orbits. The eccentricity of the orbit breaks the otherwise monotonic change of the circumstellar spiral-shell pattern with respect to radius, introducing a significant dependence of the observables on the specific locations of the stars (primarily of the mass-losing star), the pericenter, and the line of nodes of the orbit. In this paper, we limit our focus on the major differences in the circumstellar features between circular and eccentric orbits, not on the variety of patterns dependent on these three position angles.

Excluding the only exceptions (Section\,\ref{sec:pas}), we fix the line of nodes of the binary orbit (inclination axis) to be coincident with the $x^\prime$-axis, passing through the pericenters and apocenters of the two stars. In addition, in the visualized snapshots, the stars are located at their apocenters. For comparison, in Section\,\ref{sec:pas}, we show the midplane density distributions with the line of nodes of the orbit along the $y$-axis and the current stellar positions, offset from the apocenters, where the structures are best exemplified.

The individual masses of binary stars and their average separation (sum of semi-major axes of orbits), which are important parameters defining the orbits, are fixed in the four models presented in this paper. For the changes resulting from the variations of these three parameters, see \citet{kim12b}.
We note that the wind morphology is affected not only by the equation of state \citep[e.g.,][]{the93,sal18}, but also by the ratio of the terminal velocity of the wind relative to orbital velocity of the mass-losing star. In contrast to the former works, we focus on the description of the flow properties for wide binaries where this velocity ratio exceeds unity. In this velocity ratio regime, the gas morphology is not significantly affected by the choice of the equation of state, although there is a tendency to inhibit turbulent flows in the case of a lower adiabatic index (see section 3.1).

\section{RESULTS}\label{sec:res}

For ease of exposition of the numerical results, we present the density distribution of the circumstellar envelope in the plane of the sky followed by the distribution of the physical quantities of the envelope in three dimensions. In order to facilitate the comparison of the simulation results with observations, the results are introduced in terms of the velocity channel maps, position-velocity diagrams, and angle-radius diagrams.

\subsection{Density distribution in the midplane slice}\label{sec:den}
\begin{figure*} 
  \plotone{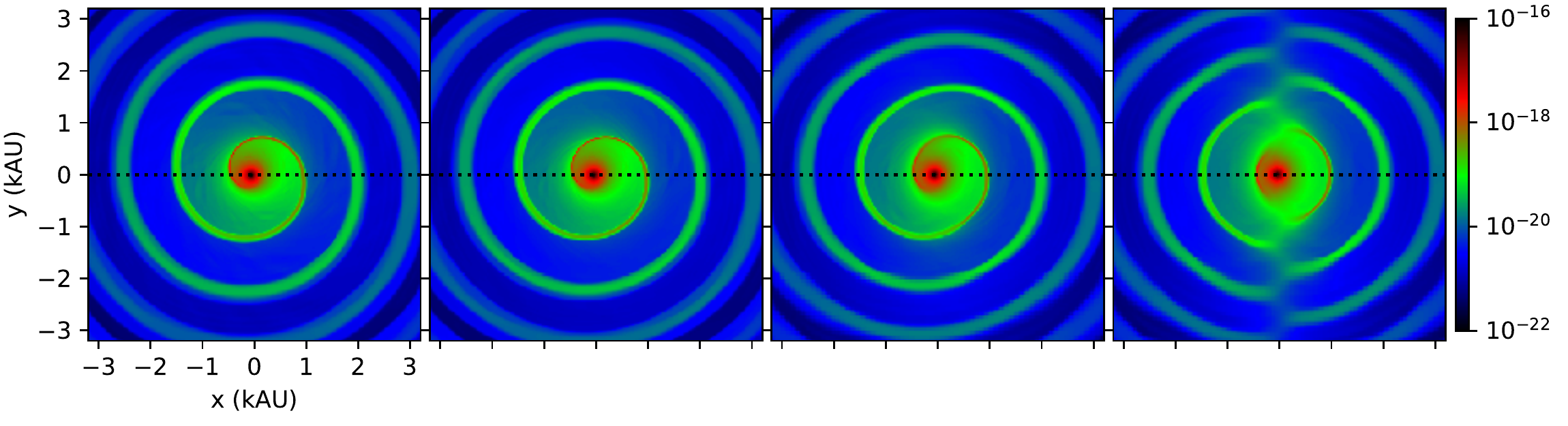}
  \caption{\label{fig:e0p}
    Model 1 ($e=0.0$, $\W=0$): density slice at the midplane in units of \dunit. Orbit inclinations with respect to the $x$-axis are 0\arcdeg, 30\arcdeg, 60\arcdeg, and 90\arcdeg\ from left to right. The horizontal dotted line ($y=0$) indicates the line of nodes of the binary orbit. The displayed image domain is 6.4\,kAU$\times$6.4\,kAU.
  }
\end{figure*}

\begin{figure*} 
  \plotone{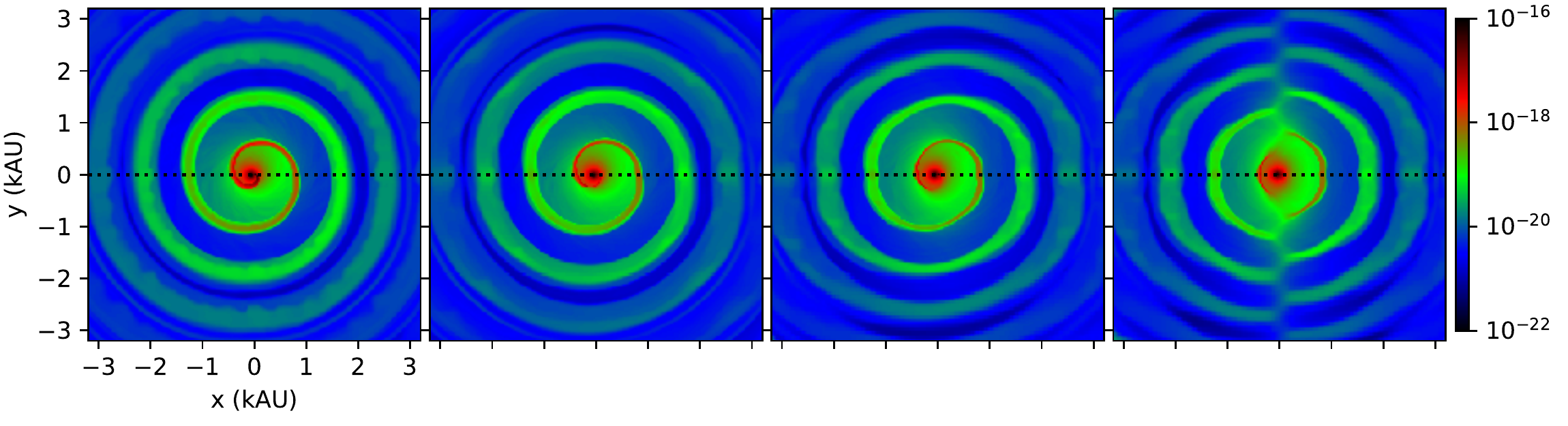}
  \caption{\label{fig:e0s}
    Model 2 ($e=0.0$, $\W=1$): density slice at the midplane in units of \dunit. See Fig.\,\ref{fig:e0p} caption for details.
  }
\end{figure*}

\begin{figure*} 
  \plotone{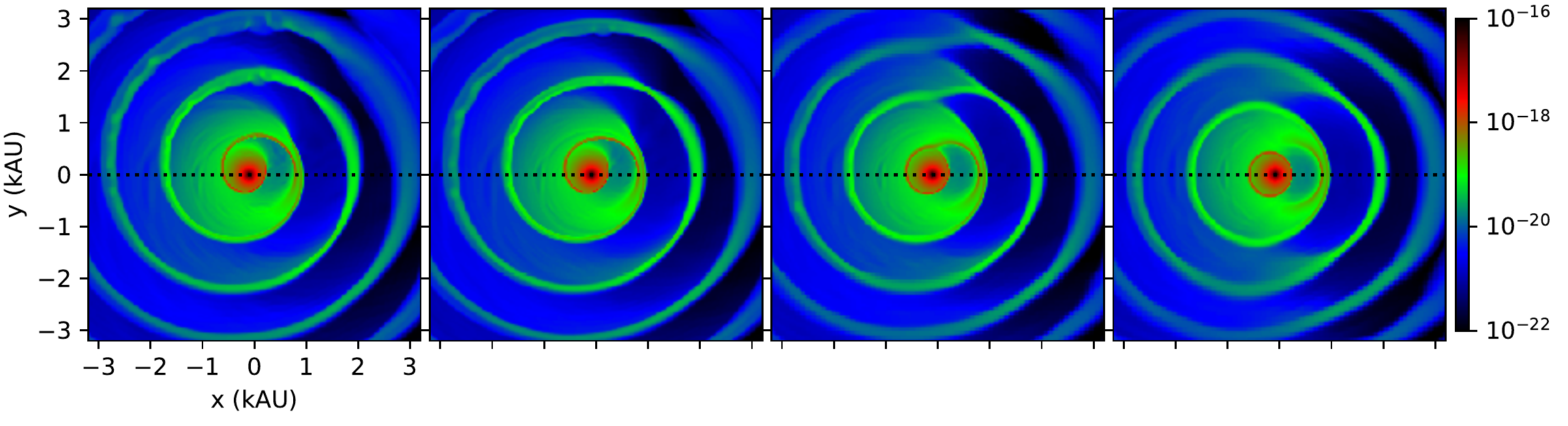}
  \caption{\label{fig:e8p}
    Model 3 ($e=0.8$, $\W=0$): density slice at the midplane in units of \dunit. See Fig.\,\ref{fig:e0p} caption for details.
  }
\end{figure*}

\begin{figure*} 
  \plotone{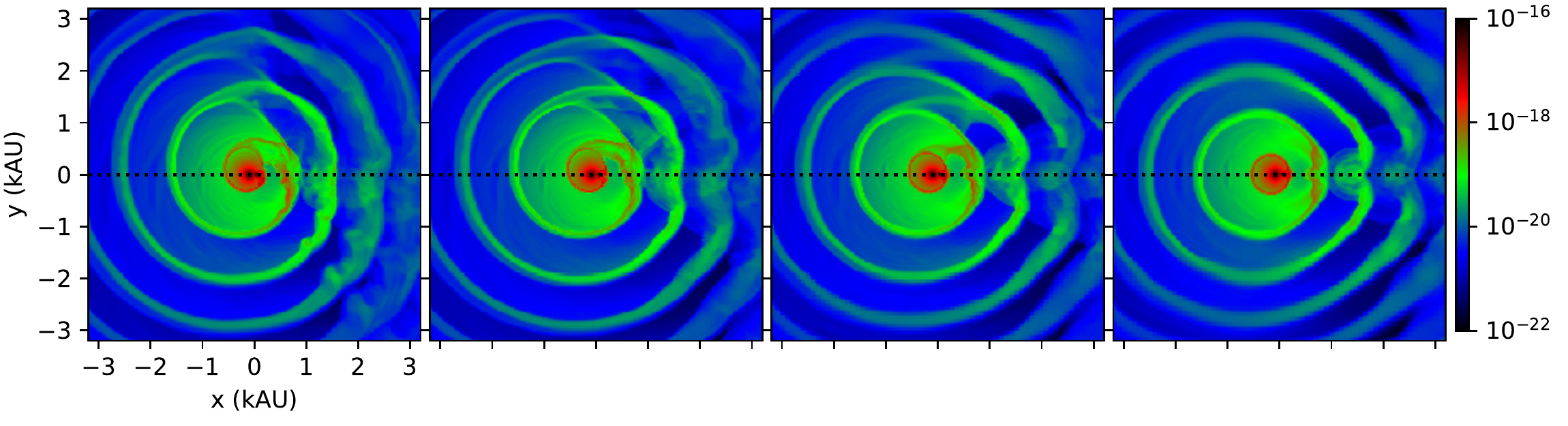}
  \caption{\label{fig:e8s}
    Model 4 ($e=0.8$, $\W=1$): density slice at the midplane in units of \dunit. See Fig.\,\ref{fig:e0p} caption for details.
  }
\end{figure*}

The midplane density distributions of our four hydrodynamic models are exhibited in Figures\,\ref{fig:e0p}--\ref{fig:e8s}, respectively. These representative snapshots are taken when the stars have completed eight orbits since starting from their apocenters. The initial background conditions are relaxed within a timescale of $r/V_w$, where $V_w$ is the local wind velocity and $r$ is the distance from the center of mass, which we take to be half the computational domain size equal to 3200 AU. Hence, the entire simulation domain has reached a steady state in these visualized snapshots. To illustrate the appearance of each model as a function of inclination angle $i$, the density distribution when the orbital plane being viewed at 0\arcdeg, 30\arcdeg, 60\arcdeg, and 90\arcdeg\ (left to right panels) with respect to the $x$-axis (horizontal dotted line) is shown.

In Figure\,\ref{fig:e0p}, the spiral-shell pattern induced by the anisotropy of the wind caused by the circular orbital motion of the mass-losing star is illustrated. For the analytic solution without considering hydrodynamic effects, see \citet{sok94}. For non-zero inclination angles, the spiral pattern is tilted and elongated slightly. One may compare, for instance, the left panel of Figure\,\ref{fig:e0p} with the panel where $i = 60\arcdeg$, where this feature is clearer\footnote{The degree of modification of the spiral shape increases with the ratio of the orbital velocity (i.e., the maximum orbital velocity achieved at the pericenter, in the case of eccentric orbit binary) of the mass-losing star to its local wind velocity \citep{kim12b,che17,sal18}. In our Model 1, this velocity ratio is only 0.3, leading to insignificant tilt and elongation of the spiral pattern with inclination angles 30\arcdeg\ and 60\arcdeg\ as shown in Figure\,\ref{fig:e0p}. For comparison, Figure\,7 in \citet{kim12b} presents a model with the velocity ratio of 1.0, showing significant modification of the spiral pattern parallel to the inclination angle. Note that while the models in \citet{kim12b} are based on an isothermal condition, different from the adiabatic equation of states employed in this paper, the overall shape of the spiral pattern is insensitive to the equation of states.  This is due to the fact that the major cause for the formation of the spiral-shell pattern is the reflex motion of the mass-losing star as shown in an analytic solution for the reflex motion without hydrodynamical effects \citep{sok94} and a piston model for sticky particles \citep{he07}.}. The pattern retains its spiral shape except very close to $i\sim90\arcdeg$. When viewed edge-on, the pattern is described as half-circles with their centers located at $+x$ and $-x$ repeated every second arc in radius.

In Model 2, the binary and wind parameters are the same as in Model 1, but the direct gravitational effect of the companion star is included. The midplane density distributions of Model 2 are shown in Figure\,\ref{fig:e0s} at different inclination angles. It can be seen that there is an overall shrinkage of the spiral-shell pattern, which is due to the gravitational attraction toward the companion star located in the very central region of the images. The position of the companion star is clearly identified in the face-on image at the head of the spiral, which was absent in Figure\,\ref{fig:e0p}. Local concentrations of gas in the orbital plane are apparent in the inclined views as revealed by the knotty density excess in the pattern along the line of nodes of the orbit (horizontal dotted line). \citet{kim12b} noted that this density excess near the orbital plane is due to the overlay of a gravitational density wake of the companion. A thickening and fluttering of the spiral pattern in the face-on view are also noticeable. Physically, the former is due to the overlay of the companion wake on the spiral induced by the reflex motion, and the latter results from the turbulence generated by a vortex (see below).

For the eccentric-orbit model (Model 3), a split of the spiral pattern is evident toward the orbital pericenter of the mass-losing star (i.e., $+x$-axis) as shown in Figure\,\ref{fig:e8p}. This split exists for all windings, but is most prominent at the first winding. Since the split is due to the difference in wind speeds at different orbital phases, its existence is a natural consequence for all windings. However, the width of the spiral pattern increases as it propagates outward \citep{kim12b}, giving an appearance for some model parameters that the two branches overlay after several windings. \citet{kim17} referred to this split as a bifurcation of the spiral pattern and used it as strong evidence to support an eccentric orbit for AFGL 3068 as a binary system, hitherto unknown because of its long $\sim800$-yr orbital period. Figure\,\ref{fig:e8p} shows that the split is prominent at all inclination angles and, therefore, is a robust feature signifying an eccentric orbit independent of the inclination angle.

By comparing the midplane density maps from the inclination angle of 30\arcdeg\ to 90\arcdeg, a gradual change of the pattern shape from a spiral to rings can be seen. The connected rings in the edge-on view ($i=90\arcdeg$) of the eccentric-orbit model are distinct from the half-circles in the circular-orbit model. The pattern shape seen in the $i=60\arcdeg$ panel of Model 3 has, in particular, both characteristics of a spiral and rings, which can be applied to the observed patterns around many AGB stars \citep[e.g., IRC+10216;][]{gue18}.

An additional conspicuous characteristic of the eccentric-orbit model is the lack of matter in the interarm regions near the $+x$-axis. This feature is due to the rapid motion of the mass-losing star near its pericenter and its pattern propagates outwards. We note that this characteristic was previously pointed out by \citet{kim15} in their interpretation of the lack of CO interarm emission of CIT 6 on the western side. The arm-to-interarm density contrast significantly differs by about an order of magnitude between the $+x$ and $-x$ directions in Figure\,\ref{fig:e8p}. Since the differential contrast trend propagates outwards, this specific feature for an eccentric-orbit binary would be easily identified in the observed images at any winding. For example, it is possible that it can be discovered in the outer arms after the star evolves to the planetary nebula phase. We point out that this feature is also found to be independent of the inclination angle of the binary orbit, as shown in Figure\,\ref{fig:e8p}. 

Model 4 is the most general case among our model simulations and corresponds to a binary in a highly eccentric orbit with the gravitational force of the companion star included. The numerical results reveal that the gas flows are turbulent, rather than laminar as in Model 3, exhibiting the most complex structures in the circumstellar envelope among the four models (see Figure\,\ref{fig:e8s}). Due to the turbulent flows, the anisotropic distribution of interarm density found in Model 3 is less evident along the $x$-axis. However, the depression of the density in the interarm region persists near the $+x$-axis.  Similar to the comparison between Model 2 and Model 1, the overall pattern size of Model 4 is reduced compared to that of Model 3. The outer branch of the bifurcated spiral in Model 3 now approaches the inner branch of the next winding. This leads to the formation of prominent rings even for small inclination angles. We note that Model 4 is an example where the mass of main-sequence companion is much greater than the mass of the mass-losing star ($M_2/M_1\sim3$). In the cases of moderate mass ratios, the density distribution and its shape could be interpolated between the numerical results of Model 3 and Model 4 to determine the degree to which the features are turbulent.

\citet{kim09} and \citet{kim10,kim11} described that the formation of turbulent flows takes place during the oscillation process of a buoyant expansion of vortices within a bow shock as a part of the gravitational density wake generated around the companion star. Although their assumption of a static background fluid velocity field differs from the radially expanding velocity field adopted in this paper, their explanation carries over for the origin of turbulent flows in Model 4 (also Model 2). We note that inclusion of radiative cooling processes in the simulation (or adopting a smaller $\gamma$) likely reduces the gas pressure gradient surrounding the companion star, tending to inhibit the formation of turbulent flows. Thus, the actual case lies between the results of Model 3 and Model 4. 

The above results, taken as an aggregate, reveal several distinguishing characteristics in the features of the circumstellar envelope that point to an eccentric binary model.  Specifically, patterns exhibiting bifurcation of the spiral structure, asymmetric interarm density contrasts, and a combined spiral/ring property are all particular signatures of the eccentric binary model.

\subsection{Density, temperature, velocity distributions in 3-D}\label{sec:dtv}

\begin{figure*} 
  \plotone{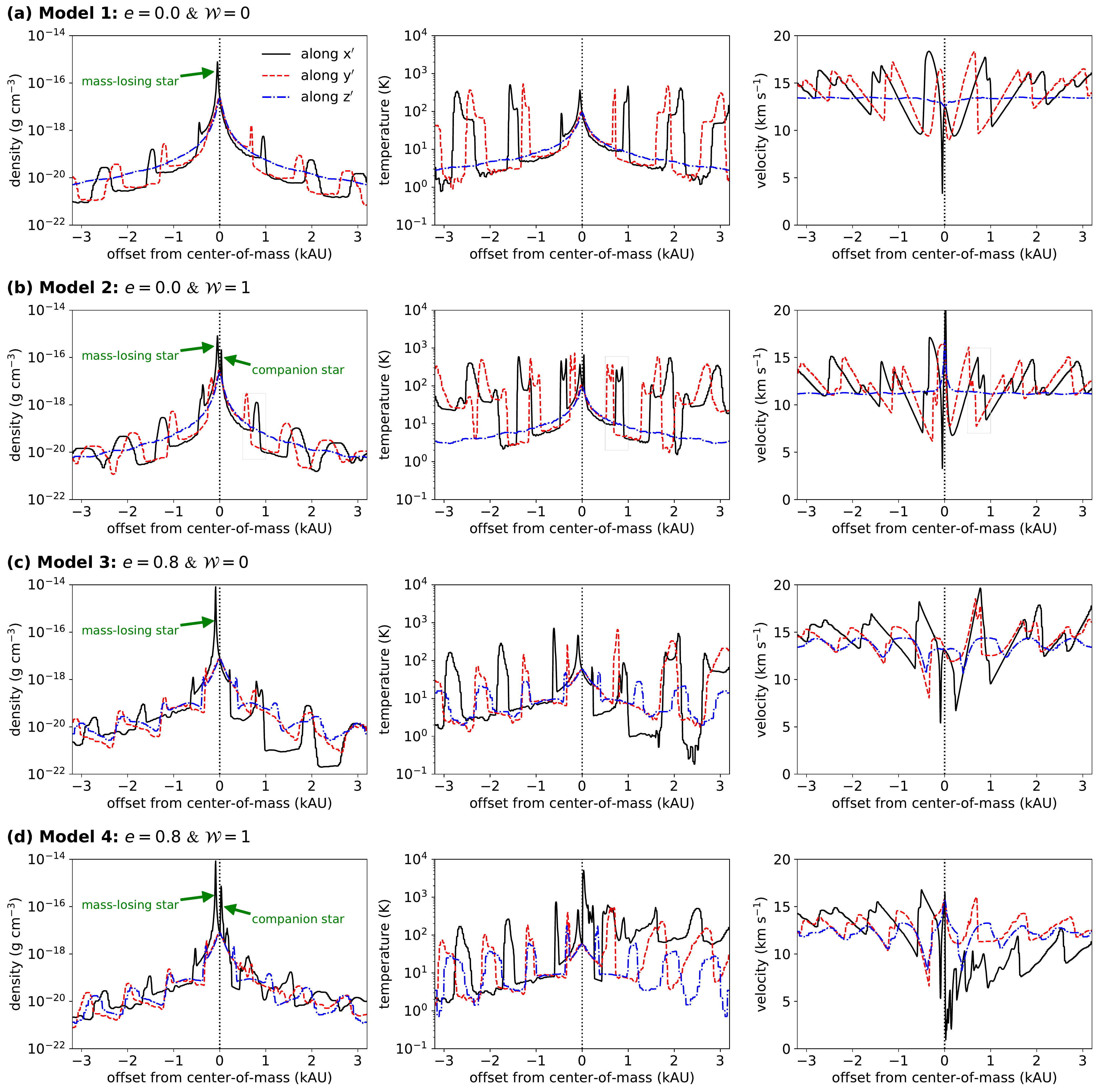}
  \caption{\label{fig:dtv}
    Density (left), temperature (middle), and velocity (right) profiles along $x^\prime$-, $y^\prime$-, and $z^\prime$-axes (black solid, red dashed, blue dash-dotted lines, respectively) for Model 1 to 4 (from top to bottom). The dotted vertical black line, labeling the origin, is to provide a reference to facilitate a measurement of the offset. The density peaks surrounding the stars are annotated with green arrows. The profiles of Model 2 along $x^\prime$-axis within gray boxes are presented in Fig.\,\ref{fig:shc}. 
  }
\end{figure*}

In order to examine the 3-D structures of the spiral-shell pattern in the orbital plane of the binary system ($x^\prime y^\prime$-plane, and orbital axis $z^\prime$-axis), we present the density, temperature, and velocity profiles along the $x^\prime$-, $y^\prime$-, and $z^\prime$-axes in Figure\,\ref{fig:dtv}. As the pattern azimuthally rotates with orbital phase, we illustrate the profiles at a particular time when the two stars are located along $x^\prime$-axis corresponding to the apocenter of the system (see Section\,\ref{sec:pas} for the pattern distribution at a different orbital phase). The highest density peaks in the black solid profiles in all models mark the positions of the mass-losing star and the second highest density peaks in Model 2 and Model 4 (at $+x^\prime$) mark the positions of the companion star. The remaining outer peaks reveal the individual windings of the spiral-shell patterns.

A remarkable characteristic of the circular-orbit models is the monotonic changes of density, temperature, and velocity profiles along the orbital ($z^\prime$) axis (blue dash-dotted lines). In particular, the nearly constant velocities of the flow along the orbital axis reveal the terminal velocities of $\sim13.4$\,\kmps\ and $\sim11.2$\,\kmps\ for Model 1 and Model 2, respectively (the blue dash-dotted lines in the rightmost panels of Figure\,\ref{fig:dtv}a-b). The absence of local peaks in the density, temperature, and velocity profiles implies that the effects of the binary orbital motion do not propagate to the orbital axis. This can be seen in the rightmost panels of Figures\,\ref{fig:e0p}--\ref{fig:e0s}, which show that the spiral-shell pattern induced by circular-orbital motions has a very large scale-height but it does not extend to the orbital axis. In contrast, the eccentric-orbit motion affects the entire solid angle and the spiral-shell pattern extends to the orbital axis, completing the rings in the edge-on density maps (see rightmost panels of Figures\,\ref{fig:e8p}--\ref{fig:e8s}). We also note that in Figure\,\ref{fig:dtv}a, showing Model 1, the temperature at the inner edge of the arm is significantly lower than the outer-edge temperature. In contrast, the temperature profile in Figure\,\ref{fig:dtv}b for Model 2 shows clear double peaks of the individual arm windings. That is, the temperature at the inner edge of the arm region becomes comparable to that at the outer edge. With the increase of temperature within an arm compared to that of Model 1, the arm width (defined as the distance from the inner edge to the outer edge of the arm) in Model 2 becomes larger.

The characteristic of eccentric-orbit models described in Section\,\ref{sec:den}, (viz., a lack of matter toward the pericenter of the mass-losing star), is clearly exhibited by a significantly lower interarm density (and also temperature) in the $+x^\prime$ direction as compared to that in $-x^\prime$ direction (see black solid profiles in Figure\,\ref{fig:dtv}c, at $x^\prime \sim$ 0.2-0.8 kAU, 1.0-1.8 kAU, and 2.1-2.6 kAU). Concurrently the velocity jumps in the $+x^\prime$ direction (toward the pericenter) are substantially larger than those in $-x^\prime$ toward the apocenter. The arm-over-interarm density ratio is about an order of magnitude in the orbital plane, but for the $+x^\prime$ direction toward the pericenter of mass-losing star the density contrast is about 2-3 orders of magnitudes. The arm-over-interarm temperature ratio is about two orders of magnitude, but about three for $+x^\prime$ direction. Also note that, while the interarm densities in the eccentric-orbit models are significantly modified from those of the corresponding circular-orbit models, the absolute values at the density peaks are similar between models regardless of the orbital shape. This indicates that the density peaks cannot be used to distinguish between the case of an eccentric orbit from a circular orbit. However, the interarm density contrasts can be used as a probe to provide support for non-circular orbital motions of the stellar components.

In the eccentric-orbit model, as shown by the black solid lines in Figure\,\ref{fig:dtv}c, the density and temperature profiles within one interarm region are flat in the direction toward the pericenter of the orbit of the mass-losing star (i.e., $+x^\prime$). In contrast, these profiles in the opposite interarm regions (i.e., $-x^\prime$) have an overall decreasing trend with an increasing radius, similar to the cases for a circular-orbit. The density and temperature profiles along the $-y^\prime$-axis (the axis which the mass-losing star passes from its apocenter to its pericenter) drawn as red dashed lines are also flat, or rising with radius, within the interarm regions. The interarm density profile along $z^\prime$ similarly shows a rising trend (see blue dash-dotted lines). 

In Model 4, the depression of the interarm density does not appear noticeable in $+x^\prime$. However, the increasing trend of the interarm density in $-y^\prime$ and $z^\prime$ may provide evidence for an eccentric-orbit binary and a hint for the position angle of the pericenter of the mass-losing star. The middle panel of Figure\,\ref{fig:dtv}d shows that the turbulent features filling the interarm regions in the $+x^\prime$ direction have high temperatures, which are nearly comparable to the temperatures in the spiral arm regions. Here, the dissipation of the turbulence on small scales can lead to the increase in temperatures in the interarm regions. The net wind velocity in the $+x^\prime$ direction significantly drops as a result of the energy cascade into small-scale eddies (right panel of Figure\,\ref{fig:dtv}d).

\begin{figure*} 
  \plotone{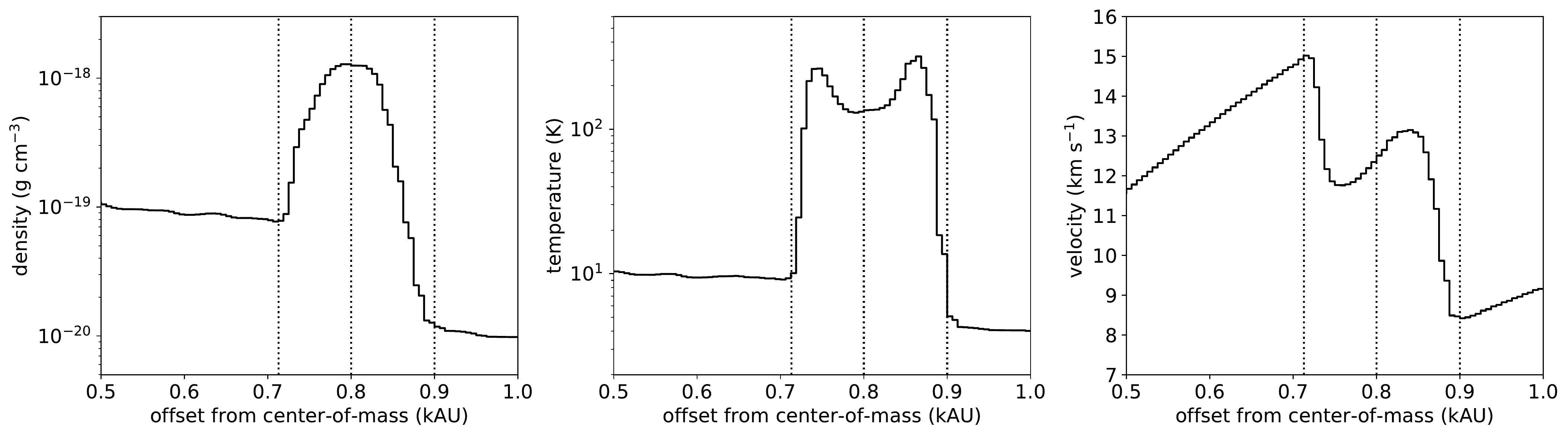}
  \caption{\label{fig:shc}
    Density, temperature, and velocity profiles across one sample spiral arm region in Model 2 along the $x^\prime$-axis within the gray boxes in Fig.\,\ref{fig:dtv}b. Vertical dotted lines denote the local velocity maximum ({\it leftmost}) and minimum ({\it rightmost}) representing the inner and outer edges of the arm region, respectively, and the density maximum (which coincides with the temperature minimum, {\it middle}).
  }
\end{figure*}

In order to elucidate the structure within the spiral arm region, the density, temperature, and velocity profiles across one sample arm region in Model 2 are illustrated in Figure\,\ref{fig:shc}. It can be seen that the arm region is enclosed by the inner and outer shocks (marked by the leftmost and rightmost vertical lines). This leads to the rapid decline of the fluid velocities with respect to the velocities of the shock fronts. The temperature peaks near the edges where the velocity profiles are non-monotonic. In contrast, the density is found to peak in the middle of the arm region, where the temperature is locally at a minimum. The difference between the positions of the density and temperature peaks implies that the fluid streamline departs from the radial direction. Such position offsets (between the density and temperature peaks) can be observationally revealed by the intensity distribution of molecular line emissions with different optical depths and/or critical densities.

\subsection{Velocity-Channel maps}\label{sec:chm}

\begin{figure} 
  \epsscale{0.6}
  \plotone{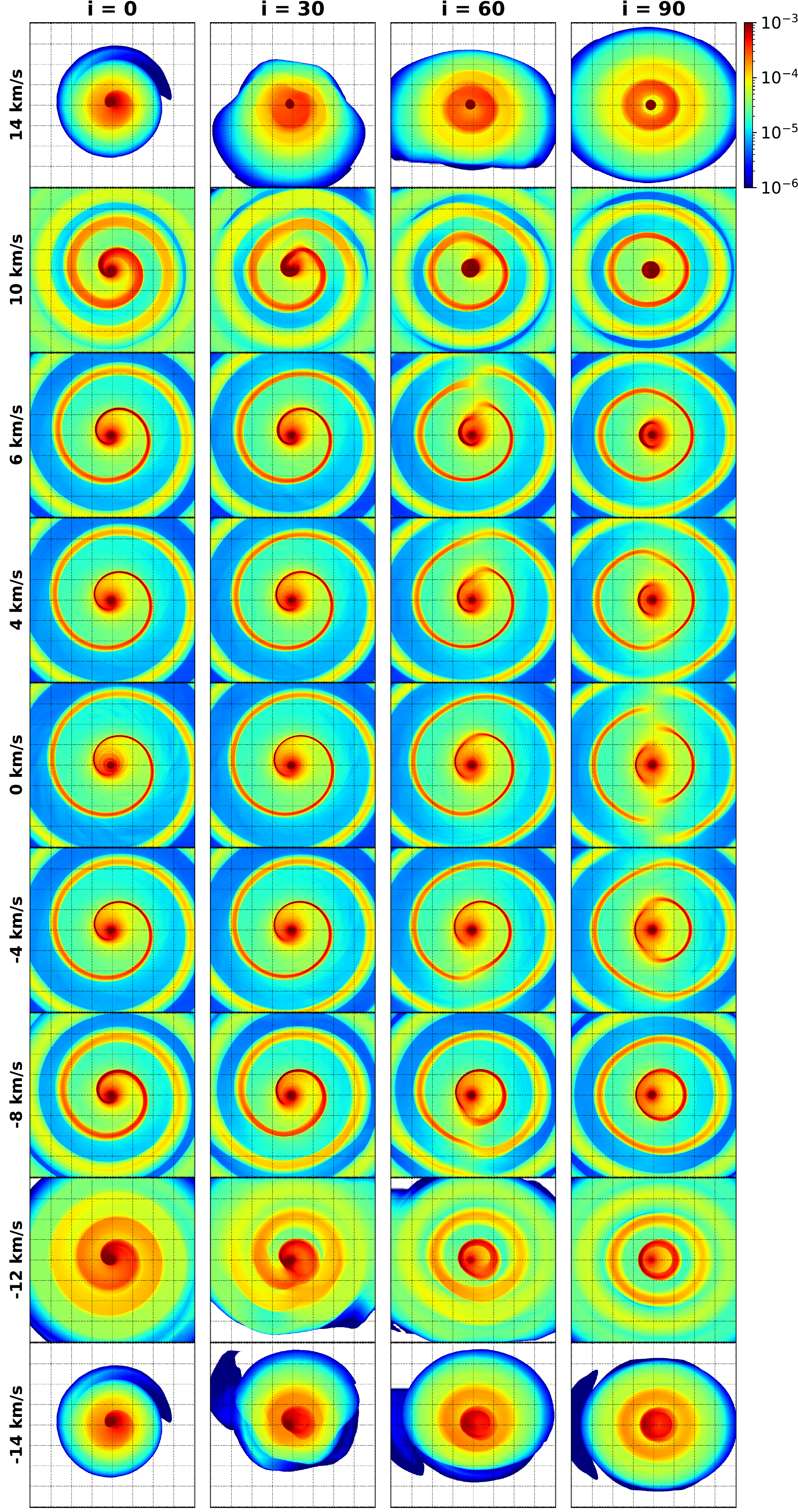}\\
  \plotone{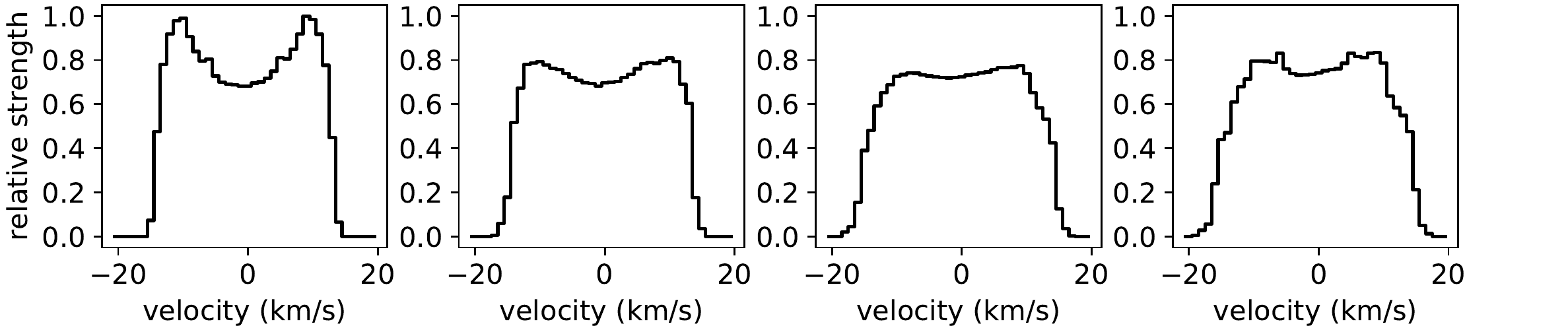}%
  \epsscale{1}
  \caption{\label{fig:px0}
    Channel maps and integrated spectra of Model 1 for inclination angles $i$ of 0\arcdeg, 30\arcdeg, 60\arcdeg, and 90\arcdeg\ from left to right. The inclination is along the $x$-axis. The representative velocity at each channel center is labelled at the left end; $-14$, $-12$, $-8$, $-4$, 0, 4, 6, 10, and 14\,\kmps, respectively, from bottom to top. In each channel, the horizontal axis ranges from $-x$ on the left to $x$ on the right and the vertical axis ranges from $-y$ at the bottom to $y$ at the top. The size of visualized image domain is 4\,kAU with the grids of intervals of 0.5\,kAU. Color bar labels in units of \columndensity.
  }
\end{figure}

\begin{figure} 
  \epsscale{0.6}
  \plotone{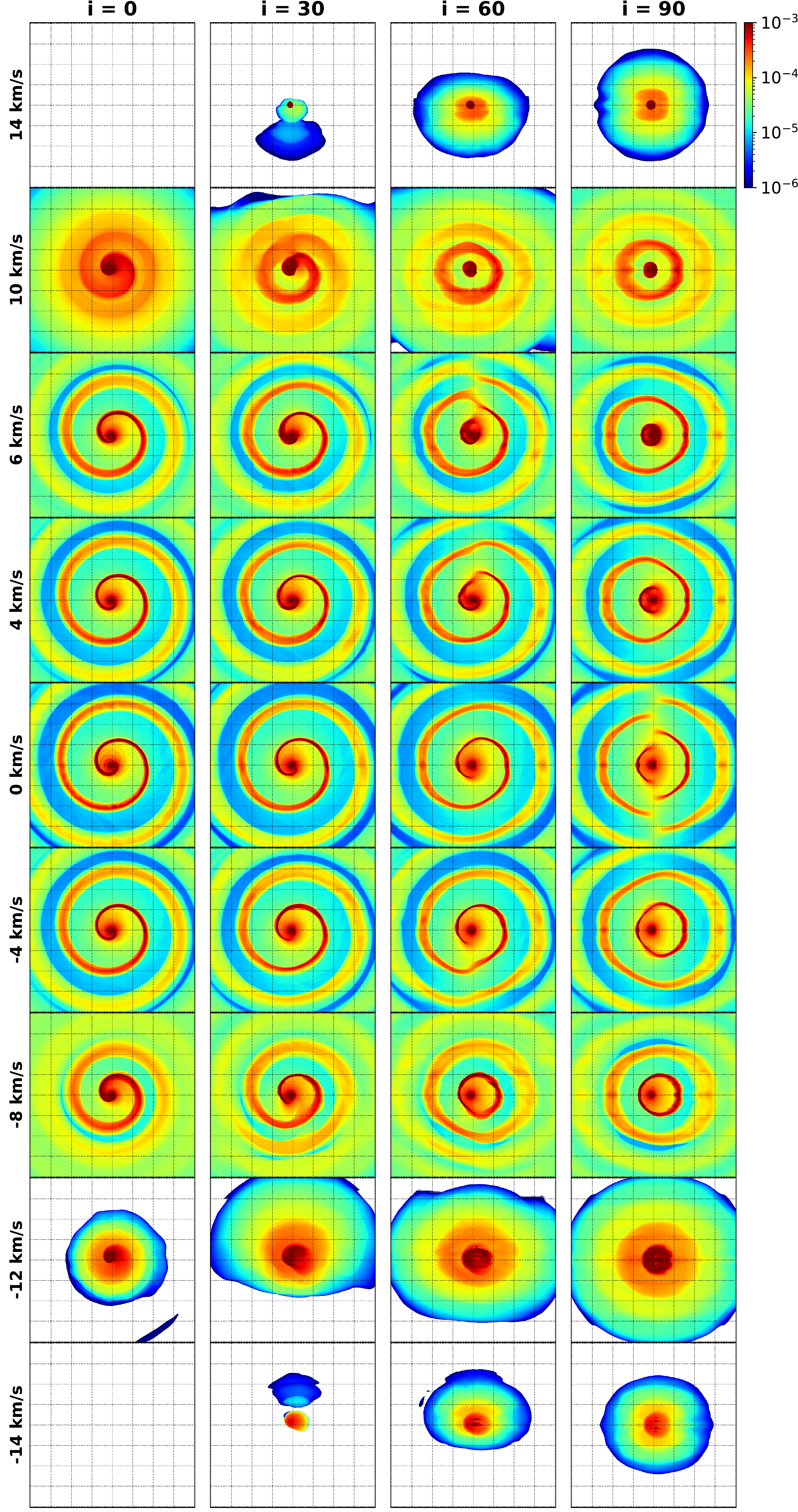}\\
  \plotone{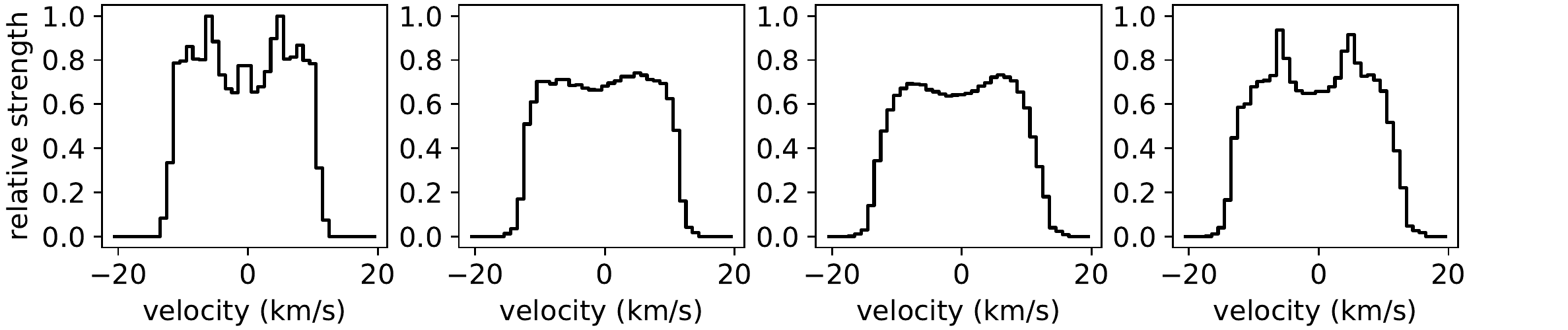}%
  \epsscale{1}
  \caption{\label{fig:ps0}
    Same as Fig.\,\ref{fig:px0} but for Model 2.
  }
\end{figure}

\begin{figure} 
  \epsscale{0.6}
  \plotone{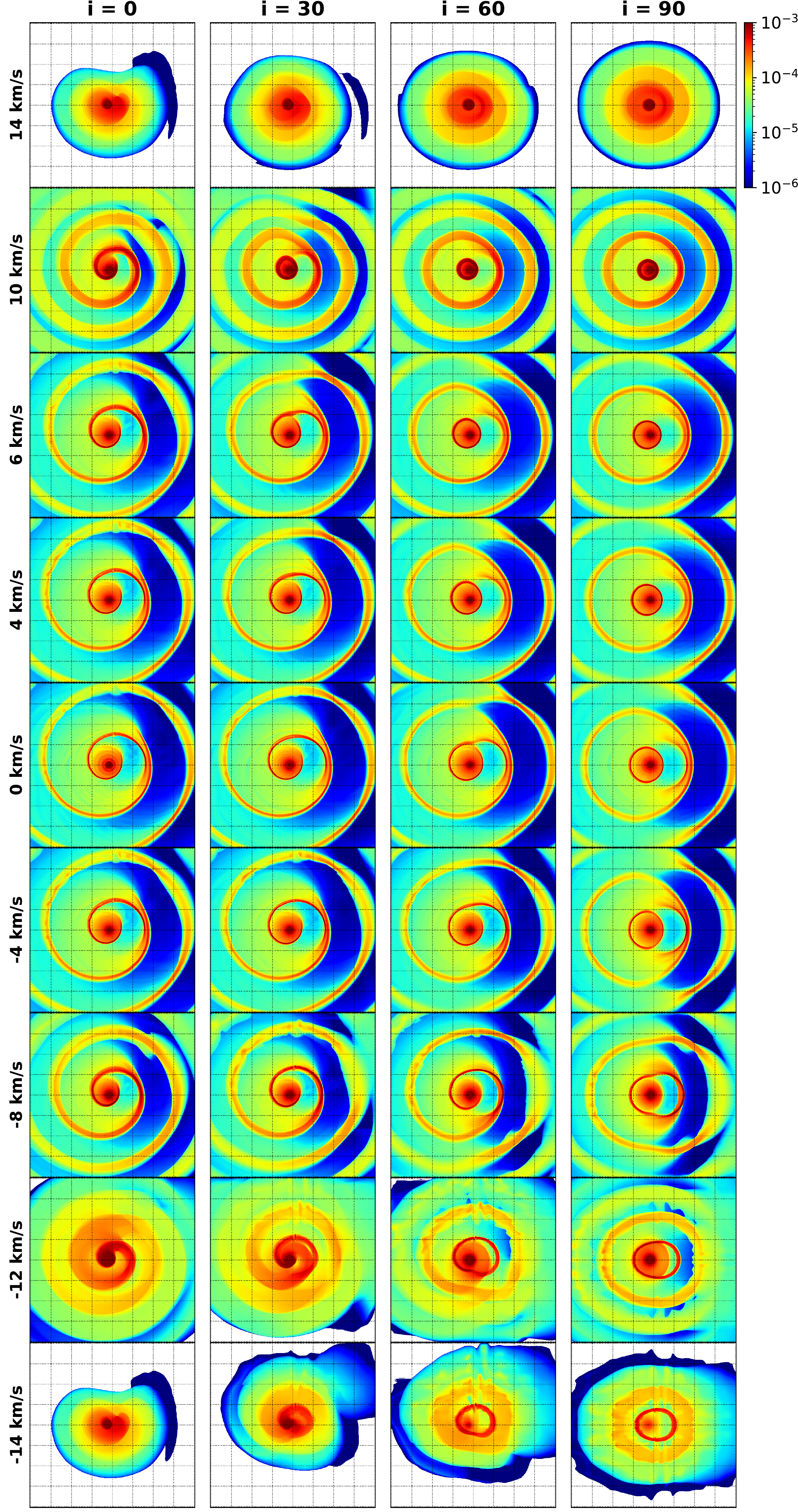}\\
  \plotone{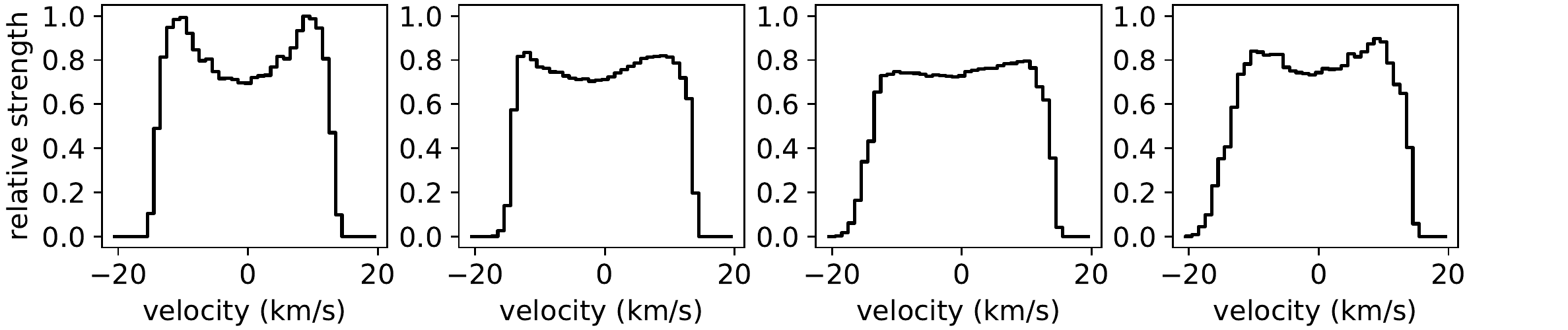}%
  \epsscale{1}
  \caption{\label{fig:px8}
    Same as Fig.\,\ref{fig:px0} but for Model 3.
  }
\end{figure}

\begin{figure} 
  \epsscale{0.6}
  \plotone{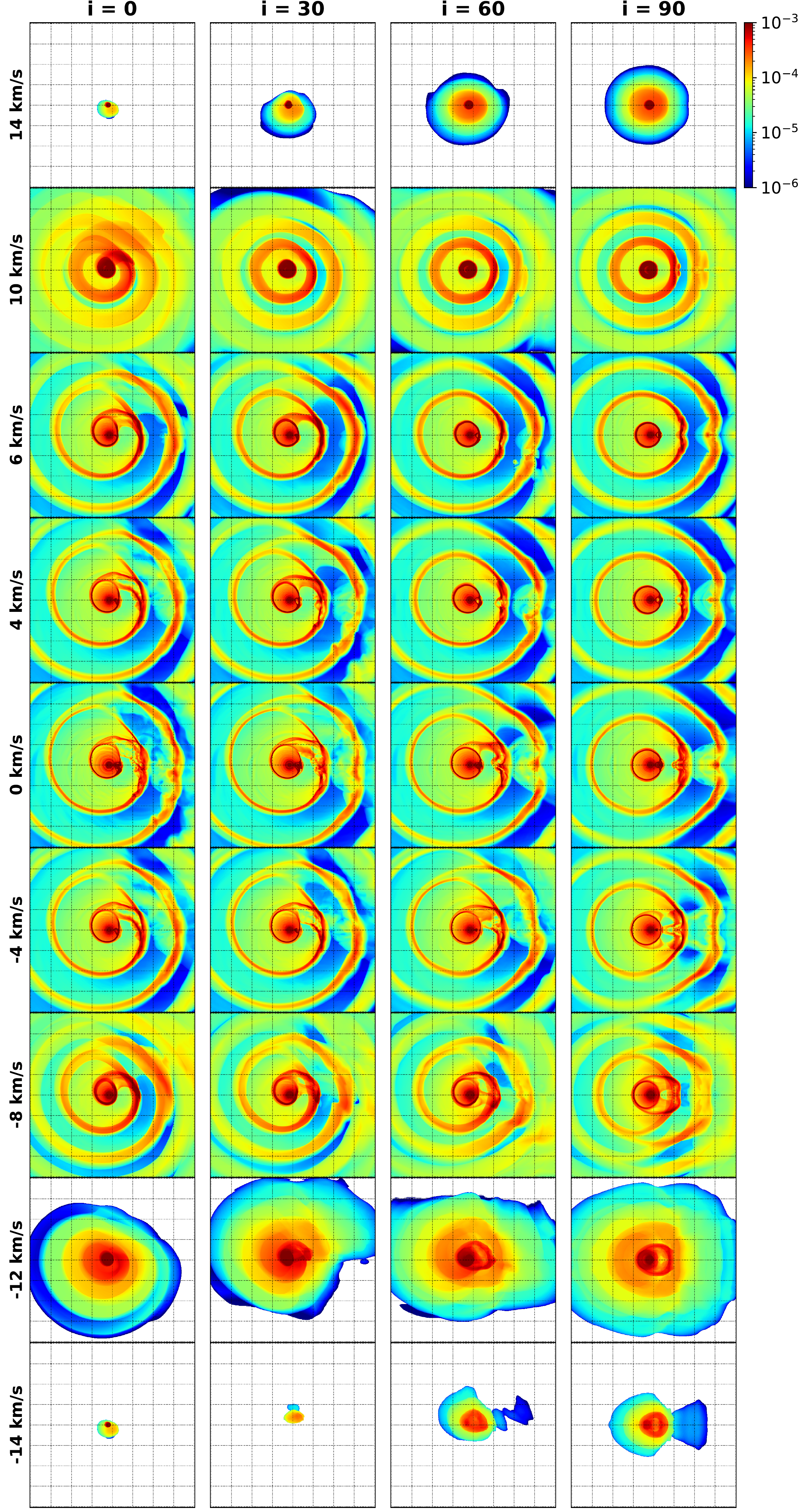}\\
  \plotone{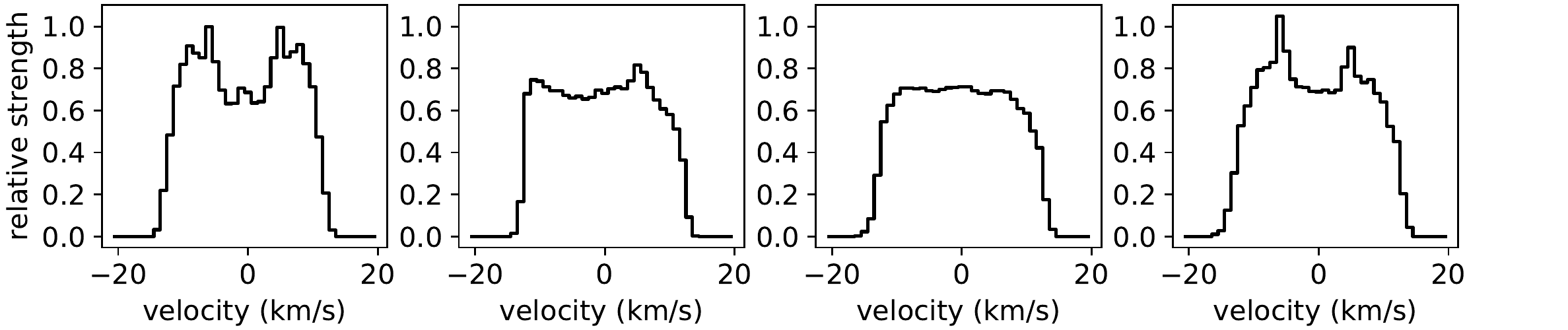}%
  \epsscale{1}
  \caption{\label{fig:ps8}
    Same as Fig.\,\ref{fig:px0} but for Model 4.
  }
\end{figure}

The velocity-channel column-density maps (see Figures\,\ref{fig:px0}--\ref{fig:ps8}) of the four simulated models are produced by integrating the density along the line of sight within the velocity range for individual channels. Radiative transfer effects are not explicitly considered in order to ignore complications due, for example, to the optical depth and temperature dependence of the molecular emission. Such channel column-density maps, nevertheless, well mimic the channel intensity maps for line emissions from thermalized optically-thin low-$J$ rotational transitions of diatomic molecules under high excitation temperature conditions.

Molecular line observations acquired with single dish telescopes, due to their limited angular and, hence, spatial resolutions, often refer to the radiation integrated over a significant portion of the circumstellar envelope. Parameterized model fitting to the observed spectral features, therefore, are conventionally carried out with prescribed profiles in order to infer the systemic velocities and wind velocities \citep[e.g.,][]{mor75,kna82}. These profiles, which depend on the optical depths and the coupling between the source sizes and telescope beam sizes, could be parabolic, flat top, or double-peak in shape, but all are symmetric. If the observed spectral line profile departs from those prescribed symmetric shapes, the systemic velocity is often defined as the middle of the zero-intensity wings of the spectral line.

From high resolution interferometric observations in which the spiral-shell pattern induced by a binary motion can be clearly resolved, our models show that the velocity of the channel with the maximum spacing of the pattern can provide a better estimate for the systemic velocity. For example, in the face-on view of Model 1 (leftmost panel of Figure\,\ref{fig:px0}), the arm spacing is a maximum at the systemic velocity ($v=0\,\kmps$) and decreases with increasing velocity of the channel for both $v>0\,\kmps$ and $v<0\,\kmps$. This follows from the fact that the transverse (projected) velocity component of the radially expanding spiral-shell pattern is maximized on the mid-plane where the mass-losing star is located, corresponding to the systemic velocity channel. Therefore, this probe can be useful in an observational data cube of molecular line emission to accurately determine the systemic velocity.
Moreover, as shown at the bottom of Figures\,\ref{fig:px0}--\ref{fig:ps8}, the integrated spectra are characterized by different shapes dependent on the inclination angle. In particular, the non-equal intensity levels between the blueshifted and redshifted peaks and the asymmetric broadening toward the blueshifted and redshifted wings of the spectrum at high inclinations would contribute to the uncertainty in the traditional estimates of the systemic velocity.

In the edge-on view of Model 1 (rightmost panel of Figure\,\ref{fig:px0}), in which the binary orbit is projected onto the $x$-axis, the spiral-shell pattern appears as half-circles at the systemic velocity and as connected rings at off-center velocities. In the intermediate inclination cases (middle columns of Figure\,\ref{fig:px0}), the multiple central channels present spiral-like patterns while the channels toward both sides of the line-edges show multiple rings. As an example, from the channel maps for $i=60\arcdeg$, it can be seen that the spiral is distorted at the 4\,\kmps\ channel. The spiral starts to be torn at the 6\,\kmps\ channel, and its outer branch connects to the inner branch of the next arm between 6 and 10\,\kmps. The channel maps for $i=30\arcdeg$ exhibit a similar transition at a velocity beyond 12\,\kmps. The velocity at which the apparent pattern changes shape from a spiral to rings decreases with increasing inclination.

Figure\,\ref{fig:ps0} displays the channel maps of Model 2. Due to the gravitational force exerted by the companion star, the net expansion velocity of the fluid in Model 2 is reduced compared to that of Model 1. The decrease of wind velocity along the line of sight is evident from the shrinkage of the spectral width and, consequently, the absence of wind material in the channels at extreme velocities. For example, the $\pm14\,\kmps$ channels are devoid of emission in the leftmost column of Figure\,\ref{fig:ps0}, while the same channels still reveal density patterns in 
Figure\,\ref{fig:px0}. 
The decrease of wind velocities in the transverse directions is evident from the overall shrinkage of the pattern spacing for each channel map. On the other hand, due to the overlay of the gravitational wake attached to the companion star, the inner edge of spiral-shell pattern in Model 2 becomes more prominent compared to that in Model 1 \citep[see also Figure\,10 of][]{kim12b}. The high temperature at this arm inner edge, leading to a double-peak temperature profile across the arm region, results in a wider arm width in Model 2 as compared to Model 1, as alluded to earlier.

In the integrated spectrum of Model 2 for $i=0\arcdeg$, the appearance of third peak near the systemic velocity is also notable. After checking the change of spectral shapes by blocking the region inside various trial radii $r$, we conclude that the origin of the central peak is not limited to the innermost region. \citet{kim12b} argued that the density contrast in the orbital plane can be decomposed into the density variation due to the reflex motion of the mass-losing star and the density fluctuation due to the gravitational wake of its companion star. \citet{kim12a} showed that the effect of the latter extends to a vertically-limited region near the orbital plane. The central peak in the face-on spectrum likely reveals the companion's density wake, and its velocity components are spread out when viewing the orbital plane at different inclination angles.

Figure\,\ref{fig:px8} illustrates the channel maps of Model 3. It reveals the off-plane behaviors of the bifurcated spiral and anisotropic interarm matter distribution described previously in Section\,\ref{sec:den}. That is, these are the unique characteristics of the circumstellar pattern induced by an eccentric-orbit binary. For example, at $i=30\arcdeg$ (second column of the figure), two branches of the bifurcated spiral are clearly seen on the right ($+x$) side in the 0\,\kmps\ channel. They gradually approach each other toward the higher velocity channels in both the blueshifted and redshifted directions, implying that the inner branch has a higher propagation speed than the outer branch. Among the displayed channels, it can be seen that the two branches are very close in the 6\,\kmps\ channel map and have merged in the $-8\,\kmps$ channel map. Beyond the transition velocity ($v\la-8\,\kmps$ or $v\ga10\,\kmps$), the order of two branches is reversed. Because the different propagation speeds of the bifurcated spiral are caused by the different speed of the mass-losing star with orbital phase, the velocity channel at which the transition occurs is nearly independent on the inclination angle.

Figure\,\ref{fig:px8} also shows that the lack of matter in the $+x$ direction is seen in almost all channels, which enables one to distinguish from observational artifacts such as side-lobes of a telescope beam pattern. We can be assured that it is due to the eccentric orbital motion and that the $+x$ direction is toward the pericenter of the orbit of the mass-losing star.

The wings of the integrated spectrum of Model 3 become wider and more visible with increasing inclination. The face-on spectrum plummets immediately beyond the velocities of the peaks ($|v|\sim10\,\kmps$) reaching zero amplitude at $|v|\sim15\,\kmps$. In contrast, the highly inclined cases ($i=60\arcdeg$ and 90\arcdeg) show slowly declining profiles reaching zero amplitude at $v\sim-20\,\kmps$ and 16\,\kmps. The different velocities at zero amplitude of the blueshifted and redshifted wings and, consequently, the overall asymmetric shapes of the spectral profiles are especially noteworthy, which are not apparent in the circular-orbit cases.
The top of the spectrum depends on the inclination angle in a similar manner to the other models. Specifically, the spectrum is clearly double peaked at $i=0\arcdeg$, is flatter at intermediate inclinations presenting a plateau-top near $i=60\arcdeg$, and is slightly double peaked again at $i=90\arcdeg$. In all cases where $i>0$, the absolute velocities of the two peaks are unequal.

Finally, Figure\,\ref{fig:ps8} shows the turbulent structure in the channel maps of Model 4. In comparison to Model 3, the wind velocities of Model 4
are reduced in both the radial and transverse propagation directions, causing the shrinkage of the spiral-shell patterns in all channel maps and of the spectral line width. The bifurcation pattern appears to be more prominent with the spreading of the angular extents of both the inner and outer branches, facilitating their reconnection to the branches from the adjacent windings. In particular, the patterns shown in the channel maps for the face-on case have both spiral and ring features ($i=0\arcdeg$, leftmost column of Figure\,\ref{fig:ps8}), in contrast to the face-on case of all the other models which maintain a spiral shape in most of the channels.

\subsection{Position-velocity diagrams}\label{sec:pvd}

\begin{figure*} 
  \plotone{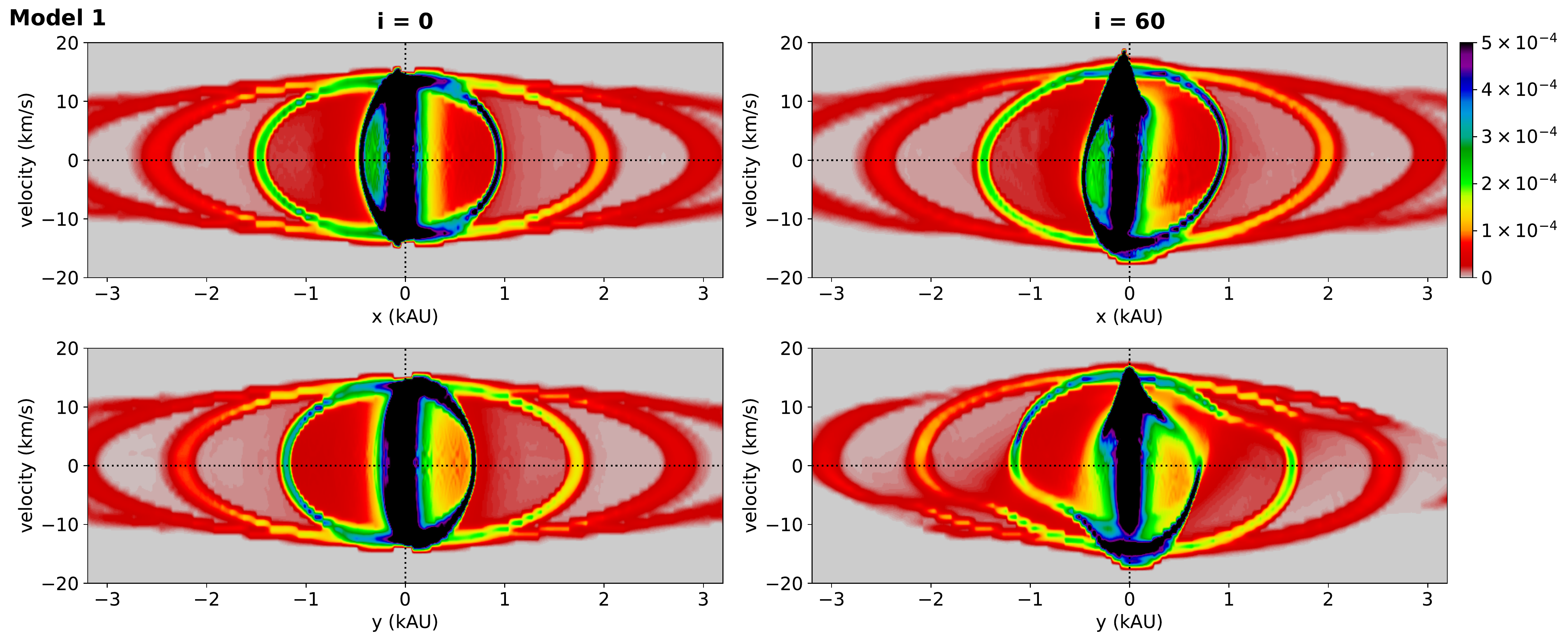}
  \caption{\label{fig:pv1}
    Position-velocity diagrams of Model 1 (circular binary orbit). From top to bottom, the diagrams along the $x$- and $y$-axes are displayed for inclination angles $i$ of 0\arcdeg\ (left) and 60\arcdeg\ (right). In these snapshots, the mass-losing star and its companion are located at their apocenters on $-x$ and $+x$ axes, respectively. The full sets are displayed in the Appendix. Color bar labels in units of \columndensity.
  }
\end{figure*}

\begin{figure*} 
  \plotone{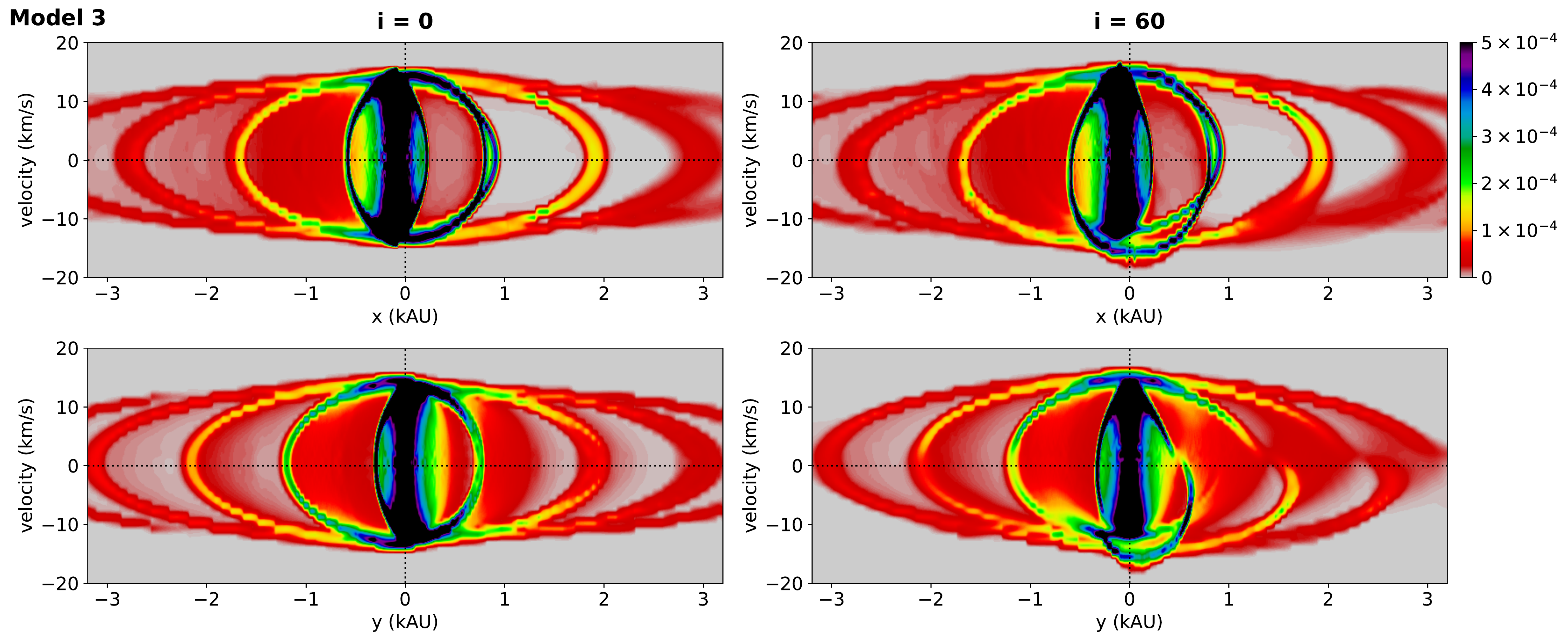}
  \caption{\label{fig:pv3}
    Same as Fig.\ref{fig:pv1} but for Model 3 (eccentric binary orbit).
  }
\end{figure*}

The position-velocity (P-V) diagram is a useful analysis tool for examining the kinematic structure along the line of sight. In Figures\,\ref{fig:pv1} and \ref{fig:pv3}, we present, for illustrative purposes, the P-V diagrams of Model 1 and Model 3 respectively, along the $x$- and $y$-axes at two different inclinations (e.g., 0\arcdeg\ and 60\arcdeg). Note that the $x$-axis is the line of nodes of the orbit, i.e., the inclination axis.

The spiral-shell pattern, induced by a circular-orbit motion, when viewed face-on ($i=0\arcdeg$; left panels of Figure\,\ref{fig:pv1}), has symmetric P-V diagrams along the velocity axis in all position angles (represented by the $x$- and $y$-axes in this figure; see Figure\,\ref{fig:pv0x00} for all position angles\footnote{A position angle (PA) is defined as an angle measured from North ($+y$) to the East ($-x$).}). In contrast, the symmetry  along the velocity axis is broken for non-zero inclinations (right panels of Figure\,\ref{fig:pv1}). Although in the inclined views the redshift-blueshift symmetry is imperfect, the arm pattern in the P-V diagram along the line of nodes of the orbit (i.e., $x$-axis) is smoothly connected in a spiral-like shape. The degree of redshift-blueshift symmetry and smoothly connected shape in the P-V diagram (when cut across the line of nodes of the orbit) result from the fact that the central stripe, corresponding to the systemic velocity, reflects the condition at the orbital plane and its immediate blue-/red-shifted parts originate from the nearly mirror-symmetric conditions above/below the orbital plane. On the other hand, the arm pattern in the P-V diagram along the $y$-axis (perpendicular to the line of nodes) is broken into regions with relatively lower densities.

In the analysis of high resolution (interferometric) observations, we can make use of the properties described above and determine the line of nodes of the orbit to be perpendicular to the position angle at which the P-V diagram presents the most distorted and most disconnected pattern (see Figure\,\ref{fig:pv0x60} for the P-V diagrams at all position angles in the $i=60\arcdeg$ case of Model 1). Meanwhile, the velocity corresponding to pattern breaks changes with respect to the inclination angle (compare e.g., the first panels of Figures\,\ref{fig:pv0x00}--\ref{fig:pv0x90}), from which one can obtain a hint for the orbital inclination of the binary. The variation of the P-V shapes among all different position angles can also provide some clues about the inclination, as the face-on case has the least (zero) variation across the position angle.

The eccentric binary orbital motion creates additional structures in the P-V diagrams, as shown in Figure\,\ref{fig:pv3}. In the face-on case ($i=0\arcdeg$; left panels), the major difference from the corresponding circular-orbit case is the split of the pattern at $x\sim0.8$\,kAU near $v\sim0$\,\kmps. This has also been noted in the previous subsections as a signature of an eccentric orbit (see also Figure\,\ref{fig:pv8x00} for all position angles). In the $i=60\arcdeg$ case (right panels), the split occurs over almost all (especially negative) velocities with an intertwinement occurring at $v\sim-5\,\kmps$, which we now identify as a signature of an eccentric orbit with a non-zero inclination. The velocity difference between the two branches near $x\sim0$ and $y\sim0$ is about 3.5\,\kmps\ (the difference between $-12.0\,\kmps$ and $-15.5\,\kmps$) in the $i=60\arcdeg$ case of Model 3. Such a velocity difference at the coordinate center increases with higher inclination (compare e.g., the middle panels of Figures\,\ref{fig:pv8x00}--\ref{fig:pv8x90}). Therefore, this characteristic can be used in the analysis of observational data to determine the inclination angle of the binary orbit. In the P-V diagram along the $y$-axis (perpendicular to the line of nodes) for $i=60\arcdeg$ of Model 3, detachment of the pattern near $v\sim0$\,\kmps\ is also found. Hence, the velocity at which the pattern is severed depends on the inclination angle (compare the first panels of Figures\,\ref{fig:pv8x00}--\ref{fig:pv8x90}) in a similar manner to the trend found in the circular-orbit model. 

Many of the other features found from the channel maps (see the previous subsection) are reconfirmed in the P-V diagrams for all position angles. For example, a comparison between Figures\,\ref{fig:pv0x00}--\ref{fig:pv0x90} and Figures\,\ref{fig:pv0o00}--\ref{fig:pv0o90} at the same inclination confirms the thicker arm width of Model 2 compared to that of Model 1. A similar arm thickening is found from Model 3 to Model 4, as can be seen in Figures\,\ref{fig:pv8x00}--\ref{fig:pv8o90}. Also found is that the P-V diagrams of Model 4, regardless the position angle for the P-V cut, present the richness of features along $+x$ (see Figures\,\ref{fig:pv8o00}--\ref{fig:pv8o90}), representing the turbulent characteristics in the direction toward the pericenter of the mass-losing star.

\begin{figure*} 
  \plotone{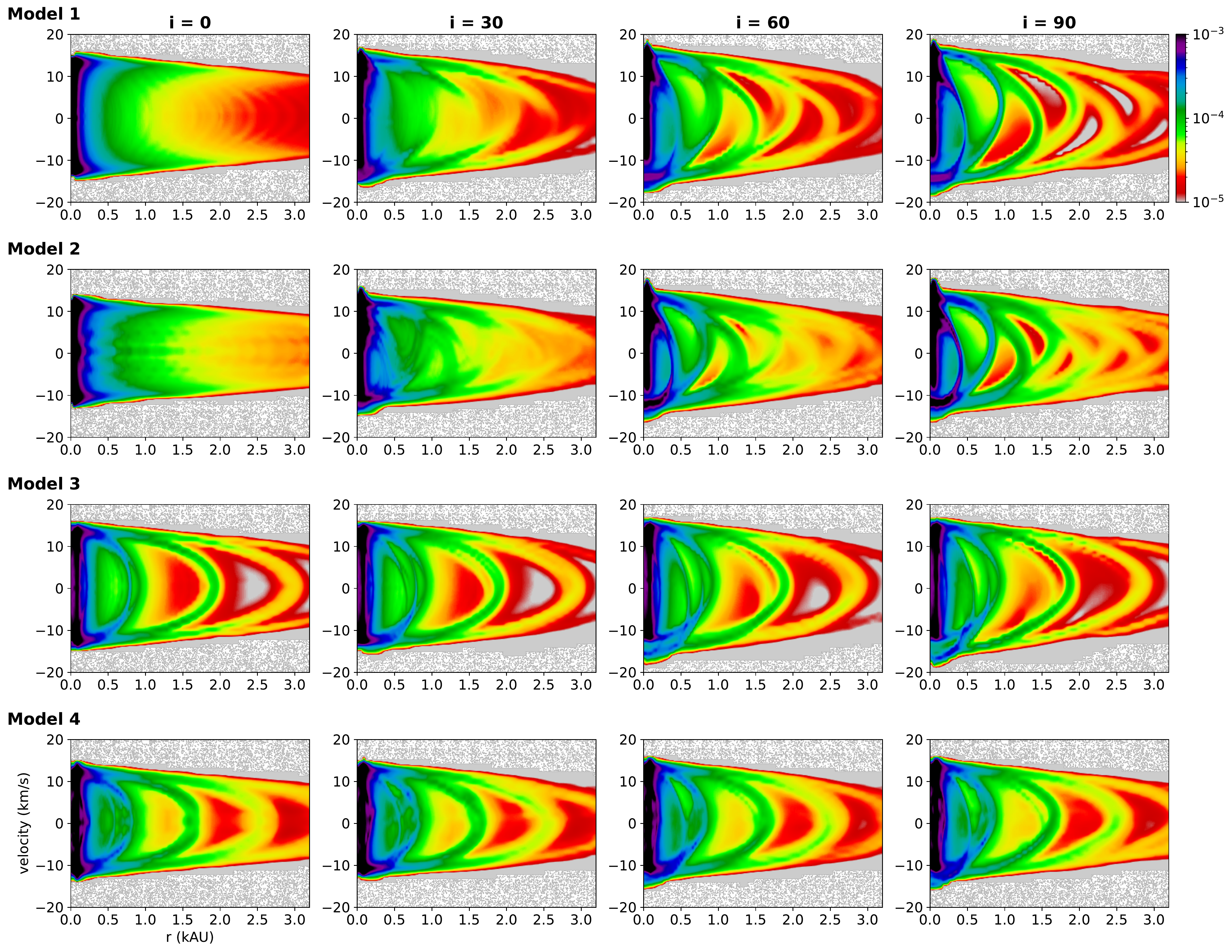}
  \caption{\label{fig:rvd}
    Azimuthally-averaged radius-velocity diagrams for Model 1 to 4 (from top to bottom) at inclination angles $i=0\arcdeg$, 30\arcdeg, 60\arcdeg, and 90\arcdeg\ (from left to right). Color bar labels in units of \columndensity.
  }
\end{figure*}

In Figure\,\ref{fig:rvd}, we present the radius-velocity (R-V) diagrams, equivalent to a P-V diagram azimuthally averaged over a 360\arcdeg-sector. The face-on case of the circular-orbit model does not show any pattern in the R-V diagram because of the regular change of the spiral in each channel. On the other hand, the eccentric-orbit binary model has repeated parabolic-shape patterns in its R-V diagram, even in the face-on view ($i=0\arcdeg$). The column density within a channel along the parabolic shape is high at the highest absolute velocity and lowest at the systemic velocity ($v=0\,\kmps$). This implies that the column density at the systemic velocity channel is the most evenly distributed along the radial distance from the center of mass of the binary system.

The R-V diagrams of the circular-orbit models (Models 1 and 2) for the inclined cases reveal thin structures in tilted parabolic-like shapes with the maximum extents at about 0.5, 1.0, 1.5, 2.0, 2.5, and 3.0 kAU at the blueshifted, redshifted, blueshifted, redshifted, blueshifted, and redshifted sides, respectively. However, the shifts of velocities at which the parabolic-like shapes reach their maximum extents are not evident in the eccentric-orbit models (Models 3 and 4). We note that the maximum extents of the parabolic-like patterns are nearly equally spaced in the circular-orbit models, but are not equally spaced in the eccentric-orbit models. This distinction is attributed to the bifurcation of the pattern shown in each channel, originating from the variation of orbital speeds of the binary stars along their orbit. As described previously for the P-V diagrams, the R-V diagrams in the highly inclined cases of Model 3 ($i=60\arcdeg$ and 90\arcdeg) also clearly show at $r=0$ the velocity difference between the split branches. This difference introduces a velocity gap between the blueshifted branch ($v\le-15.5\,\kmps$) and a major portion of the dense region ($v\ge-12.0\,\kmps$).

\subsection{Angle-radius diagrams}\label{sec:trd}

The pattern shapes depend on the orbital eccentricity and inclination of the system, as described in the previous sections. These dependences can be easily visualized using angle-radius diagrams. In particular, Figure\,\ref{fig:trd} illustrates the key features (see below) for two sample cases (namely, Model 1 in a face-on view and Model 3 at an inclination $i=60\arcdeg$) displayed at three (zero and blue/redshifted) velocity channels.

A circular-orbit model viewed face-on (Model 1; left of Figure\,\ref{fig:trd}) reveals a perfect Archimedean spiral pattern about the center of mass of the binary system at each channel, corresponding to the straight lines in the angle-radius diagrams. Note that the coordinate center is at the center of mass of the binary system, not at the position of the mass-losing star. The very central blob ($r<0.3$\,kAU), which shows the quasi-hydrostatic equilibrium region around the mass-losing star, appears wavy in the angle-radius diagram because of a mismatch of the coordinate center from the center of the blob.

In an analysis of actual observations, it is difficult to determine the center of mass of the binary system. Therefore, the position of the mass-losing star would be often set as the coordinate center. In the Appendix, we display the full sets of angle-radius diagrams for the four models in the coordinates with the origin at the position of the mass-losing star (see Figures\,\ref{fig:tr0x00}--\ref{fig:tr8o90}). In these coordinates, the central blob around the mass-losing star is located below a horizontal (instead of wavy) line in the angle-radius diagram. Even the Archimedean spiral pattern in the circular-orbit model for a face-on view has a slight undulation with respect to a straight increasing/decreasing line in the angle-radius diagram in these coordinates.

While one winding of the systemic-velocity pattern in a circular orbit model always satisfies a single-valued function in the angle-radius plot, the corresponding arm pattern in Model 3 for a binary in a highly eccentric orbit clearly departs from a single-valued function due to the bifurcation (compare the panels in the middle row of Figure\,\ref{fig:trd}). Specifically, a small gap exists between the two branches at $\Theta\sim\pi$ (and $r\sim1$\,kAU) in the systemic velocity channel of Model 3. With an increasing absolute velocity, the gap between the outer and inner branches becomes smaller and finally the two branches are entangled, as shown in the $-10\,\kmps$ channel, with a reversed order in distances from the system center (see also Figure\,\ref{fig:tr8x60}). The  different propagation speeds of these two branches have already been described in Section\,\ref{sec:chm} based on changes over channel maps in the $xy$-coordinates. Here, the reversed order of the two branches of the bifurcated pattern is better visualized in the angle-radius diagrams. The width of the gap at the systemic velocity and, consequently, the velocity for the switch of the order of the two branches depend on the binary orbital phase.

\subsection{Dependence on orbital phase and the line of nodes}\label{sec:pas}

\begin{figure*} 
  \plotone{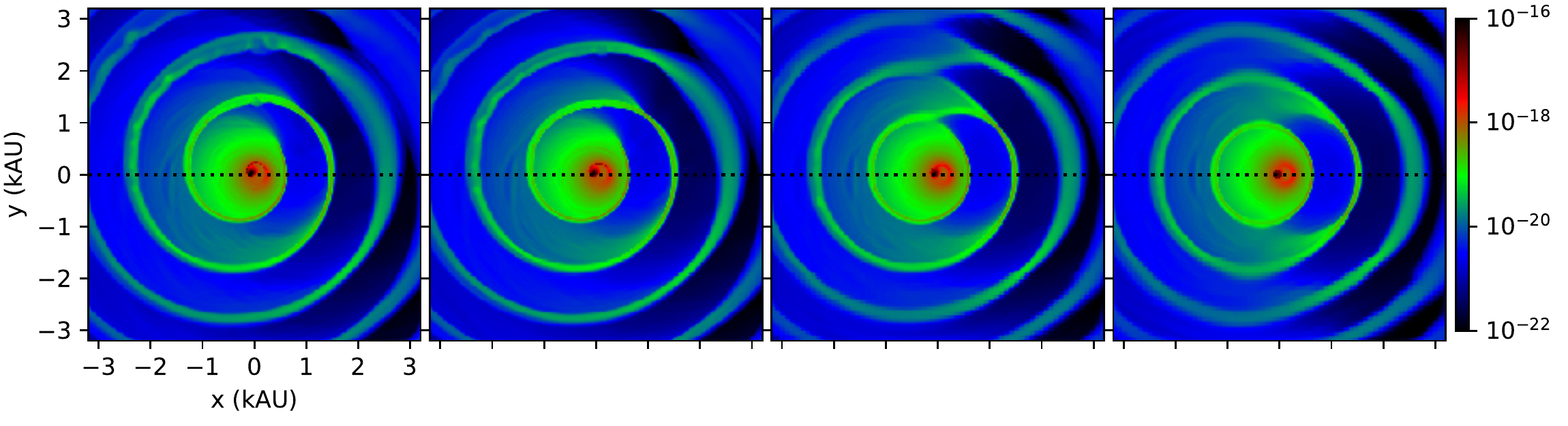}
  \plotone{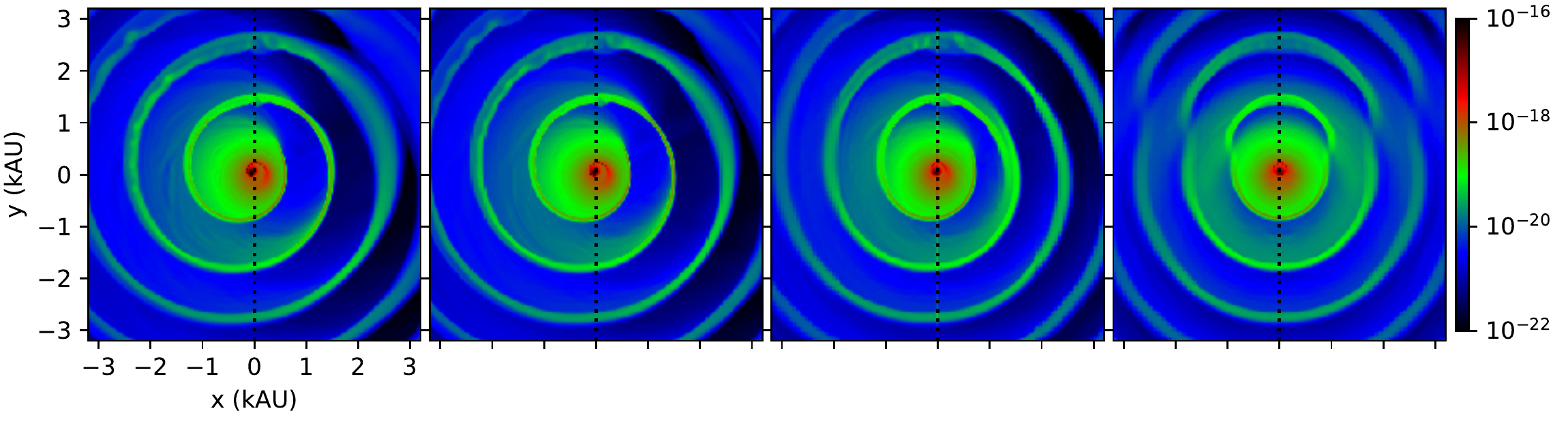}
  \caption{\label{fig:xiy}
    Midplane density slice for Model 3 with the mass-losing star located at the position angle of 55\arcdeg, on the way from its pericenter to apocenter. The coordinate center is at the center of mass of the binary system. The position angle is measured from $+y$-axis in the counterclockwise direction. Colorbar labels the density in units of \dunit\ in logarithmic scale. The size of displayed image domain is 6.4\,kAU. {\it Upper} panels exhibit the midplane density for the cases of orbit inclinations $i=0\arcdeg$, 30\arcdeg, 60\arcdeg, and 90\arcdeg\ (from left to right) with respect to the $x$-axis as the line of nodes of the binary
    orbit (horizontal dotted line). {\it Lower} panels present the corresponding cases with their line of nodes at the $y$-axis (vertical dotted line).
  }
\end{figure*}

\begin{figure*} 
  \plotone{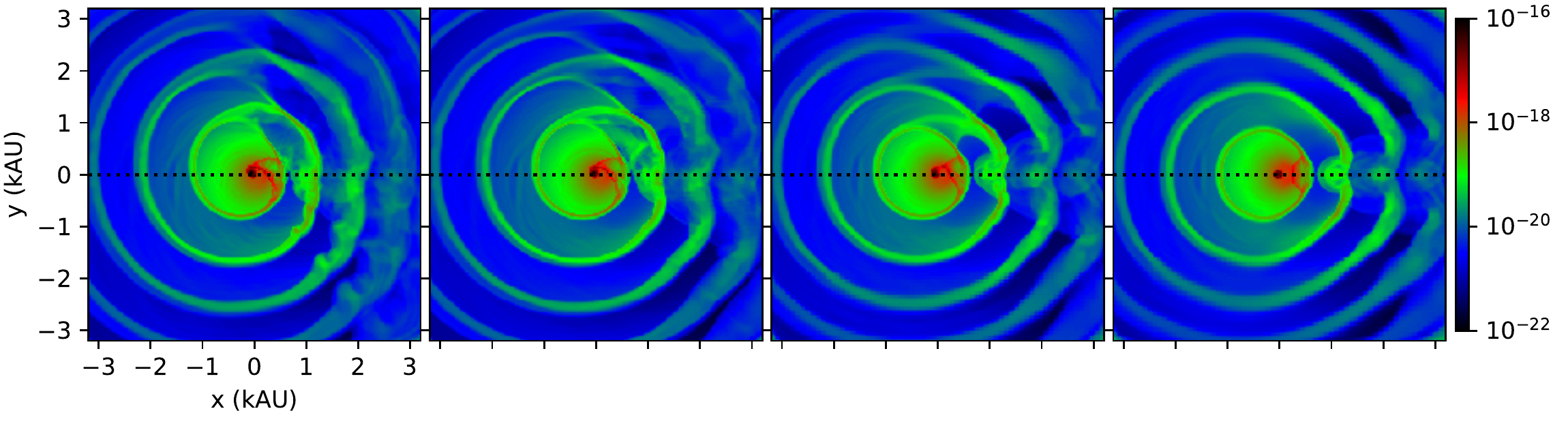}
  \plotone{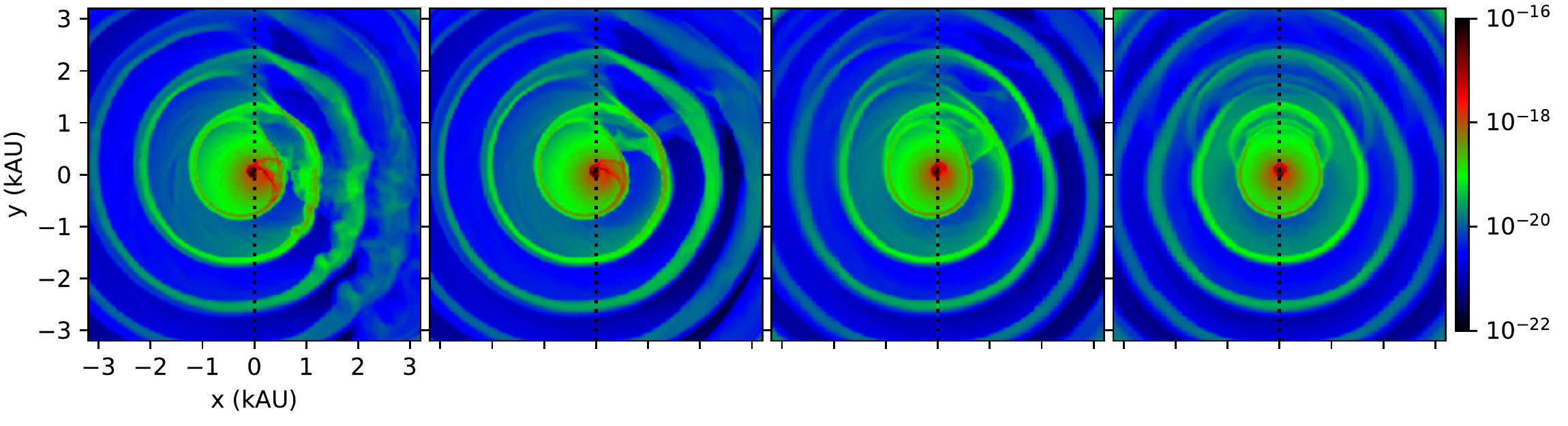}
  \caption{\label{fig:oiy}
    Same as Fig.\,\ref{fig:xiy} but for Model 4.
  }
\end{figure*}

Unlike the other figures in this paper where the snapshots correspond to the mass-losing star located at its apocenter in the $-x$-axis (at the position angle of 90\arcdeg), in this subsection we present the snapshots at a different orbital phase. In Figures\,\ref{fig:xiy}--\ref{fig:oiy}, the mass-losing star is approaching its apocenter (at the position angle of $\sim55\arcdeg$), where significant differences in the features from those at the apocenter can be noticed. The upper panels of Figures\,\ref{fig:xiy} and \ref{fig:oiy} exhibit the midplane density slices of Model 3 and Model 4, respectively, for the orbit inclinations of 0\arcdeg, 30\arcdeg, 60\arcdeg, and 90\arcdeg\ (left to right panels) with respect to the $x$-axis. This is the same as for Figures\,\ref{fig:e8p} and \ref{fig:e8s} except for the orbital phase of the star. The first spiral winding (reddish color; $\ga 10^{-18}\,\dunit$) in Figure\,\ref{fig:xiy}, equivalent to the inner branch of bifurcated spiral in Figure\,\ref{fig:e8p}, hasn't evolved to closely approach the outer branch, i.e., the innermost part of the next winding. The change of distance between the inner and outer branches confirms the different propagation speeds between the two branches. In the upper panels of Figure\,\ref{fig:oiy} for Model 4, the attachment of the outer branch of the first winding to the inner branch of the second winding is still in place, resulting in the ring-like shapes in addition to the spiral-like shapes. 

The lower panels of Figures\,\ref{fig:xiy} and \ref{fig:oiy} present the midplane views when the inclination axis (line of nodes of the orbit) is set to be the $y$-axis. The $y$-axis is perpendicular to the apocenter-pericenter line ($x^\prime$). The inclination angles remain the same; 0\arcdeg, 30\arcdeg, 60\arcdeg, and 90\arcdeg\ (from left to right). The overall spiral shape is elongated along the inclination axis, marked by dotted line (see e.g., the third column of Figure\,\ref{fig:xiy} for the $i=60\arcdeg$ views). The split of the spiral pattern in the midplane image for $i=60\arcdeg$ with respect to $y$-axis, again, in the $+x$ direction is of particular interest. It is due to the coexistence of flows with different velocities, originating from the rapid change of orbital motion of the mass-losing star near its pericenter, located at $+x^\prime$. Therefore, the direction of the orbit pericenter can be inferred from the position angle showing the aligned pattern split, as well as the line of nodes of the orbit from the position angle showing the pattern elongation.

In this paper we present only two representative examples for the inclination axis (the line connecting the pericenter and apocenter, and the line perpendicular to it), but the inclination axis can be located in an arbitrary direction in an observed image. Hence, the appearance of a  variety of pattern shapes can be understood within the framework of the binary-induced spiral-shell patterns viewed at different orientations.

\section{DISCUSSION}\label{sec:dis}

\subsection{Application to observed sources}\label{sec:obs}

\begin{figure*} 
  \plotone{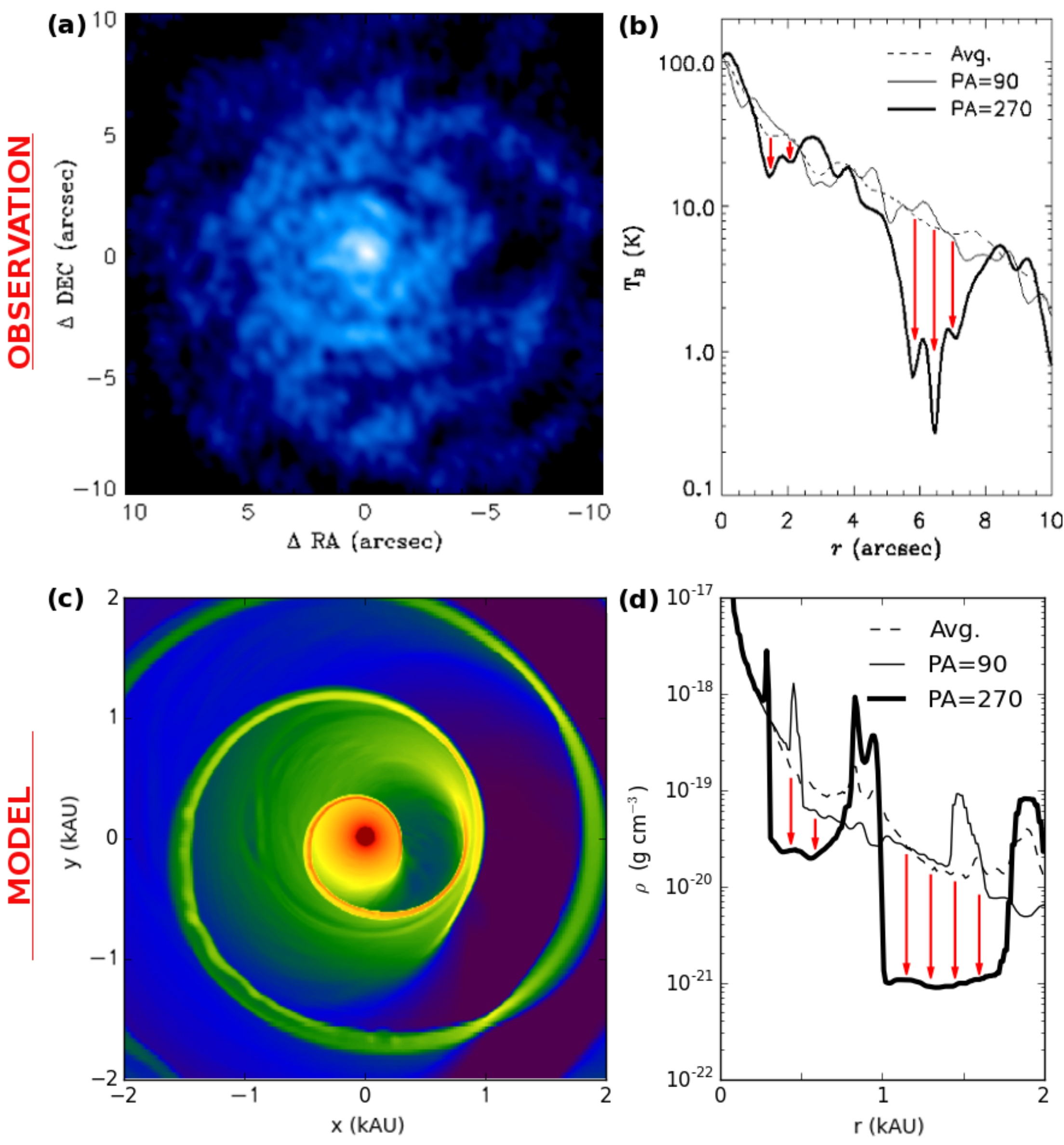}
  \caption{\label{fig:cit}
    One-sided interarm depression of CIT 6 in the CO $J=2-1$ molecular line observed using the Submillimeter Array \citep{kim15} displayed (a) in the central channel map and (b) in the brightness temperature profile at the position angle of 270\arcdeg\ compared to the opposite side. This feature is well mimicked by (c) our Model 3 viewed face-on (same as Fig.\,\ref{fig:e8p} leftmost panel but with a vertical flip) and (d) its density profiles at the same position angles as in (b).
  }
\end{figure*}

Recently, binaries in eccentric orbits, as illustrated below, have more frequently been invoked for explaining the complex recurrent structures in the outflowing circumstellar envelopes of AGB stars. \citet{kim15} suggested that the asymmetric, one-sided lack of interarm CO emission in CIT 6 hints at an eccentric orbit of the central binary stars (see also Figure\,\ref{fig:cit} for comparison with our eccentric binary model). Similiarly, the HC$_3$N emission is also asymmetrically biased toward the eastern side of CIT 6 \citep[see also][]{din09,cla11}. On the other hand, based on the spiral bifurcation pattern found in the density distribution in numerical calculations, \citet{kim17}, for the first time, proposed that the bifurcation in the circumstellar spiral pattern of AFGL 3068 is produced by an eccentric-orbit binary. In these interpretations, it was assumed that the density could be an adequate proxy for the emission. However, the fact that the  bifurcation pattern is absent in CIT 6 and a significant asymmetry in the interarm emission is lacking in AFGL 3068 indicates that additional factors enter in determining their observational appearance. For example, CIT 6 may resemble the case illustrated in Figure\,\ref{fig:xiy}, namely, the absence of a bifurcated feature can be understood if the system is at an orbital phase away from the apocenter. Alternatively, the inclination axis may be far from the axis described by the pericenter-apocenter. For AFGL 3068, \citet{kim17} employed only the long baseline interferometric observations, for which the level of extended interarm emission could not be appropriately measured. Full synthesis imaging of the system is required in order to recover the short-/zero-spacing information.

Recent high-resolution high-sensitivity ALMA observations unveiled complex and disordered, but still repeatable, circumstellar features surrounding several AGB stars. Among them, IRC+10216, in particular, clearly shows a mixture of repeated ring- and spiral-like patterns \citep[see e.g., Figure\,10 of][]{gue18}. Such a mixed appearance is similar to the pattern shape shown in Figures\,\ref{fig:e8p}--\ref{fig:e8s} (Model 3 at $i=60\arcdeg$ or Model 4 at $i<90\arcdeg$), possibly indicating an eccentric orbit for a centrally located binary. Similarly mixed structural appearances are also clearly seen in the ALMA observations of the circumstellar patterns of EP Aqr \citep{hom18}, II Lup \citep{lyk18}, OH 26.5+0.6/OH 30.1-0.7 \citep{dec19}, and GX Mon (Randall et al., {\it submitted}). The detailed modeling based on the relative arrangements of the pattern segments, interarm gaps, and turbulent features is subject to a future investigation for the individual sources.

\subsection{Measurement of expansion velocity and mass-loss rate}\label{sec:vel}

Inferring the expansion velocities and mass-loss rates of AGB winds has been the central focus of many single-dish or low-resolution interferometric observations \citep[e.g.][]{kna82,lou93,sch02,deb10}. The stellar radial velocity and wind velocity are conventionally determined through parameterized spectral profile fitting, as noted earlier, under the assumption that the outflowing wind is spherically symmetric and flowing at a constant velocity. The mass loss rate is estimated by incorporating this inferred wind velocity together with the measured extent of the gas emission (which is regulated by the interstellar radiation field) and the fractional abundance of the molecular tracer.

If the AGB star resides in a binary system, the observed spectral width of its expanding wind depends on the inclination angle of the binary orbit, as shown in the integrated spectra of our binary models (see bottom row of Figures\,\ref{fig:px0}--\ref{fig:ps8}). Given that the fluid velocity is the fastest in the orbital plane, the observed (line-of-sight) velocity width is the greatest in the edge-on view. For example, in our Model 3, the full width of the face-on spectrum is 30\,\kmps, while that of the edge-on view is 36\,\kmps. Hence, the effect of inclination itself could readily lead to a 20\% uncertainty in the traditional measurement of expansion velocity and consequently in the derived mass-loss rate.

From various line surveys for IRC+10216, which include molecular lines emanating only in the very central region, the spatial profile of the wind velocity has been explored and the increase of the velocity with radius has been interpreted on the basis of several dust acceleration models \citep[e.g.,][]{pat11,dec15}. However, if IRC+10216 is a binary system as proposed by some authors \citep[e.g.,][]{dec15,cer15,kim15,soz17}, the interpretation of its velocity profile requires an additional consideration. As seen in the rightmost column of Figure\,\ref{fig:dtv} in the central 100\,AU region, the fluid velocity can be accelerated by the gravitational potential of the companion star. Such an acceleration near the companion is local, but can affect the velocity profile derived from the above method. Moreover, in the outer region, the wind is accelerated before reaching the spiral shocked region and its speed decreases during passage across the arm (see Figure\,\ref{fig:shc}). The gas velocity variation due to the shocked regions can be determined by high-resolution interferometric observations, but would appear as irregular deviations from an average value in low-resolution observations leading to its misinterpretation. 

\section{SUMMARY}\label{sec:sum}

Building upon our previous modeling work of an asymmetric wind from a mass-losing AGB star in a wide binary system, we expand our investigation with new 3-D numerical hydrodynamical simulations. Here, we focus on the dependence of the structural characteristics and appearances of the circumstellar wind/envelope on the orbital eccentricity, companion mass, and viewing angle of the system. Specifically, four distinct model setups, covering a combination of a circular ($e=0$) or an eccentric ($e=0.8$) orbit, and with or without the gravitational effects of the companion star on the stellar wind, are inspected. This exploration of these conditions provides diagnostic templates for comparing and interpreting observations of the recurrent features in circumstellar envelopes which are now commonly found to surround post-main sequence stars. We summarize our key findings as follows:

\begin{itemize} 
\item Overall a diversity of structural features, such as spiral, (half-)circles, and rings, are produced in the projected density distributions of these models (Section\,\ref{sec:den}). The mixture/combination of spiral and ring patterns are found to result from the eccentric orbit configuration and resemble features seen in many recent observations (Section\,\ref{sec:obs}).
\item Specifically, as demonstrated in Section\,\ref{sec:den}, the presence of bifurcation/splitting of spiral patterns and the asymmetric interarm contrasts in the density distribution are the most unique discriminating signatures of a binary system in an eccentric orbit (nearly independent of the inclination angle).
\item The bifurcation/splitting patterns can also exhibit prominently in the channel maps and the position-velocity diagrams obtained through spectroimaging of molecular line emission and serve as an additional kinematic affirmation of the eccentric binary model (Sections\,\ref{sec:chm} and \ref{sec:pvd}). Such bifurcation/splitting due to the motion of stars in eccentric orbits is further manifested in the R-V diagrams as several unevenly-spaced parabolic-like patterns and in the angle-radius diagrams as multiple branches
(Sections\,\ref{sec:pvd} and \ref{sec:trd}).
\item Another key signature of the eccentric binary orbit model is the asymmetric one-sided interarm contrast/depression toward the pericenter direction of the mass-losing star. It leads to not only higher arm-to-interarm contrasts in the 3-D volume density and temperature distributions (Section\,\ref{sec:dtv}), but also high column density contrast in the channel maps (Section\,\ref{sec:chm}). Potentially such highly asymmetric interarm contrasts can be preserved and identified toward the outer envelope even after the central AGB star has evolved to a PN (Section\,\ref{sec:den}).
\item When the binary companion with a non-negligible mass in the eccentric orbit case is considered, the gas flows become turbulent and the one-sided interarm density suppression become less significant (Section\,\ref{sec:den}). Nevertheless, the bifurcation pattern of the system remains prominent in the channel maps, signifying the eccentric nature of the binary orbit (Secton \ref{sec:chm}).
\item The P-V diagrams can be instrumental for determining the orbital line-of-nodes, which would be perpendicular to the position angle at which the corresponding P-V diagram displays the most distorted and disconnected pattern (Section\,\ref{sec:pvd}). In addition, the line of nodes can also be inferred from the central channel images, as an elongation of the apparent spiral-shell pattern and an alignment of knotty structures within the pattern occurring along such an axis (Section\,\ref{sec:den}).
\item For both circular and eccentric orbit binaries, the P-V diagram perpendicular to the line of nodes shows a blueshifted velocity gap at the coordinate center and breaks in the P-V pattern in the off-center region. As the inclination angle increases, the width of the velocity gap increases and the velocity at which the P-V pattern breaks approaches the systemic velocity (Section\,\ref{sec:pvd}).
\item In eccentric binary models, the P-V diagram along the line of nodes clearly shows two branches of the recurrent pattern. Their intertwinement indicates that the binary orbital plane is inclined and the velocity at which the intertwinement is observed increases with inclination (Section\,\ref{sec:pvd}).
\item The presence of the companion and the orbital motion in a binary system modify the outflowing wind velocity and consequently the overall molecular spectral profile, which would also depend on the inclination angle of the system. It is worthwhile to bear in mind such uncertainties when deriving the wind acceleration and mass loss rates (Sections\,\ref{sec:chm} and \ref{sec:vel}). \item Considering the 3-D physical parameters, the circular-orbit model results in monotonic variation of density, temperature and velocity along the orbital axis with no particular modulation resulting from the binary orbital motion. This is in contrast to the eccentric-orbit model in which the spiral-shell pattern extends to the orbital axis (Section\,\ref{sec:dtv}).
\end{itemize}

\acknowledgments
The numerical simulations presented here have been performed using the computing resources of the Theoretical Institute for Advanced Research in Astrophysics (TIARA) in the Academia Sinica Institute of Astronomy \& Astrophysics (ASIAA) and a high performance computing cluster at the Korea Astronomy and Space Science Institute (KASI). FLASH code is developed by the DOE-supported ASC/Alliance Center for Astrophysical Thermonuclear Flashes at the University of Chicago.

\clearpage
\newpage

\begin{figure*} 
  \plottwo{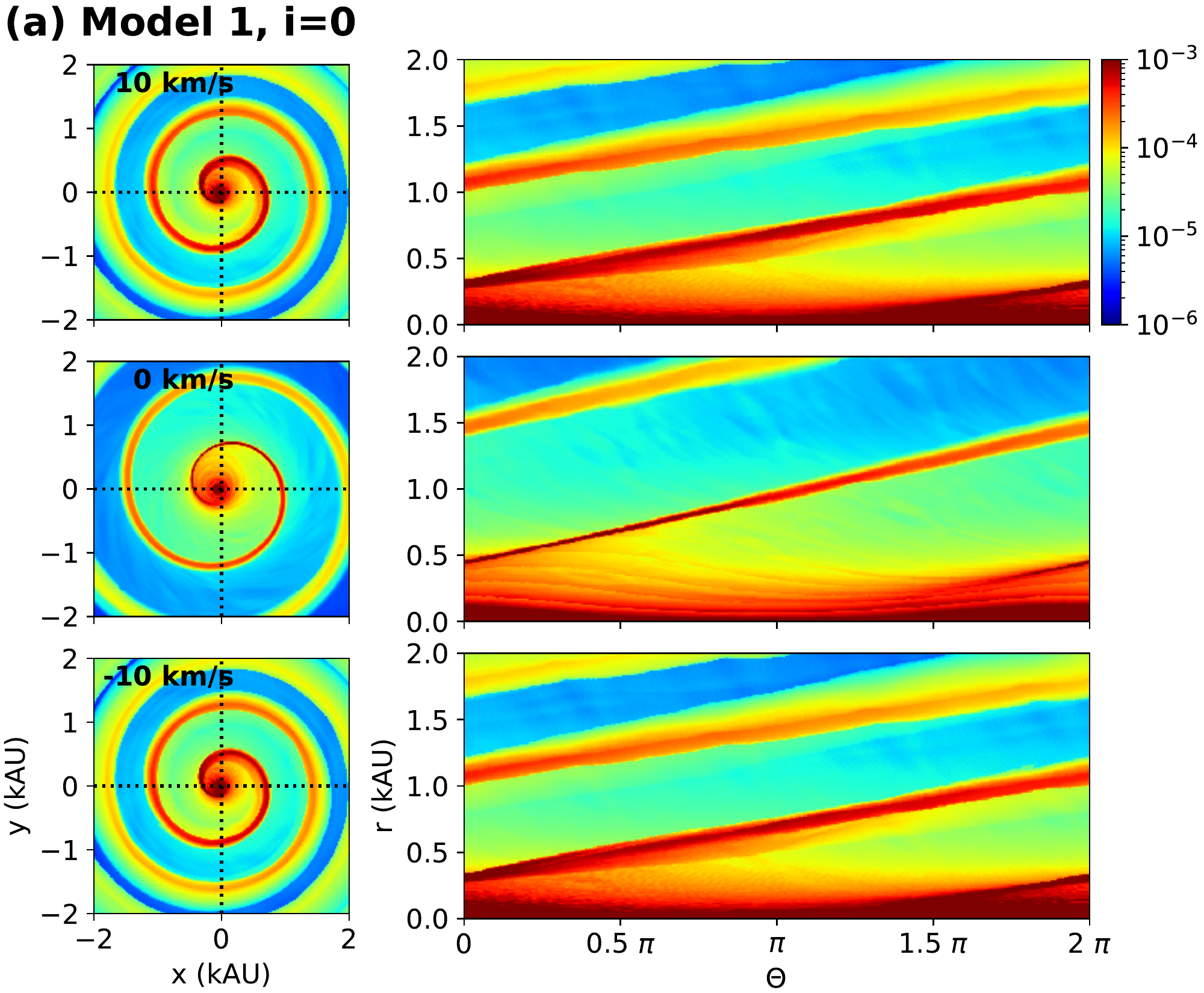}{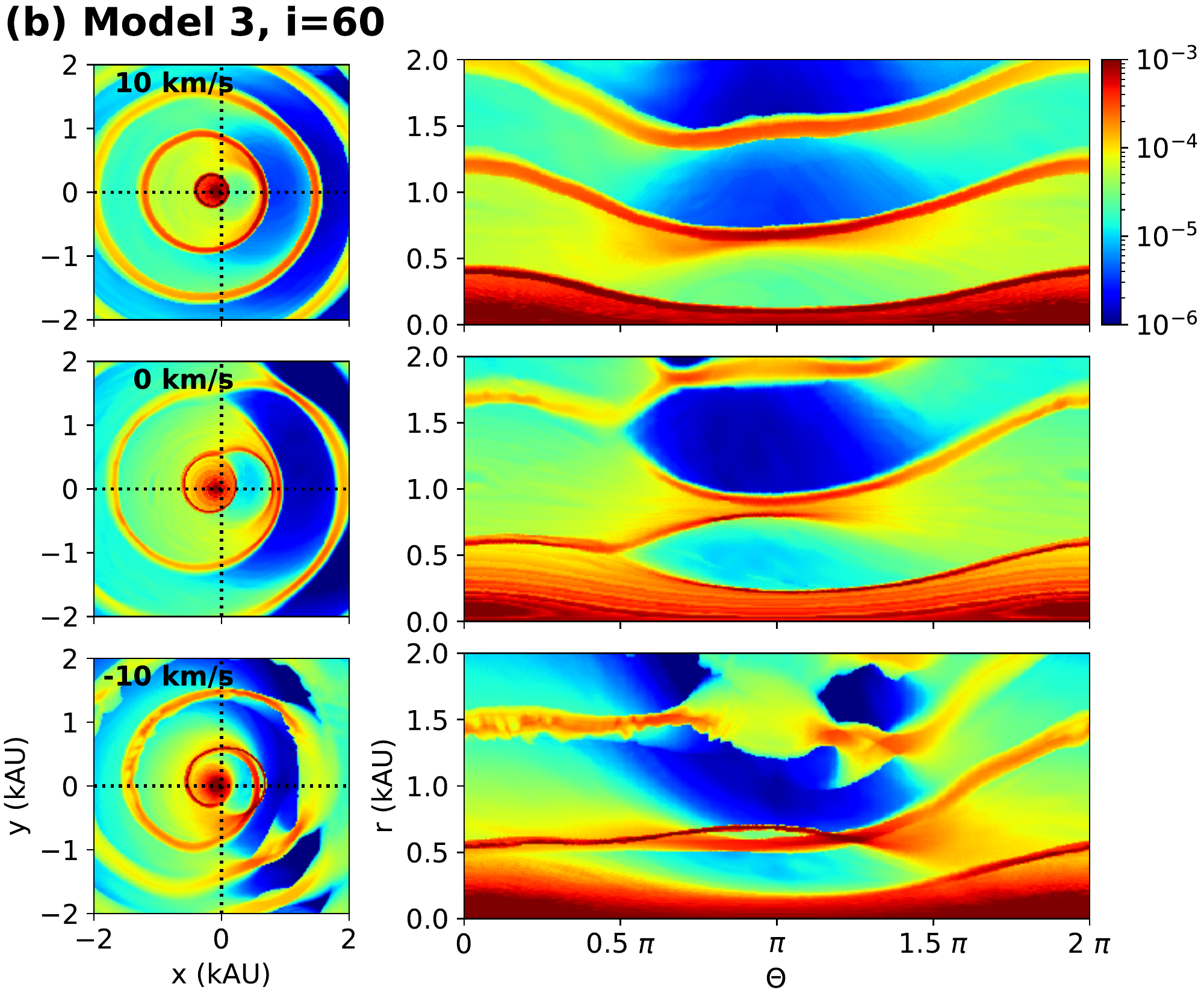}
  \caption{\label{fig:trd}
    Sample angle-radius diagrams. ({\it left}) Model 1 at inclination angle $i$ of 0\arcdeg\ for channel velocities of 10, 0, and $-10$\,\kmps\ (from top to bottom). ({\it right}) The corresponding diagram of Model 3 at an inclination angle of 60\arcdeg. Color bar labels the density in units of \dunit. The coordinate center is at the center of mass of the binary system. The full sets with the coordinate center at the mass-losing star are presented in the Appendix.
  }
\end{figure*}

\clearpage
\appendix

\section{Position-velocity diagrams in all directions}
\renewcommand\thefigure{A\arabic{figure}}
\setcounter{figure}{0}
\begin{sidewaysfigure} 
  \centering
  \includegraphics[width=\textwidth]{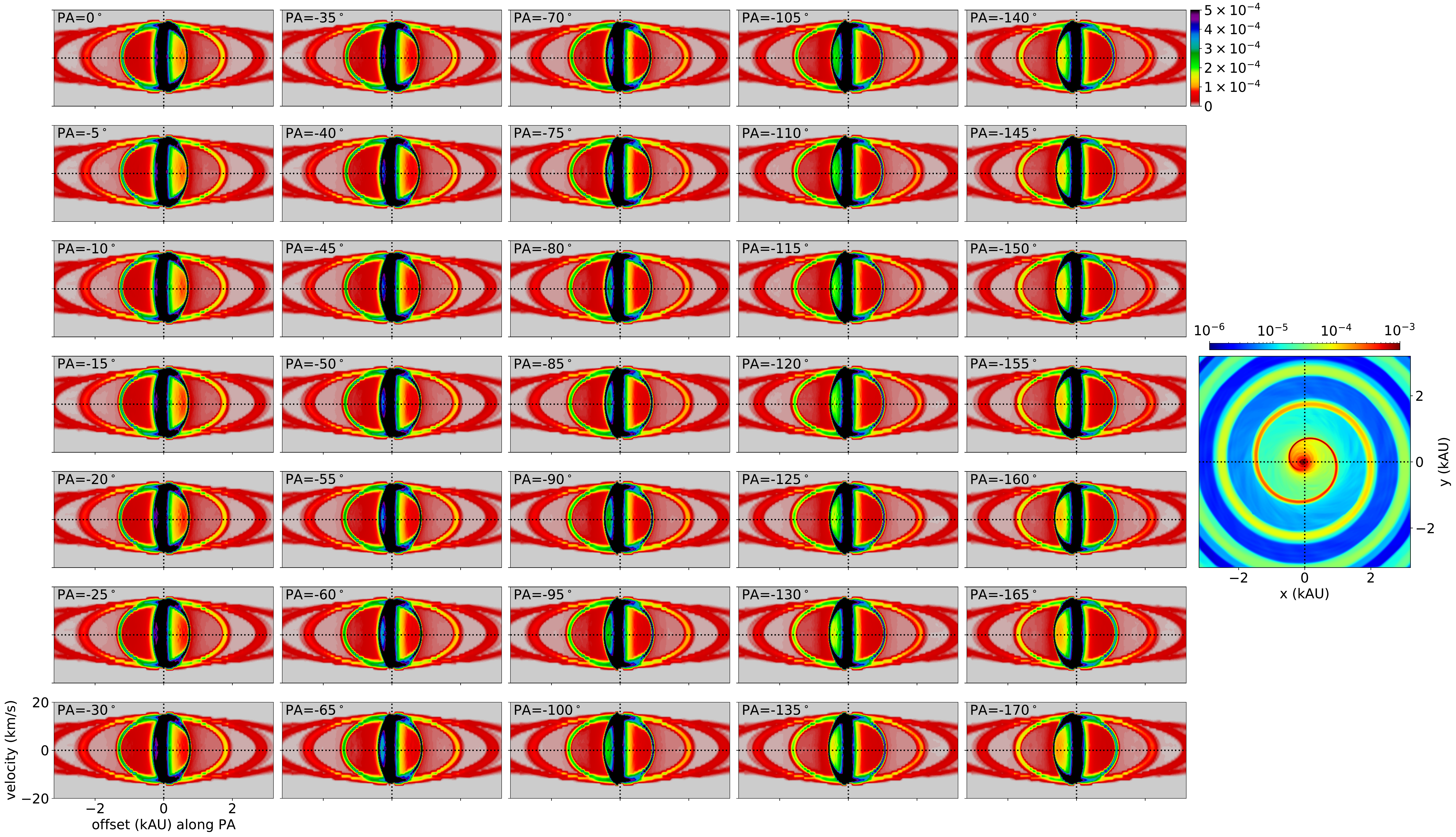}
  \caption{\label{fig:pv0x00}
    Position-velocity diagrams of Model 1 at inclination angle $i=0\arcdeg$ for position angle from 0\arcdeg\ to $-170\arcdeg$ (from top-left to bottom-right). Color bar labels in units of \columndensity. On the right of the position-velocity diagrams, the density slice at the midplane is displayed in units of \dunit.
  }
\end{sidewaysfigure}

\begin{sidewaysfigure} 
  \centering
  \includegraphics[width=\textwidth]{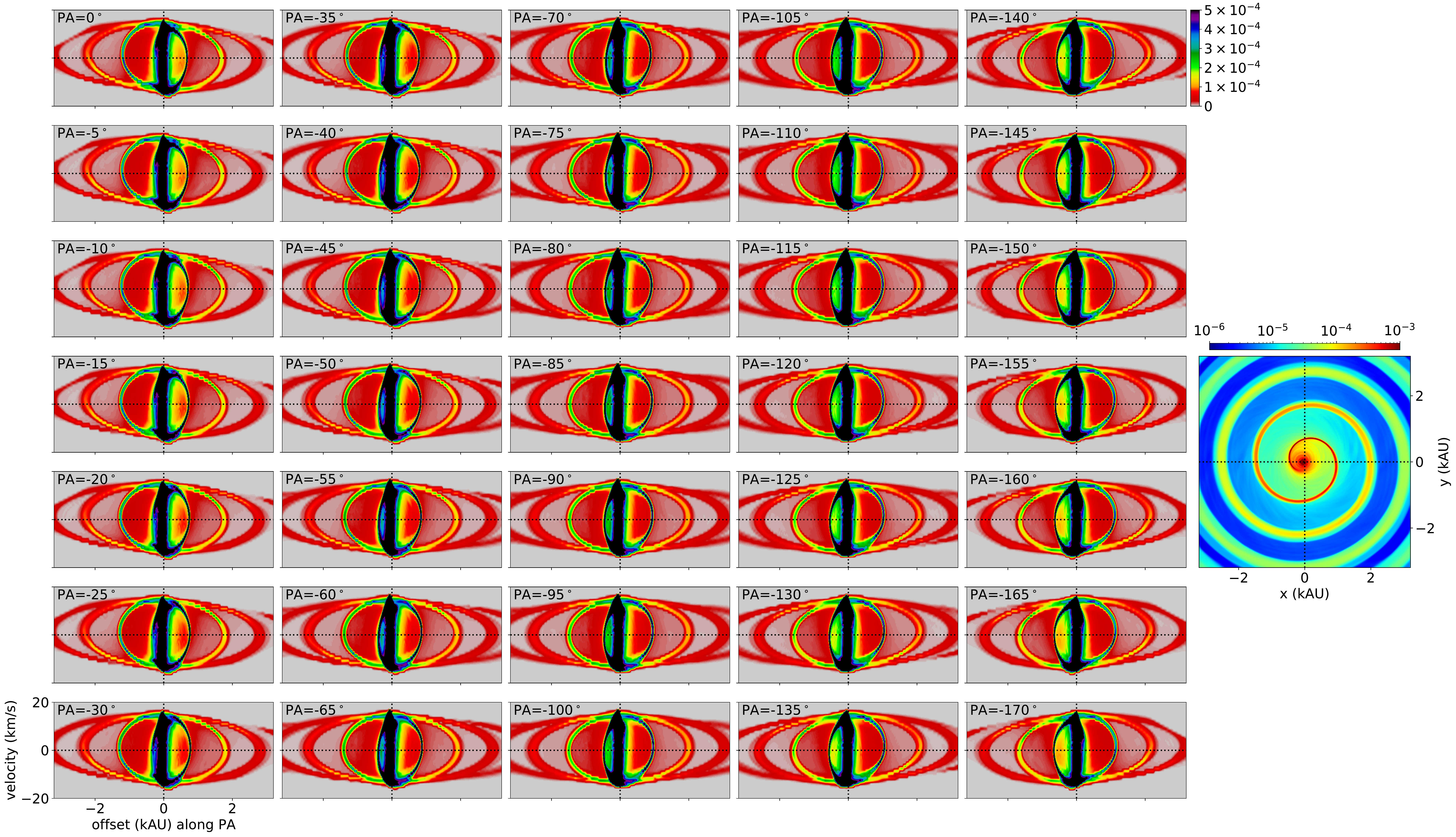}
  \caption{\label{fig:pv0x30}
    Same as Fig.\,\ref{fig:pv0x00} (Model 1) but for $i=30\arcdeg$.
  }
\end{sidewaysfigure}

\begin{sidewaysfigure} 
  \centering
  \includegraphics[width=\textwidth]{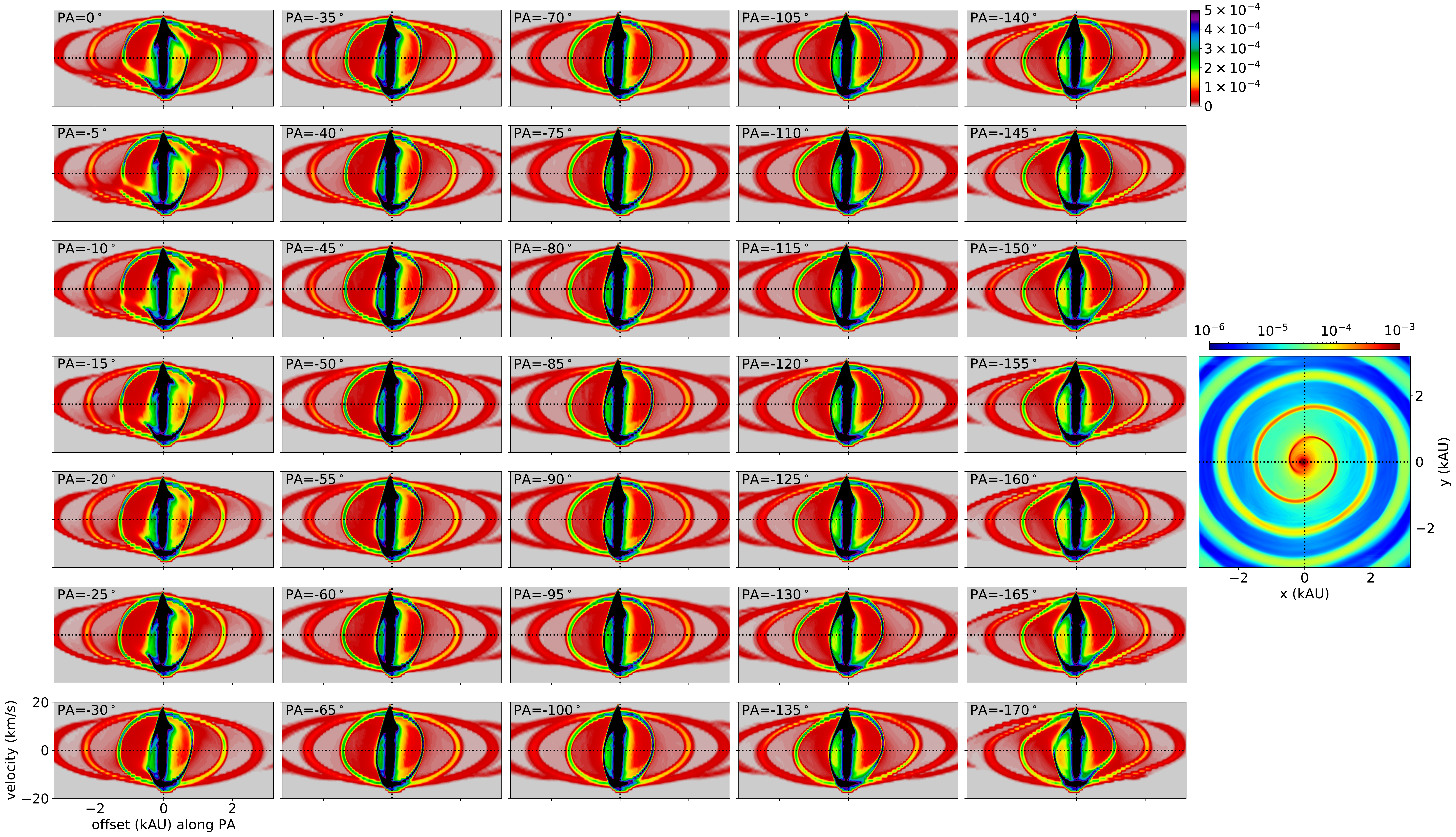}
  \caption{\label{fig:pv0x60}
    Same as Fig.\,\ref{fig:pv0x00} (Model 1) but for $i=60\arcdeg$.
  }
\end{sidewaysfigure}

\begin{sidewaysfigure} 
  \centering
  \includegraphics[width=\textwidth]{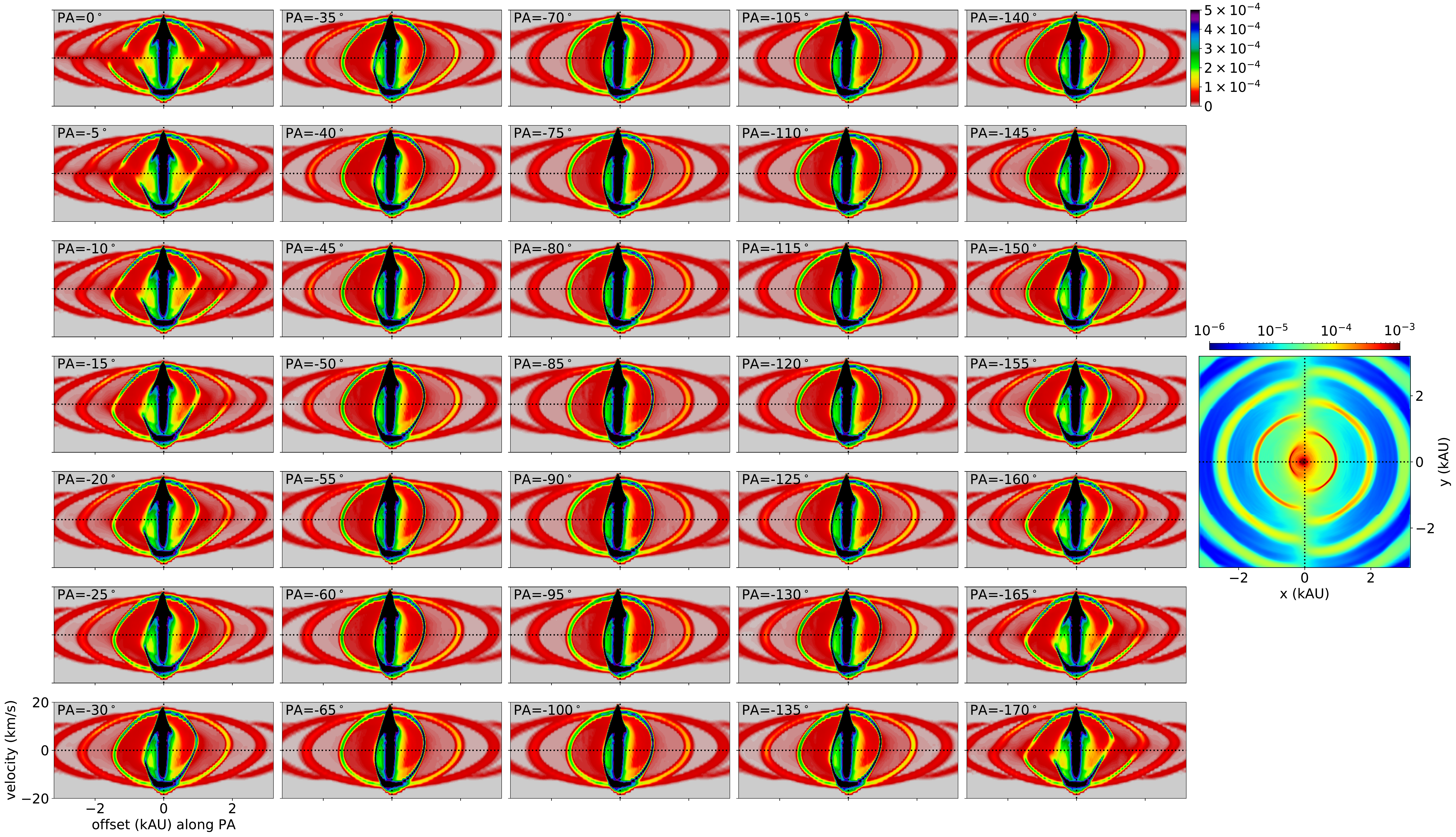}
  \caption{\label{fig:pv0x90}
    Same as Fig.\,\ref{fig:pv0x00} (Model 1) but for $i=90\arcdeg$.
  }
\end{sidewaysfigure}

\begin{sidewaysfigure} 
  \centering
  \includegraphics[width=\textwidth]{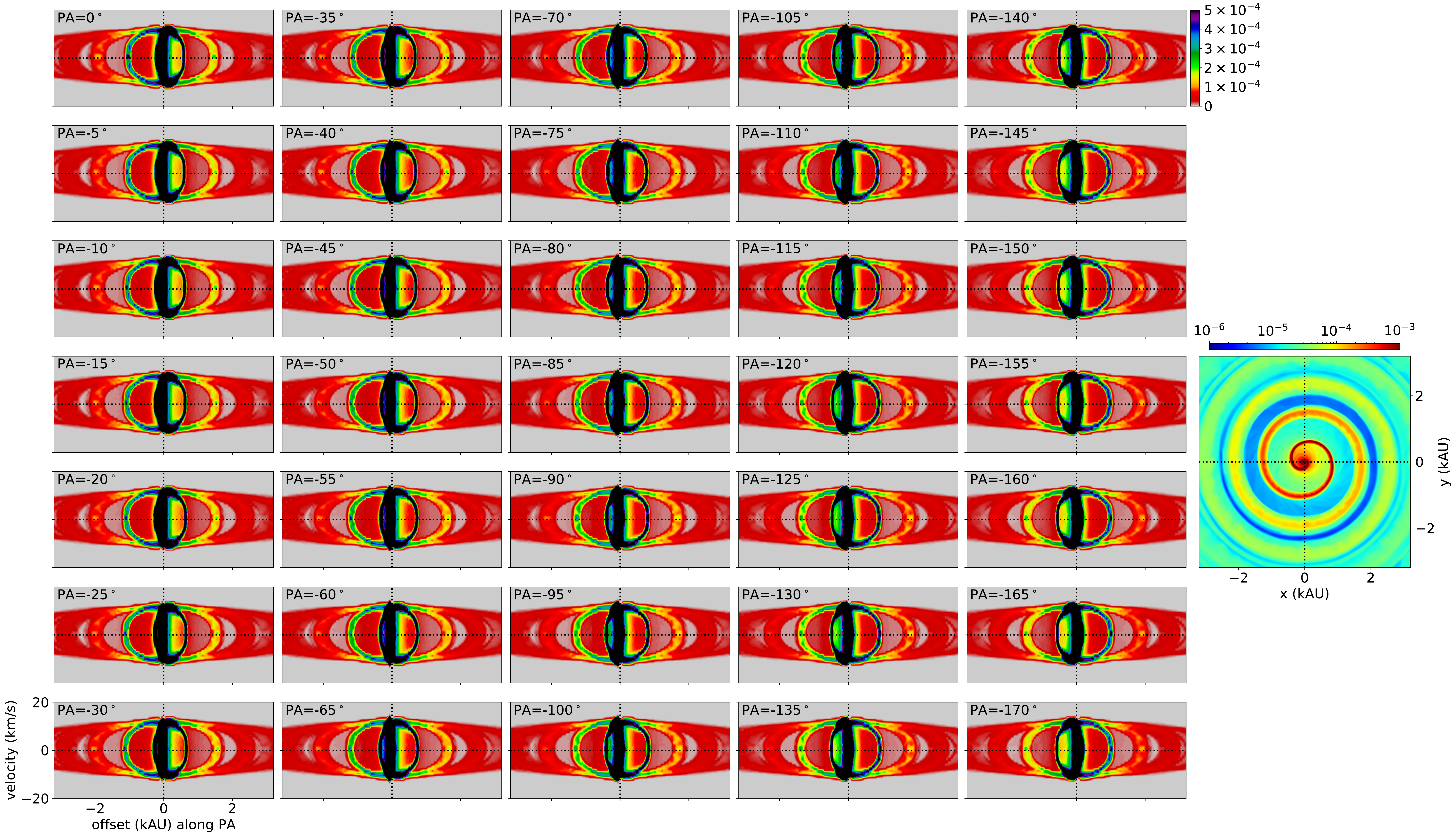}
  \caption{\label{fig:pv0o00}
    Position-velocity diagrams of Model 2 at inclination angle $i=0\arcdeg$ for position angle from 0\arcdeg\ to $-170\arcdeg$ (from top-left to bottom-right). Color bar labels in units of \columndensity. On the right of the position-velocity diagrams, the density slice at the midplane is displayed in units of \dunit.
  }
\end{sidewaysfigure}

\begin{sidewaysfigure} 
  \centering
  \includegraphics[width=\textwidth]{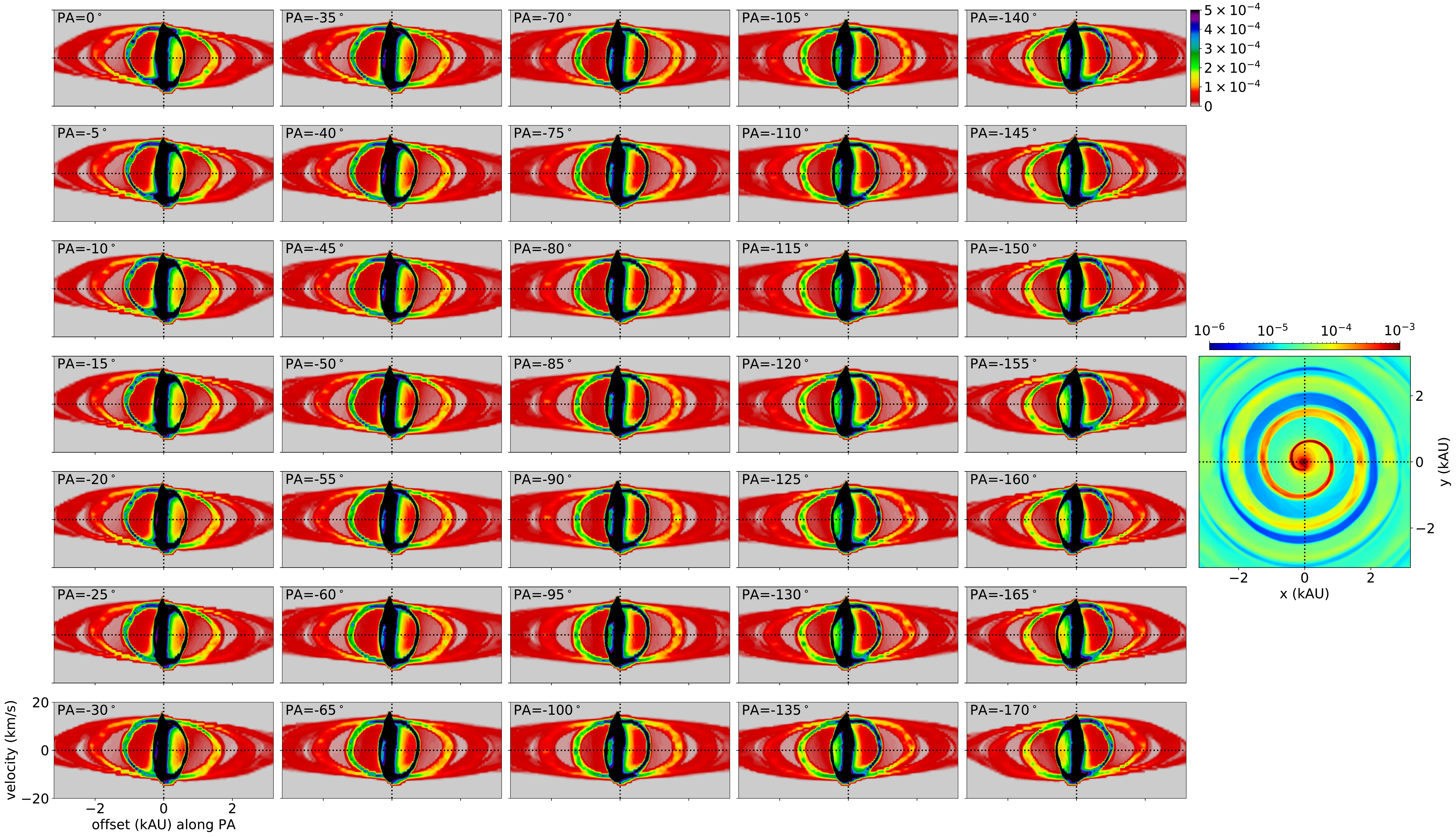}
  \caption{\label{fig:pv0o30}
    Same as Fig.\,\ref{fig:pv0o00} (Model 2) but for $i=30\arcdeg$.
  }
\end{sidewaysfigure}

\begin{sidewaysfigure} 
  \centering
  \includegraphics[width=\textwidth]{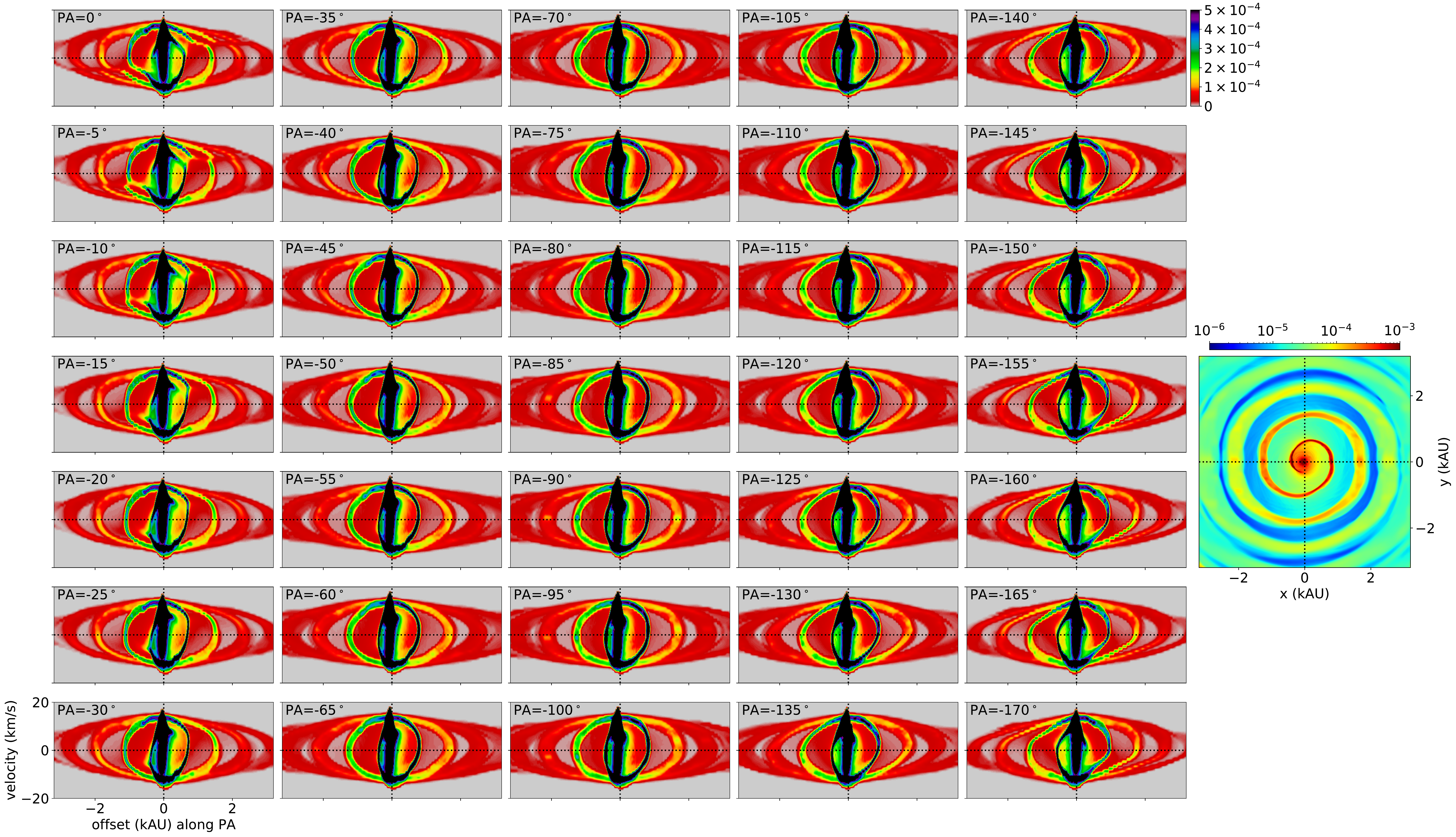}
  \caption{\label{fig:pv0o60}
    Same as Fig.\,\ref{fig:pv0o00} (Model 2) but for $i=60\arcdeg$.
  }
\end{sidewaysfigure}

\begin{sidewaysfigure} 
  \centering
  \includegraphics[width=\textwidth]{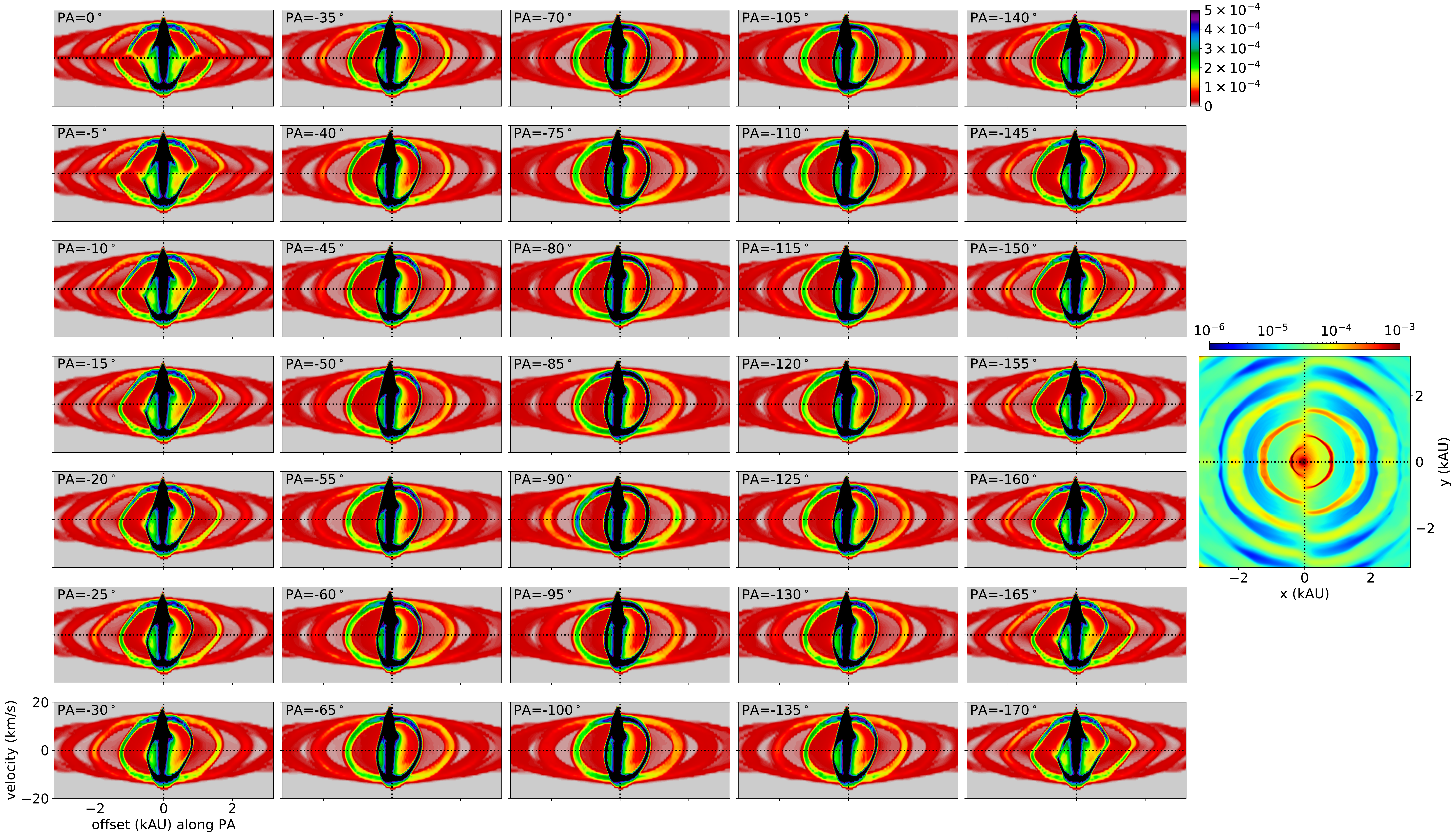}
  \caption{\label{fig:pv0o90}
    Same as Fig.\,\ref{fig:pv0o00} (Model 2) but for $i=90\arcdeg$.
  }
\end{sidewaysfigure}

\begin{sidewaysfigure} 
  \centering
  \includegraphics[width=\textwidth]{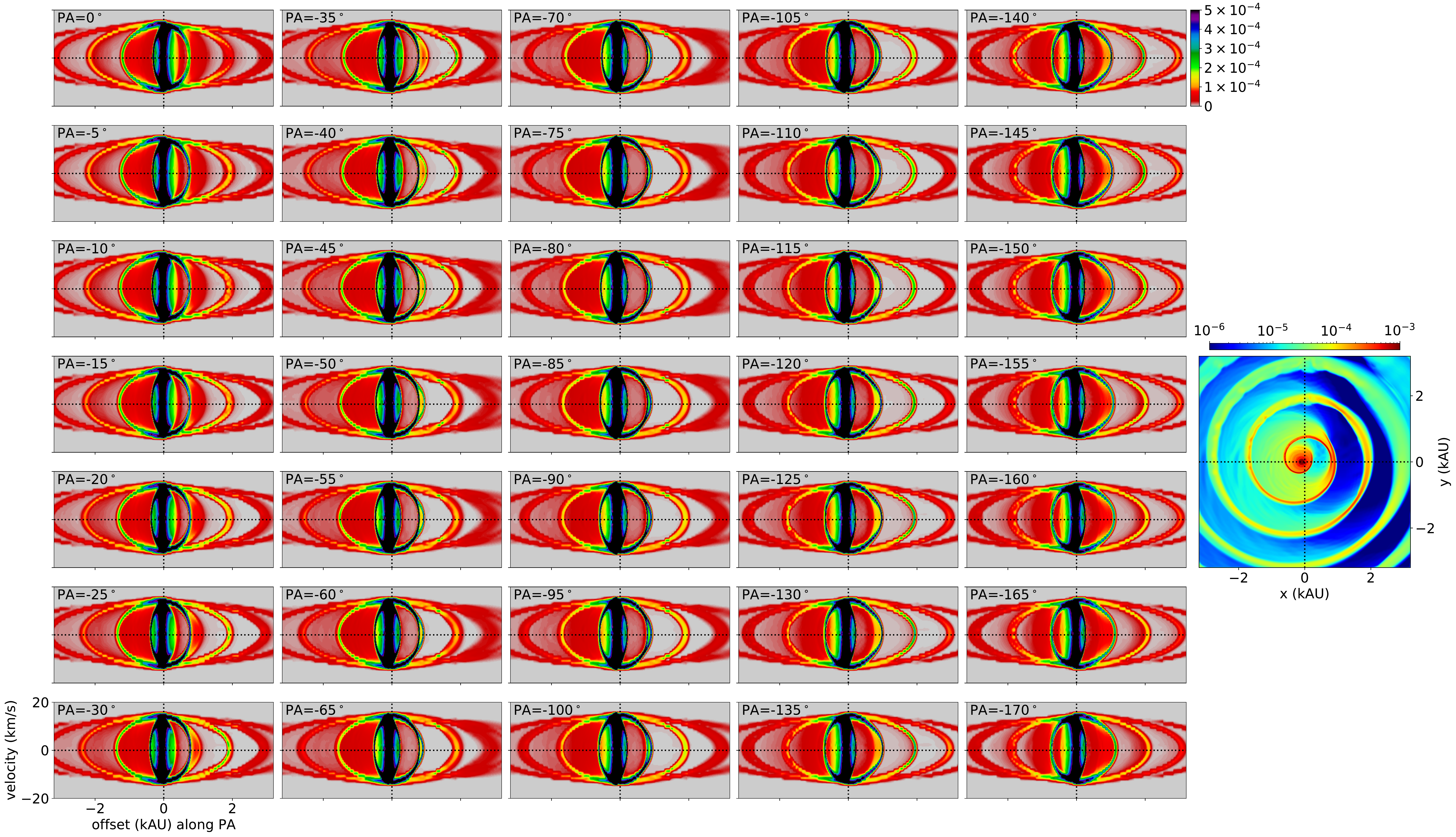}
  \caption{\label{fig:pv8x00}
    Position-velocity diagrams of Model 3 at inclination angle $i=0\arcdeg$ for position angle from 0\arcdeg\ to $-170\arcdeg$ (from top-left to bottom-right). Color bar labels in units of \columndensity. On the right of the position-velocity diagrams, the density slice at the midplane is displayed in units of \dunit.
  }
\end{sidewaysfigure}

\begin{sidewaysfigure} 
  \centering
  \includegraphics[width=\textwidth]{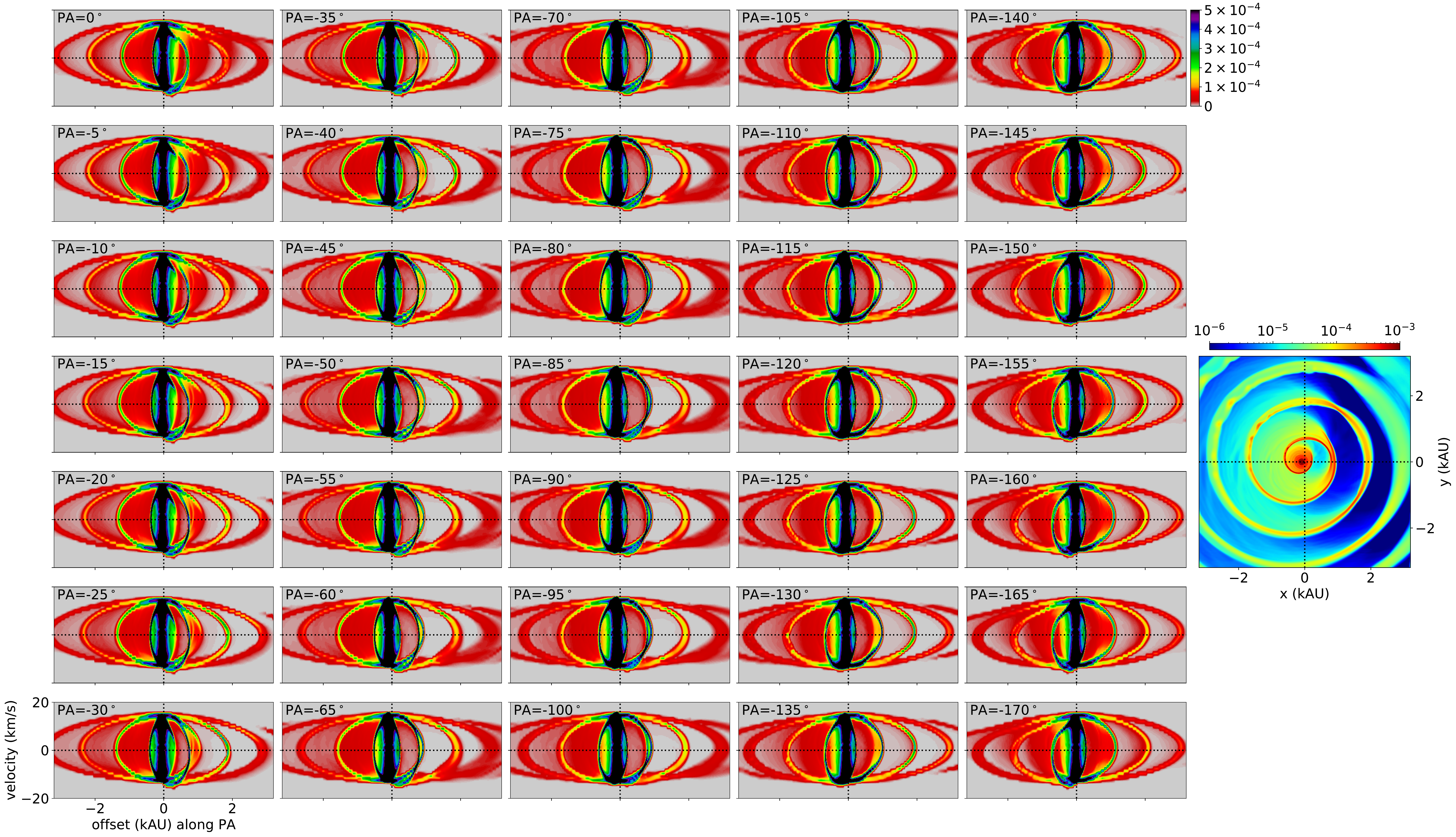}
  \caption{\label{fig:pv8x30}
    Same as Fig.\,\ref{fig:pv8x00} (Model 3) but for $i=30\arcdeg$.
  }
\end{sidewaysfigure}

\begin{sidewaysfigure} 
  \centering
  \includegraphics[width=\textwidth]{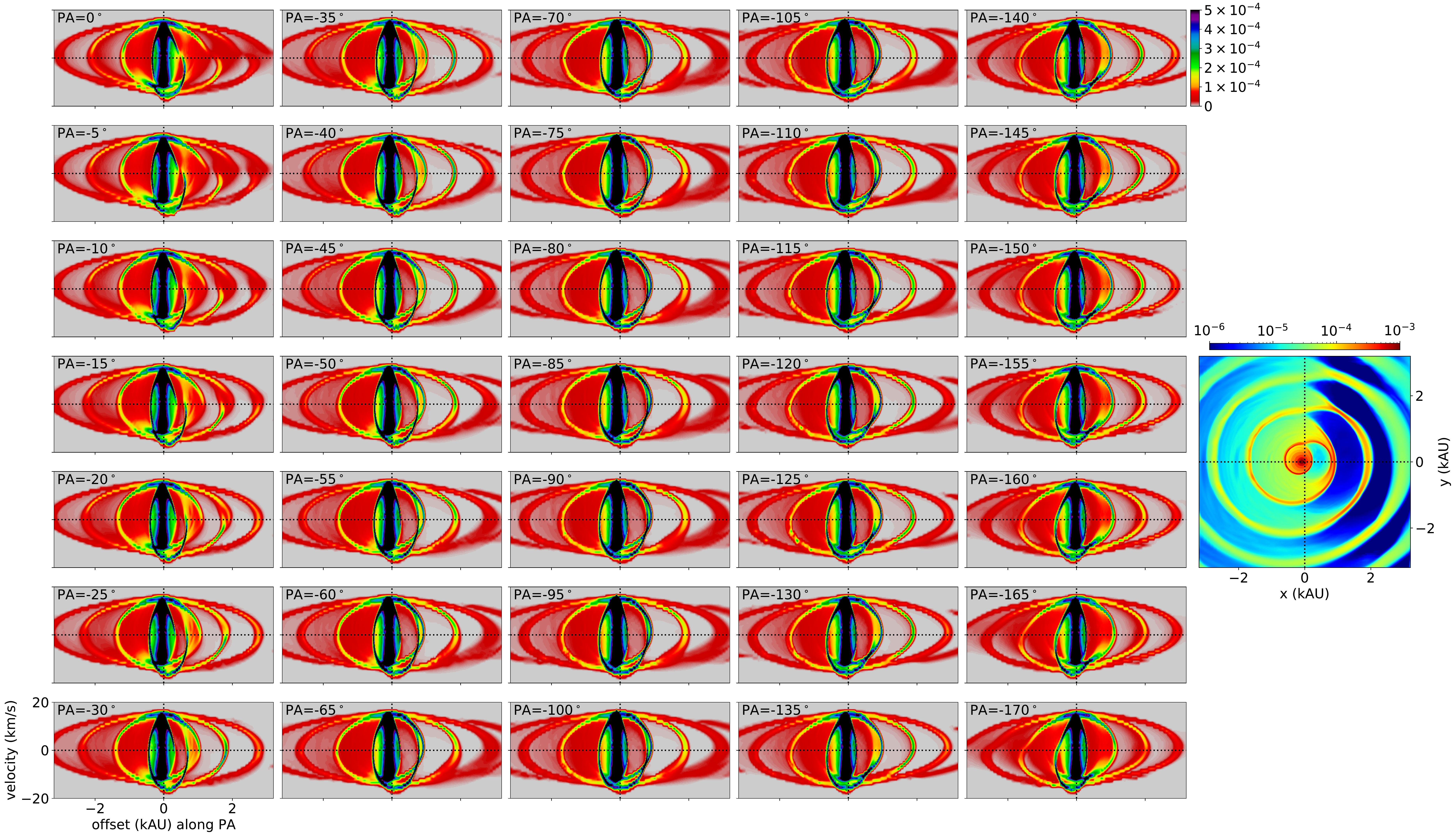}
  \caption{\label{fig:pv8x60}
    Same as Fig.\,\ref{fig:pv8x00} (Model 3) but for $i=60\arcdeg$.
  }
\end{sidewaysfigure}

\begin{sidewaysfigure} 
  \centering
  \includegraphics[width=\textwidth]{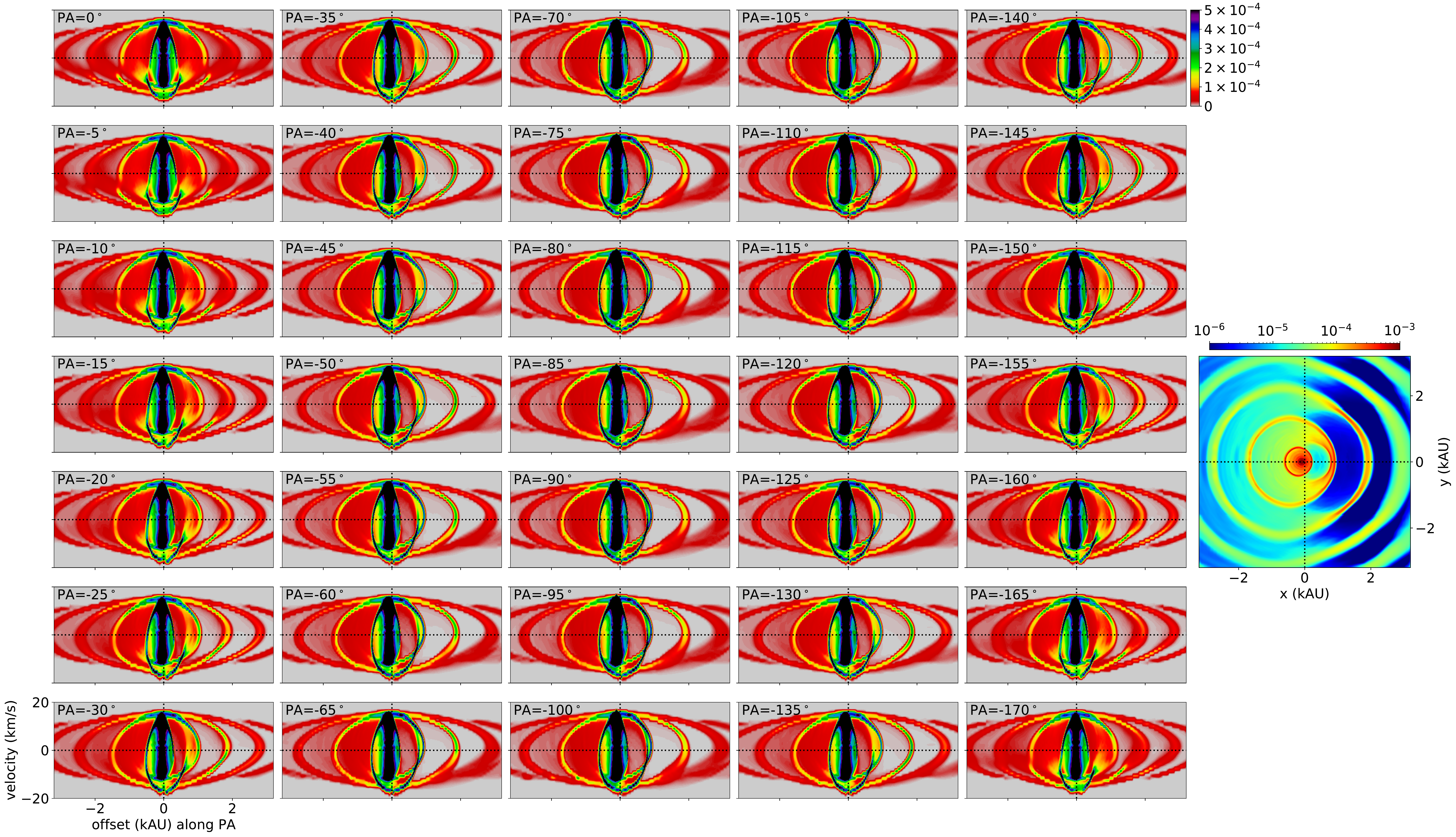}
  \caption{\label{fig:pv8x90}
    Same as Fig.\,\ref{fig:pv8x00} (Model 3) but for $i=90\arcdeg$.
  }
\end{sidewaysfigure}

\begin{sidewaysfigure} 
  \centering
  \includegraphics[width=\textwidth]{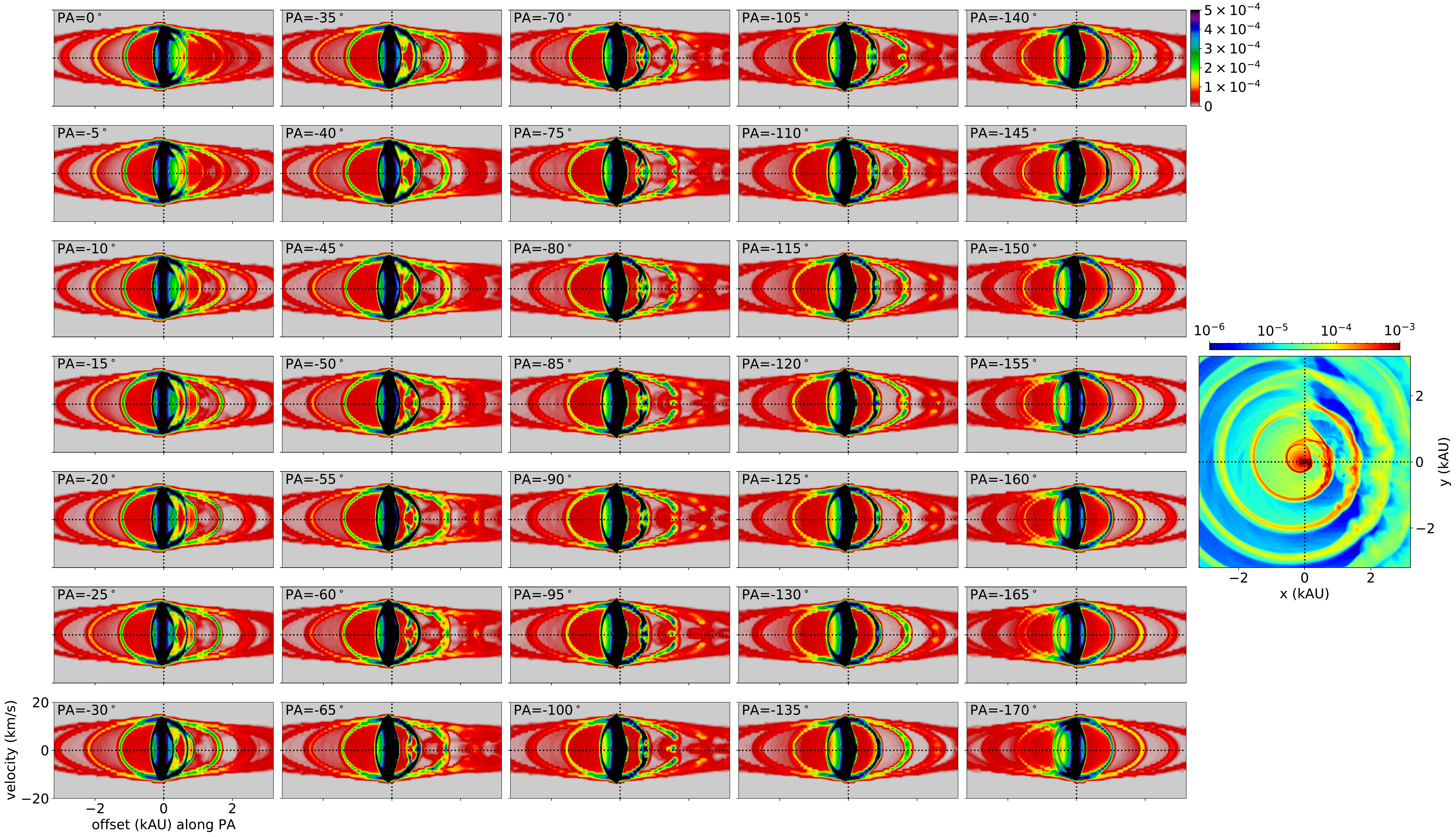}
  \caption{\label{fig:pv8o00}
    Position-velocity diagrams of Model 4 at inclination angle $i=0\arcdeg$ for position angle from 0\arcdeg\ to $-170\arcdeg$ (from top-left to bottom-right). Color bar labels in units of \columndensity. On the right of the position-velocity diagrams, the density slice at the midplane is displayed in units of \dunit.
  }
\end{sidewaysfigure}

\begin{sidewaysfigure} 
  \centering
  \includegraphics[width=\textwidth]{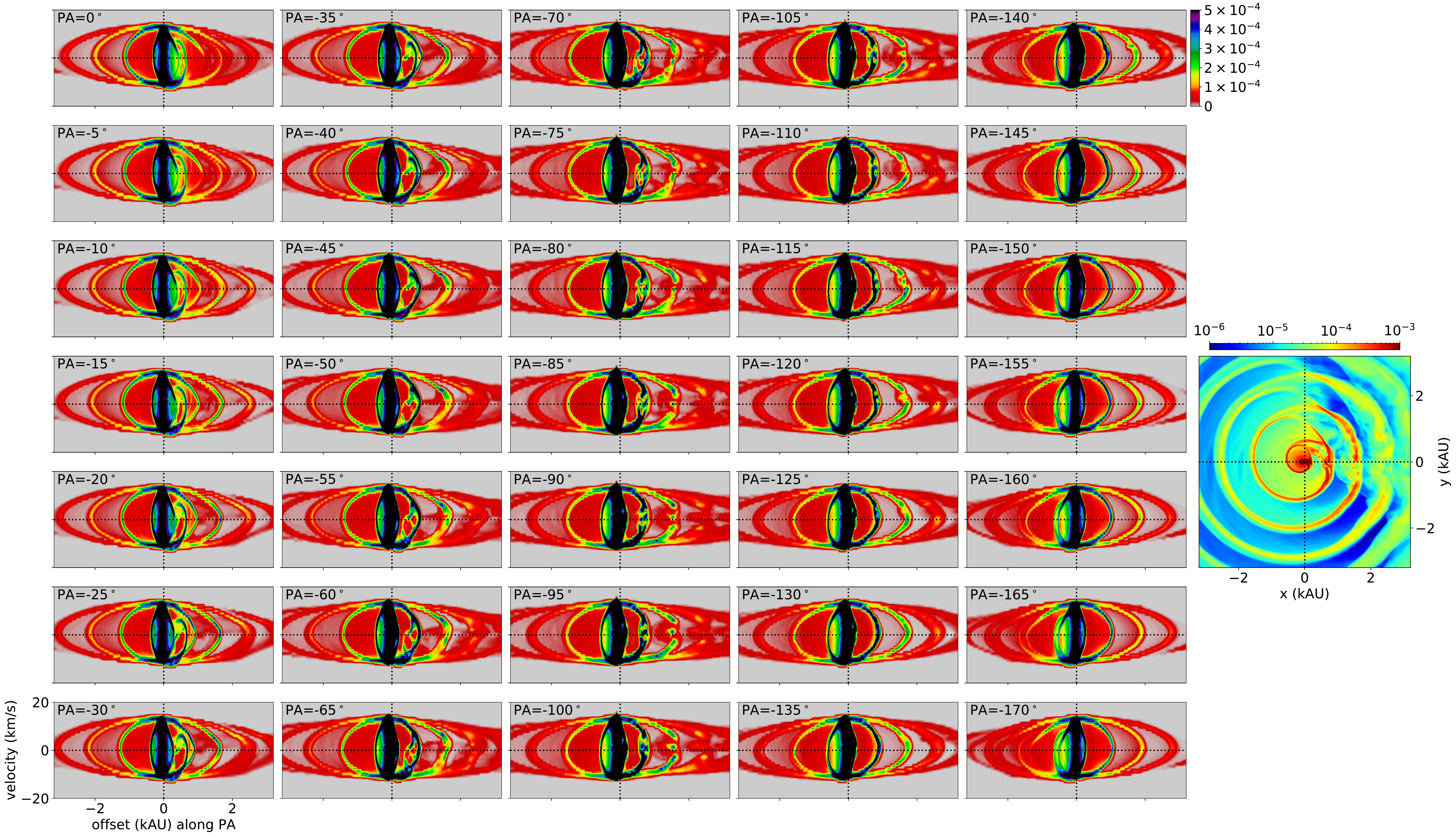}
  \caption{\label{fig:pv8o30}
    Same as Fig.\,\ref{fig:pv8o00} (Model 4) but for $i=30\arcdeg$.
  }
\end{sidewaysfigure}

\begin{sidewaysfigure} 
  \centering
  \includegraphics[width=\textwidth]{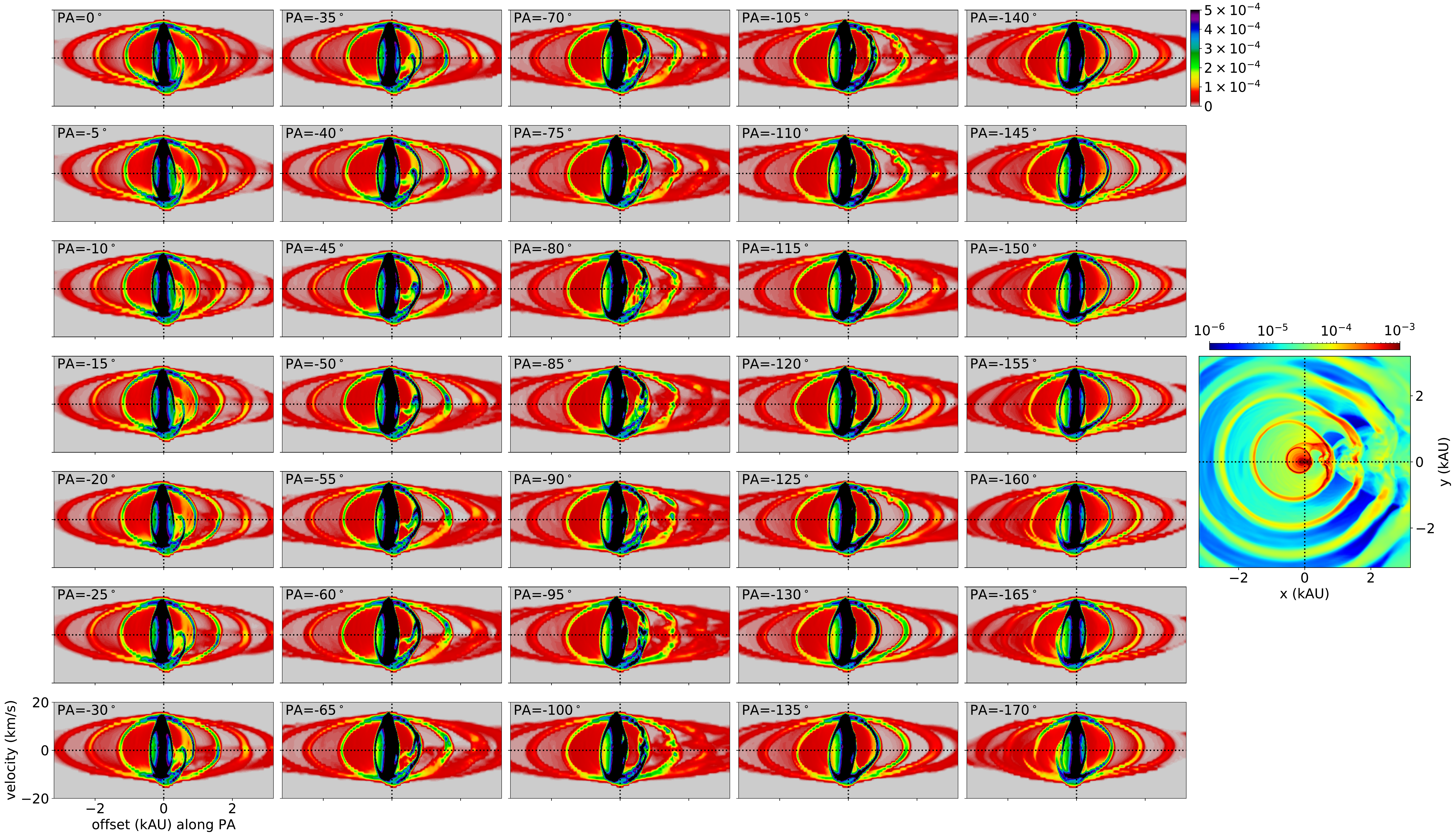}
  \caption{\label{fig:pv8o60}
    Same as Fig.\,\ref{fig:pv8o00} (Model 4) but for $i=60\arcdeg$.
  }
\end{sidewaysfigure}

\begin{sidewaysfigure} 
  \centering
  \includegraphics[width=\textwidth]{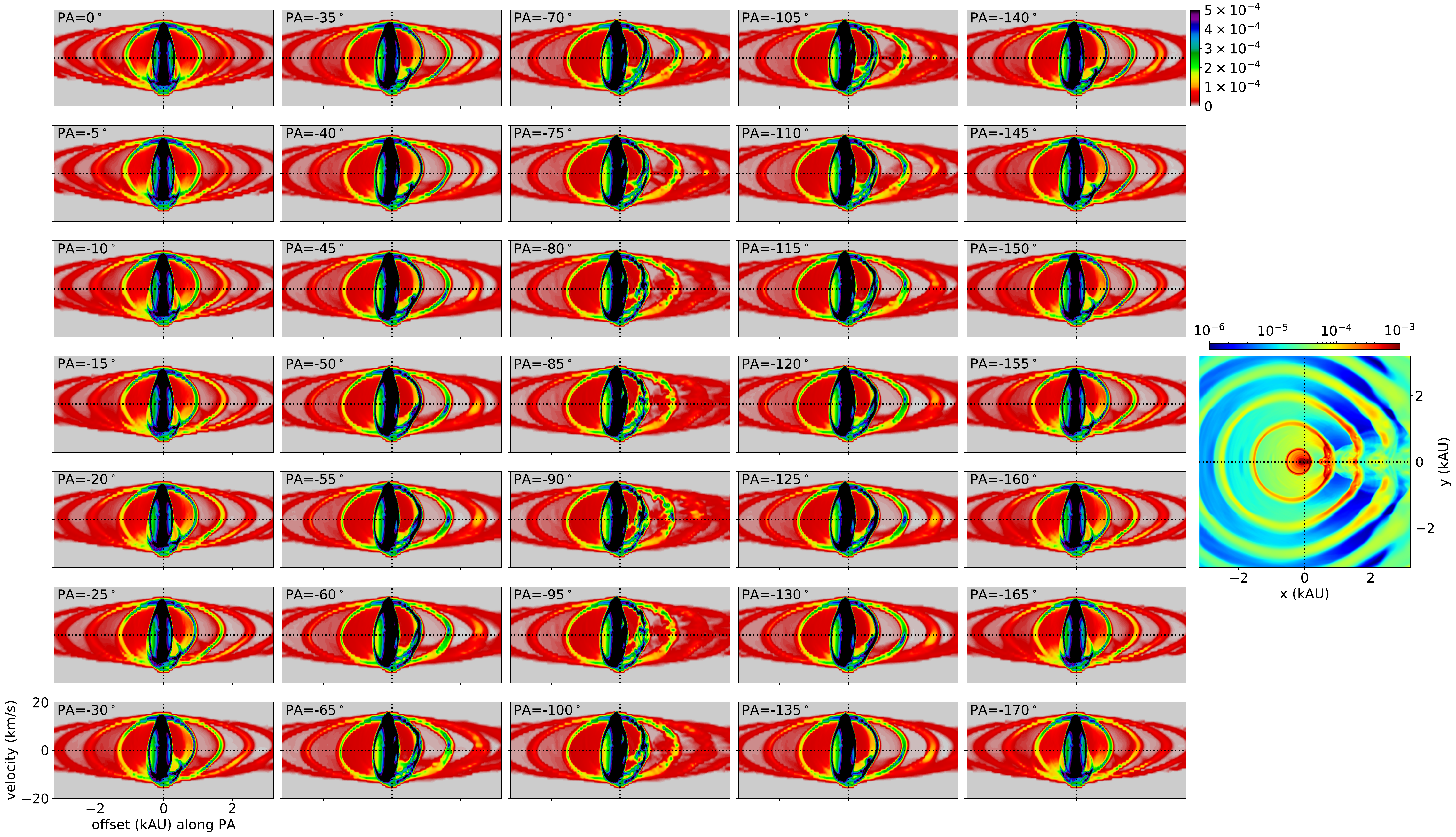}
  \caption{\label{fig:pv8o90}
    Same as Fig.\,\ref{fig:pv8o00} (Model 4) but for $i=90\arcdeg$.
  }
\end{sidewaysfigure}

\section{Angle-radius diagrams centered at the mass-losing star over velocity channels}
\renewcommand\thefigure{B\arabic{figure}}
\setcounter{figure}{0}
\clearpage
\begin{figure*} 
  \plotone{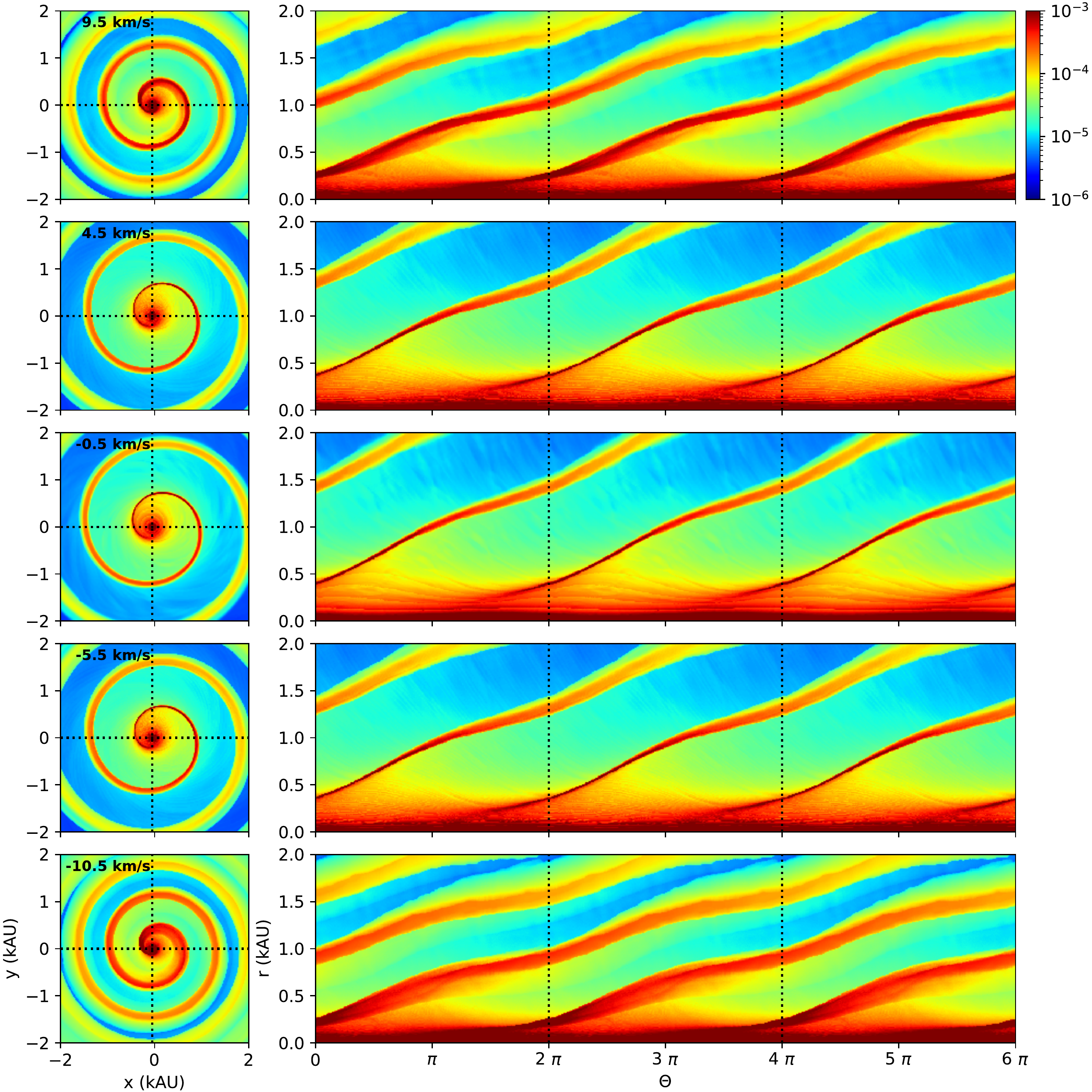}
  \caption{\label{fig:tr0x00}
    Angle-radius diagrams of Model 1 at inclination angle $i=0\arcdeg$ for channel velocities of 9.5, 4.5, $-0.5$, $-5.5$, and $-10.5$\,\kmps\ (from top to bottom of {\it right} panels). The coordinate center is at the center of the mass-losing star. {\it Left} panels show the corresponding channel maps in $xy$-coordinates for comparison. Color bar labels the density in units of \dunit.
  }
\end{figure*}

\begin{figure*} 
  \plotone{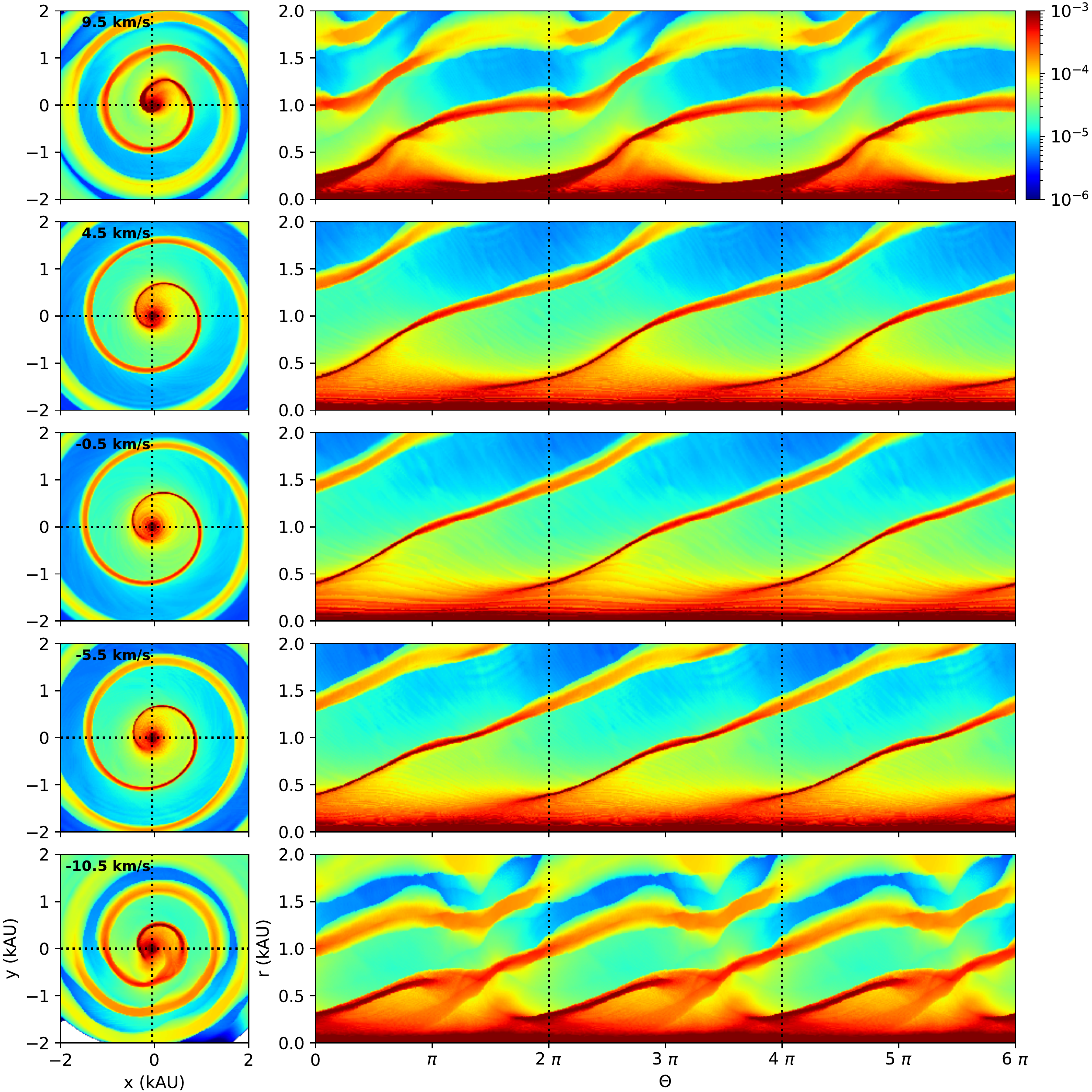}
  \caption{\label{fig:tr0x30}
    Same as Fig.\,\ref{fig:tr0x00} (Model 1) but for $i=30\arcdeg$.
  }
\end{figure*}

\begin{figure*} 
  \plotone{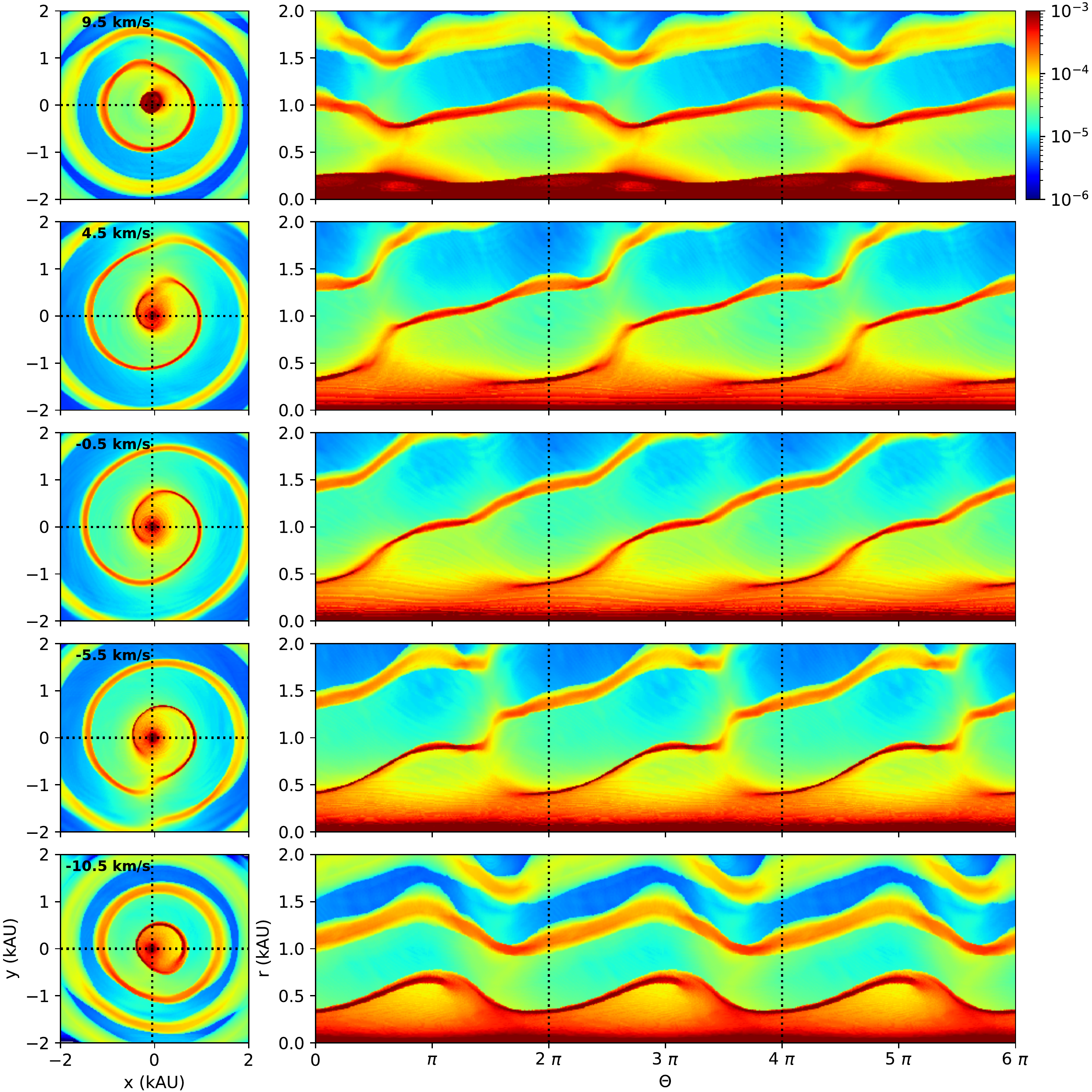}
  \caption{\label{fig:tr0x60}
    Same as Fig.\,\ref{fig:tr0x00} (Model 1) but for $i=60\arcdeg$.
  }
\end{figure*}

\begin{figure*} 
  \plotone{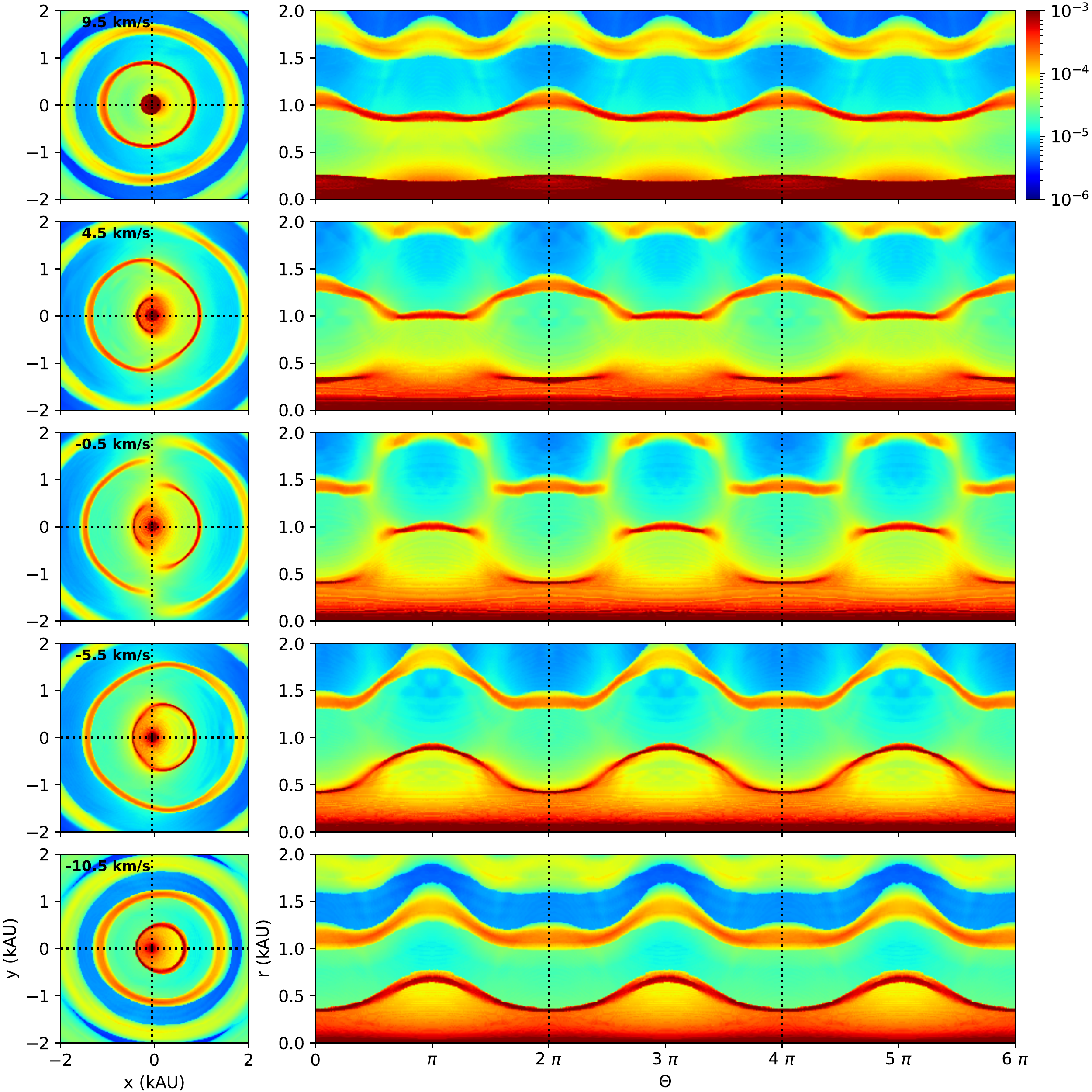}
  \caption{\label{fig:tr0x90}
    Same as Fig.\,\ref{fig:tr0x00} (Model 1) but for $i=90\arcdeg$.
  }
\end{figure*}

\begin{figure*} 
  \plotone{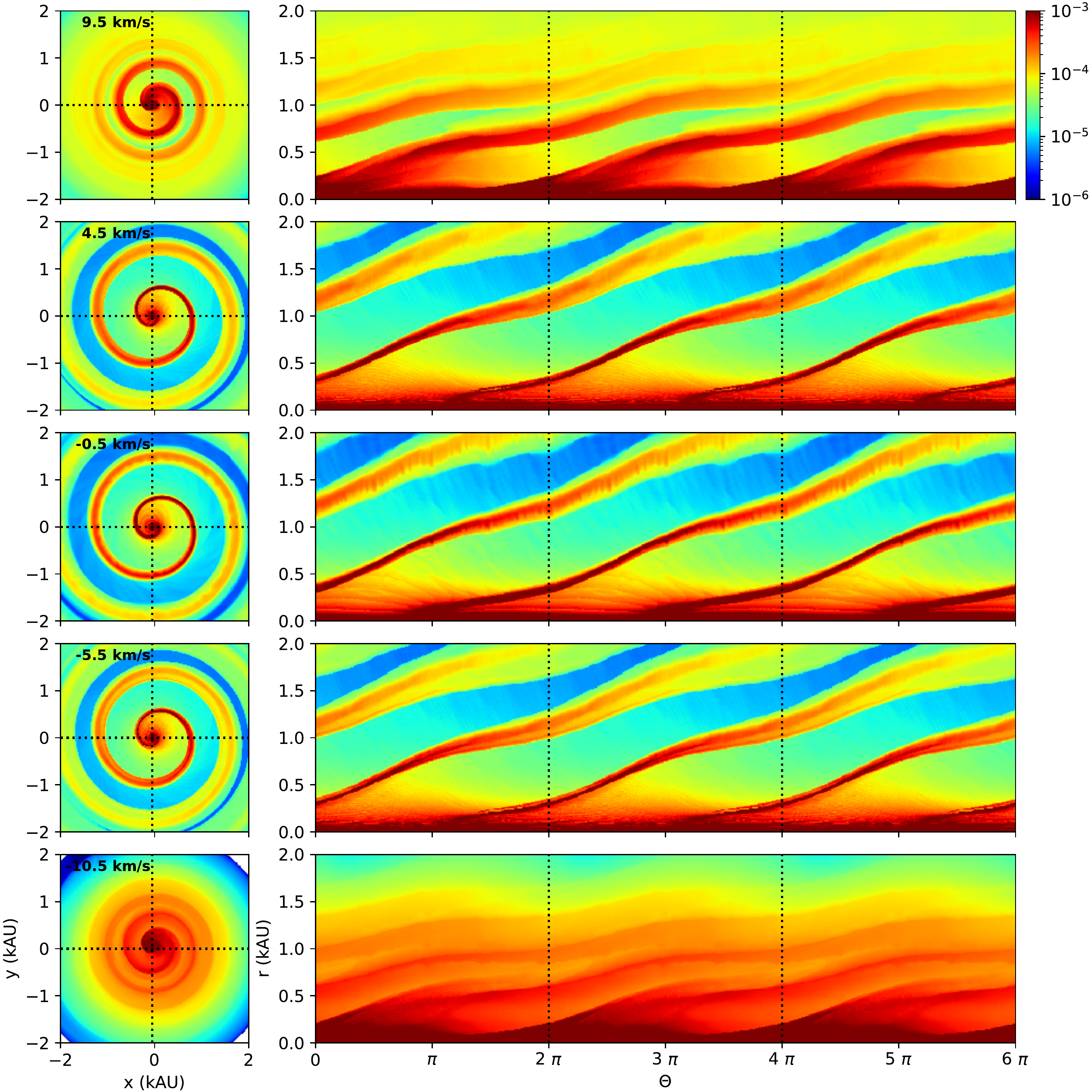}
  \caption{\label{fig:tr0o00}
    Angle-radius diagrams of Model 2 at inclination angle $i=0\arcdeg$.
    See the caption of Fig.\,\ref{fig:tr0x00} for details.
  }
\end{figure*}

\begin{figure*} 
  \plotone{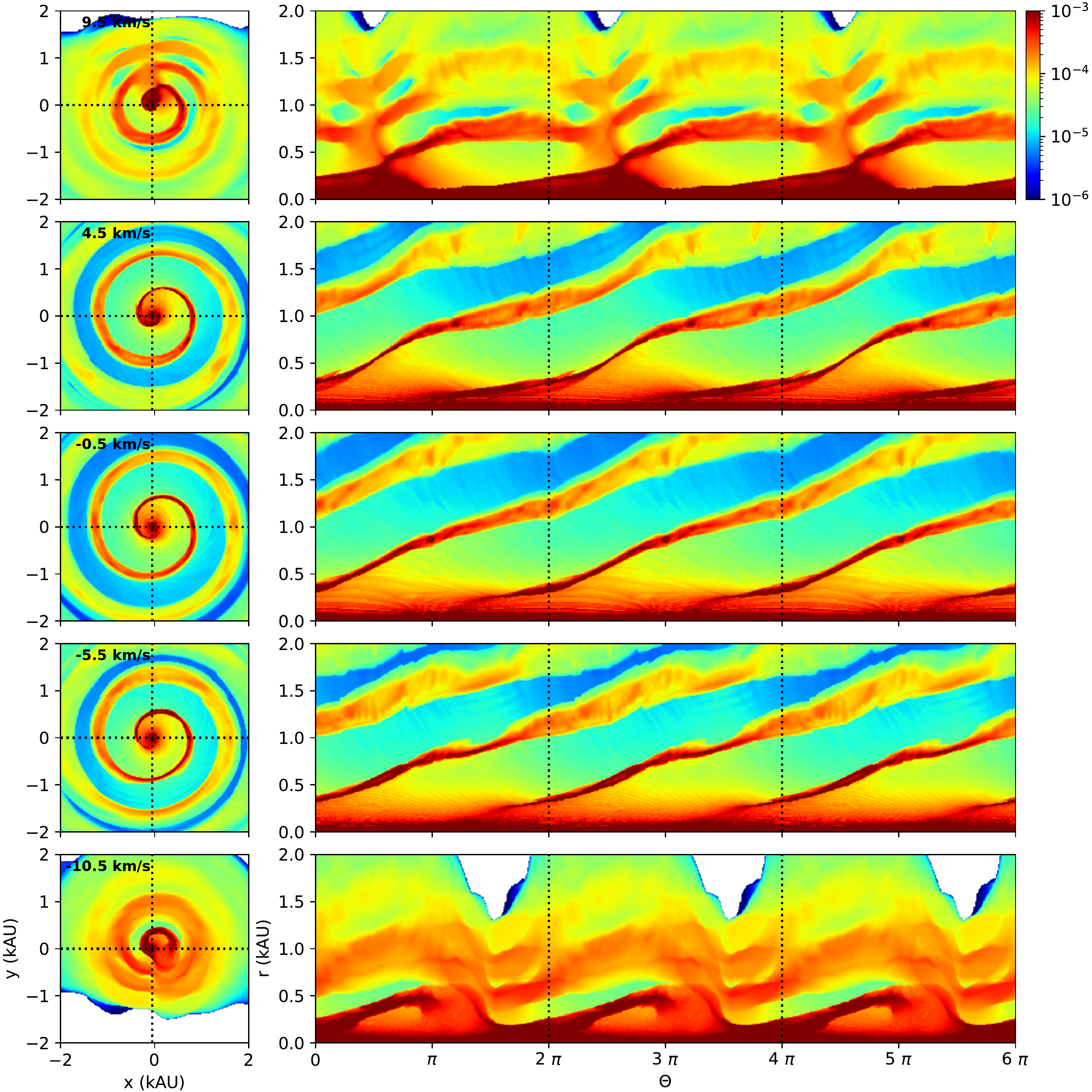}
  \caption{\label{fig:tr0o30}
    Same as Fig.\,\ref{fig:tr0o00} (Model 2) but for $i=30\arcdeg$.
  }
\end{figure*}

\begin{figure*} 
  \plotone{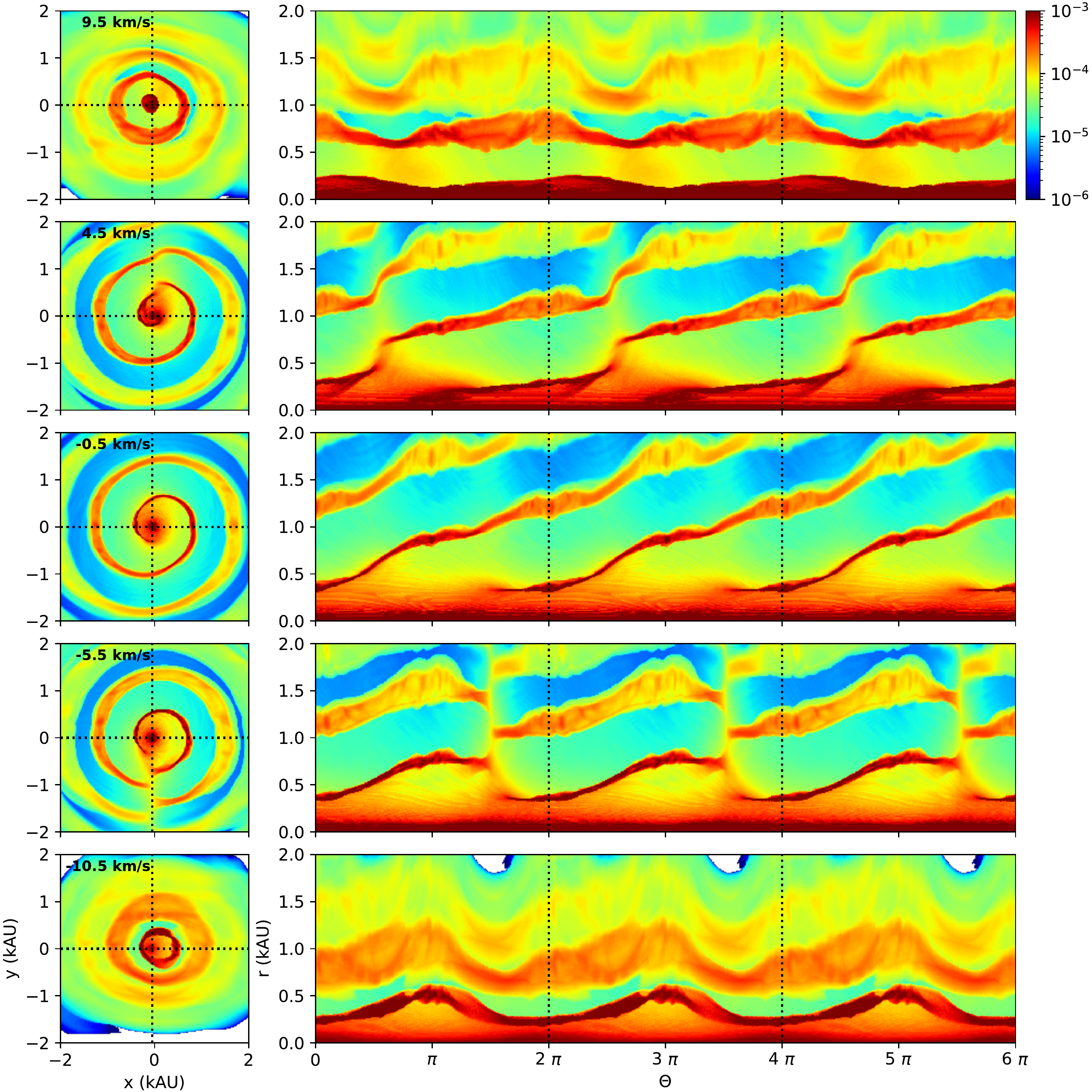}
  \caption{\label{fig:tr0o60}
    Same as Fig.\,\ref{fig:tr0o00} (Model 2) but for $i=60\arcdeg$.
  }
\end{figure*}

\begin{figure*} 
  \plotone{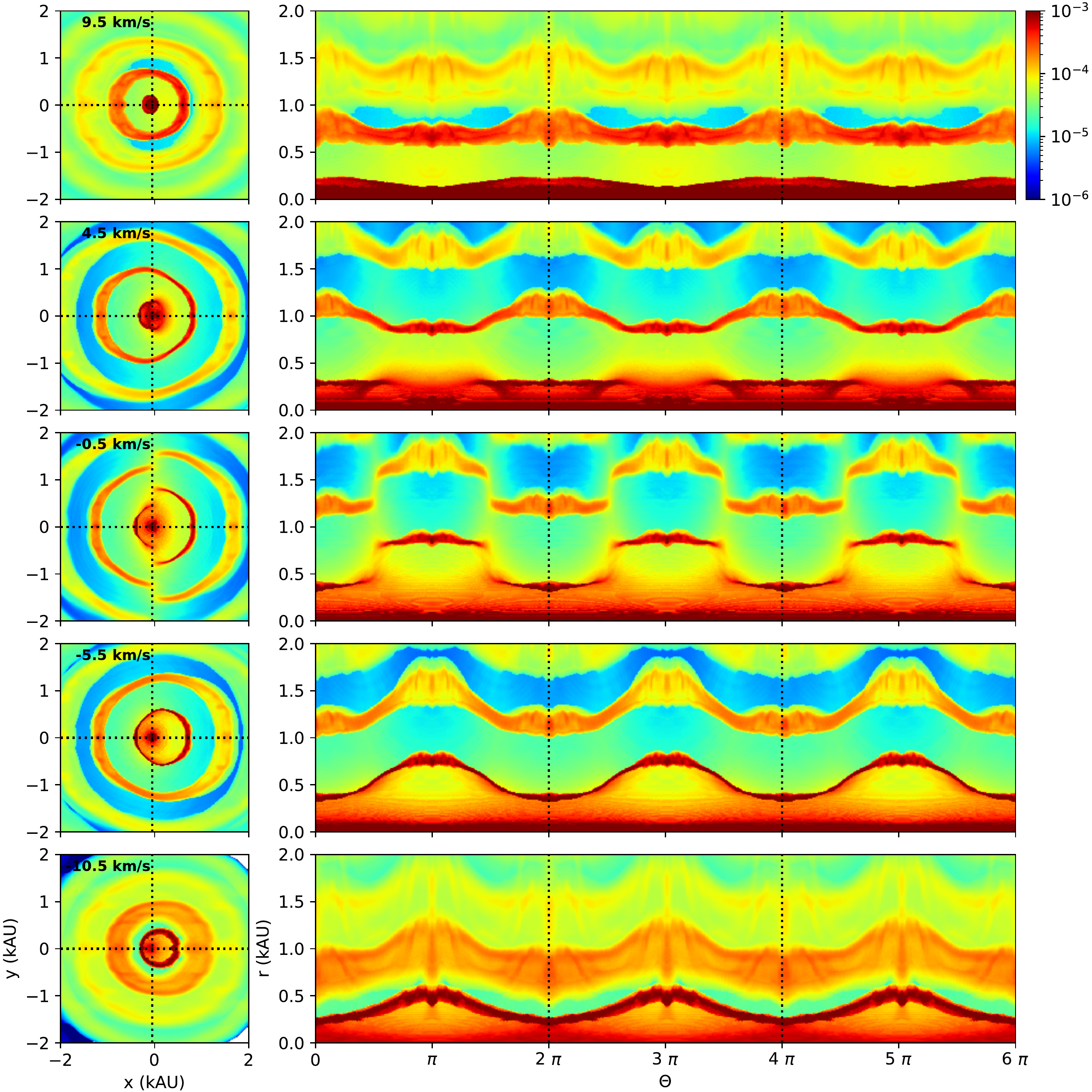}
  \caption{\label{fig:tr0o90}
    Same as Fig.\,\ref{fig:tr0o00} (Model 2) but for $i=90\arcdeg$.
  }
\end{figure*}

\clearpage
\begin{figure*} 
  \plotone{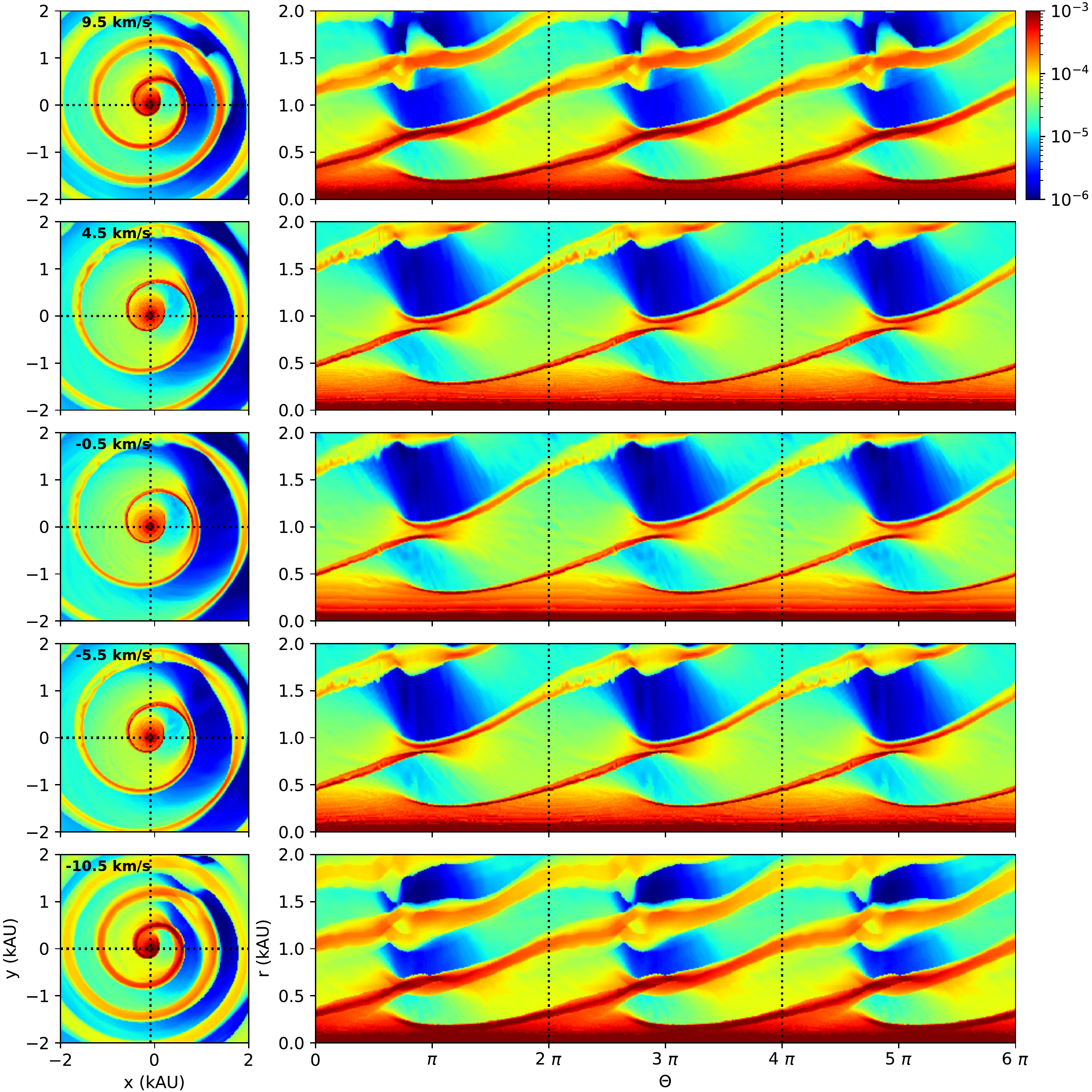}
  \caption{\label{fig:tr8x00}
    Angle-radius diagrams of Model 3 at inclination angle $i=0\arcdeg$.
    See the caption of Fig.\,\ref{fig:tr0x00} for details.
  }
\end{figure*}

\begin{figure*} 
  \plotone{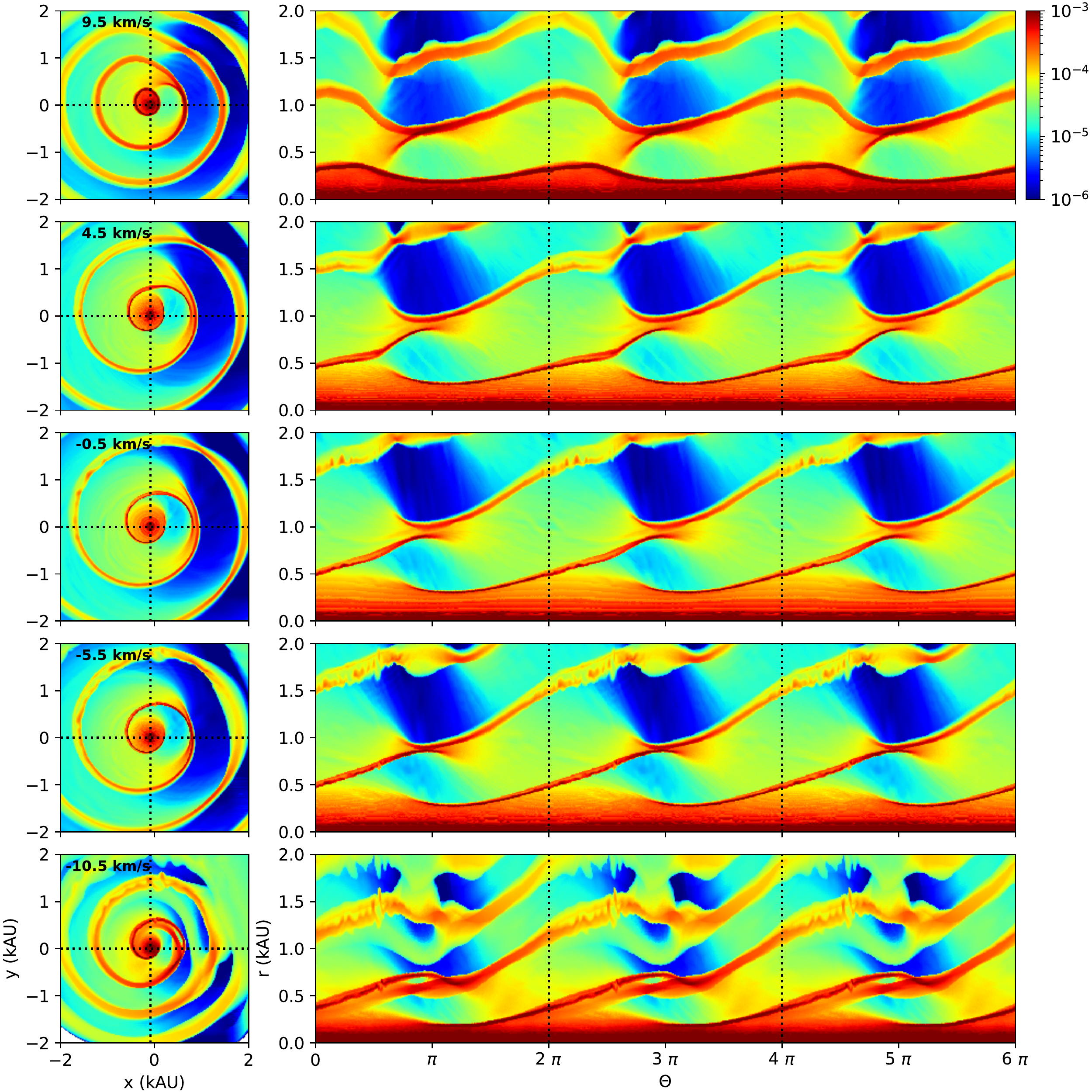}
  \caption{\label{fig:tr8x30}
    Same as Fig.\,\ref{fig:tr8x00} (Model 3) but for $i=30\arcdeg$.
  }
\end{figure*}

\begin{figure*} 
  \plotone{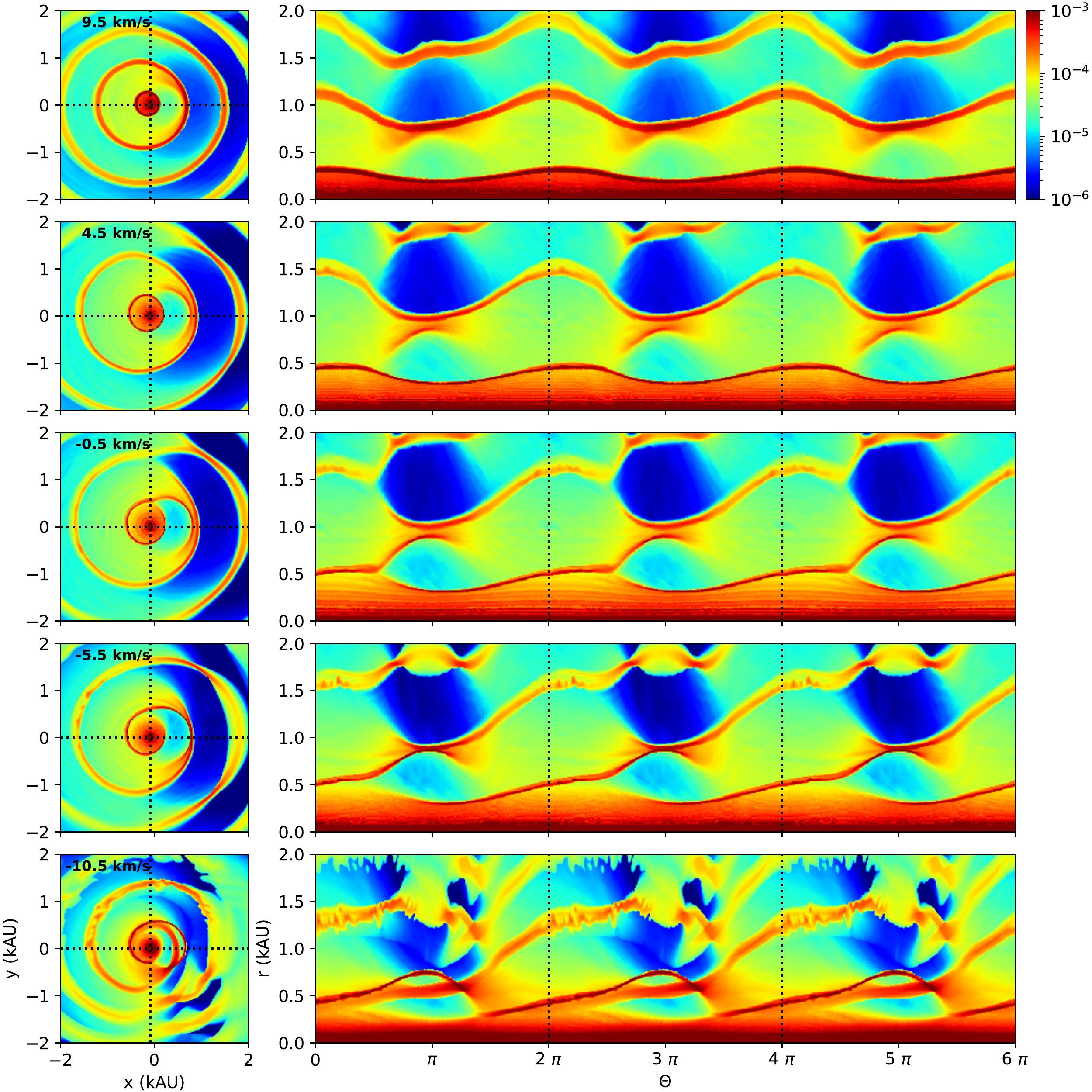}
  \caption{\label{fig:tr8x60}
    Same as Fig.\,\ref{fig:tr8x00} (Model 3) but for $i=60\arcdeg$.
  }
\end{figure*}

\begin{figure*} 
  \plotone{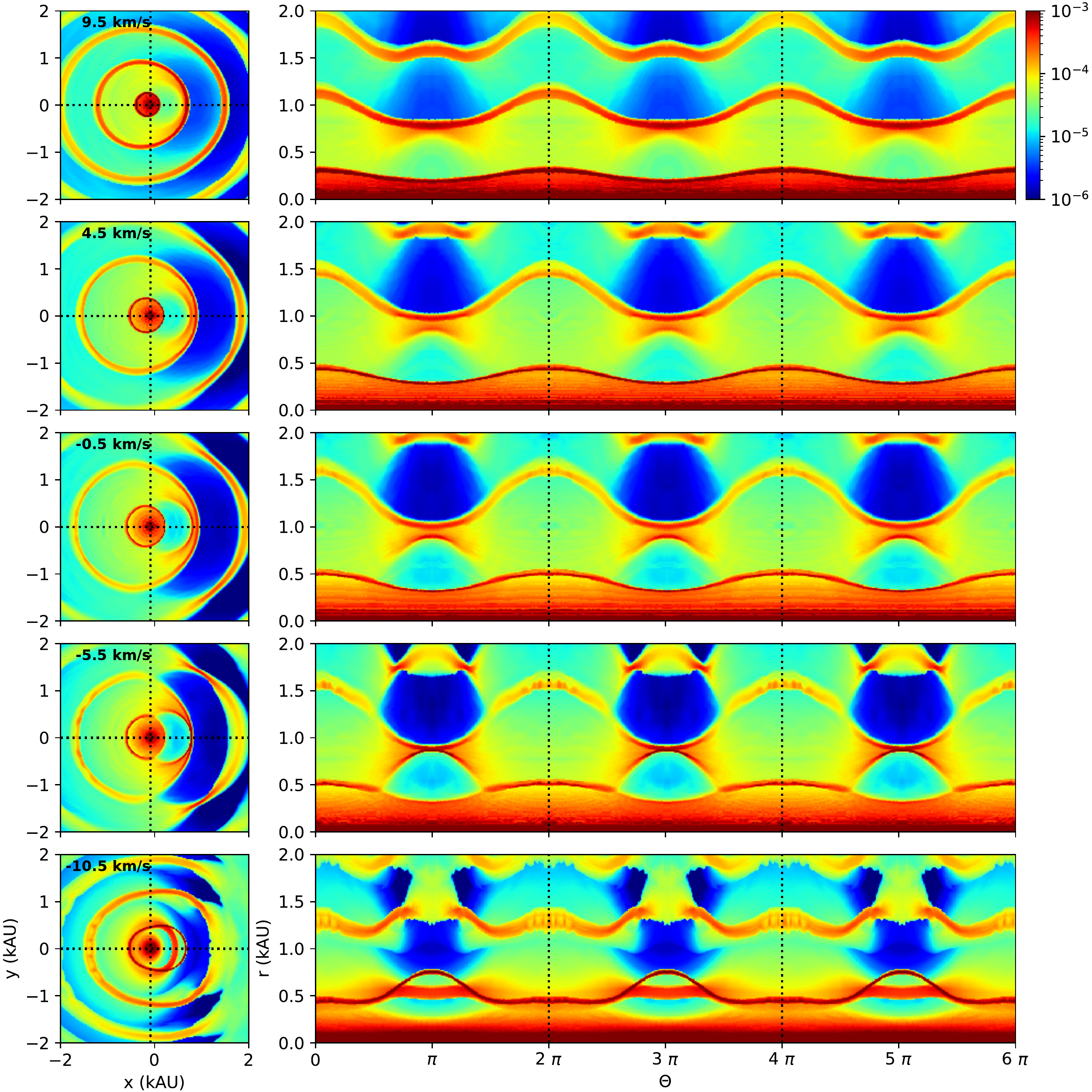}
  \caption{\label{fig:tr8x90}
    Same as Fig.\,\ref{fig:tr8x00} (Model 3) but for $i=90\arcdeg$.
  }
\end{figure*}

\begin{figure*} 
  \plotone{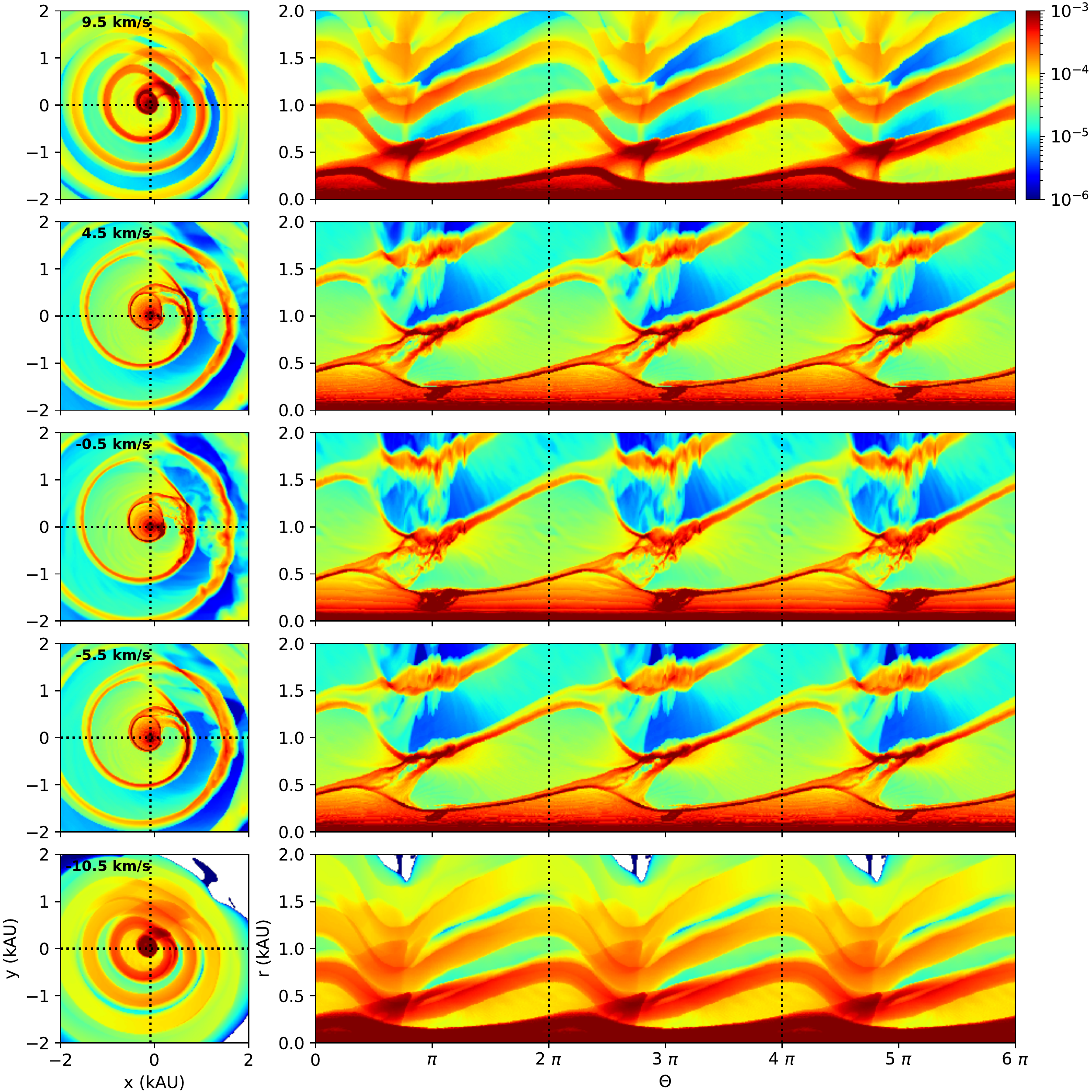}
  \caption{\label{fig:tr8o00}
    Angle-radius diagrams of Model 4 at inclination angle $i=0\arcdeg$.
    See the caption of Fig.\,\ref{fig:tr0x00} for details.
  }
\end{figure*}

\begin{figure*} 
  \plotone{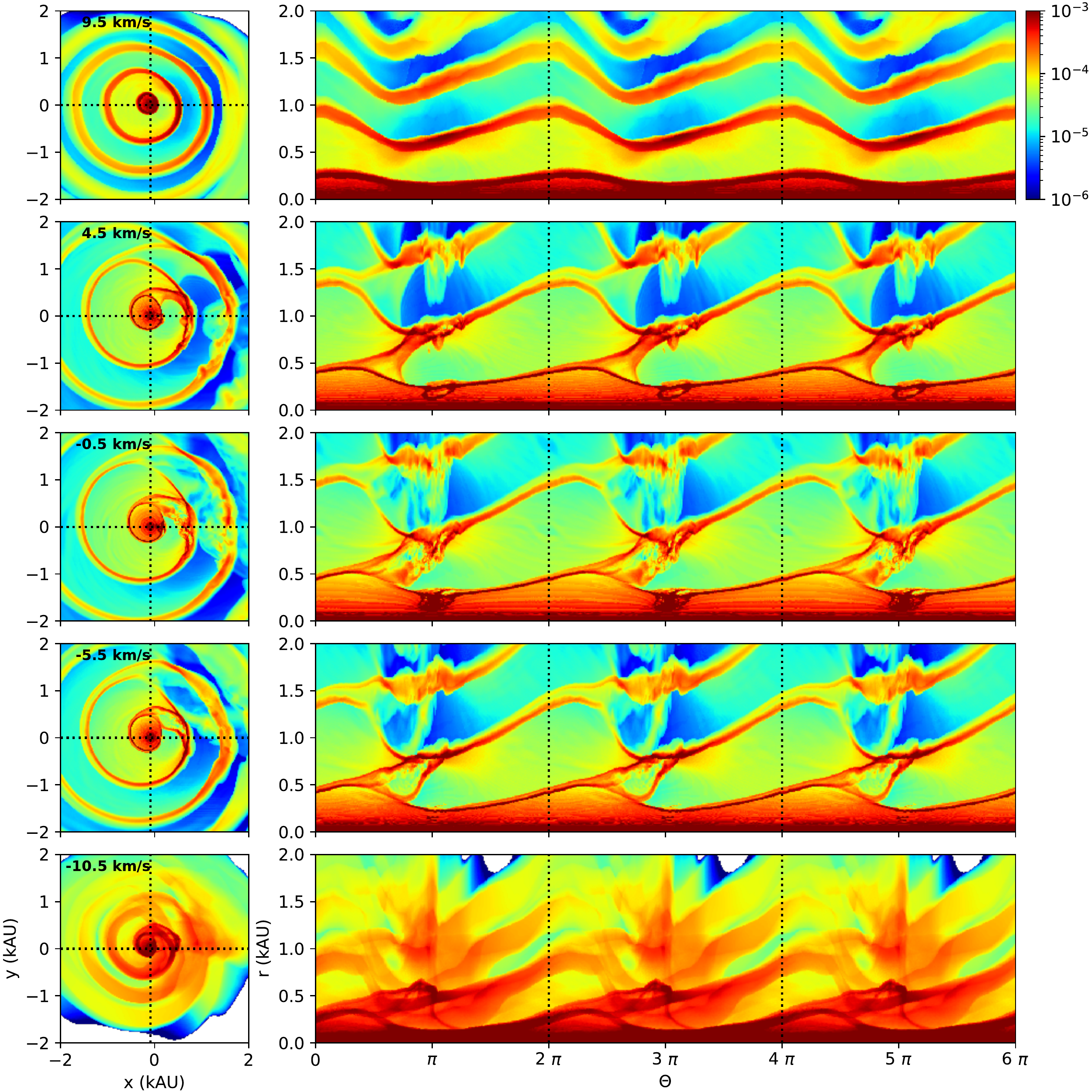}
  \caption{\label{fig:tr8o30}
    Same as Fig.\,\ref{fig:tr8o00} (Model 4) but for $i=30\arcdeg$.
  }
\end{figure*}

\begin{figure*} 
  \plotone{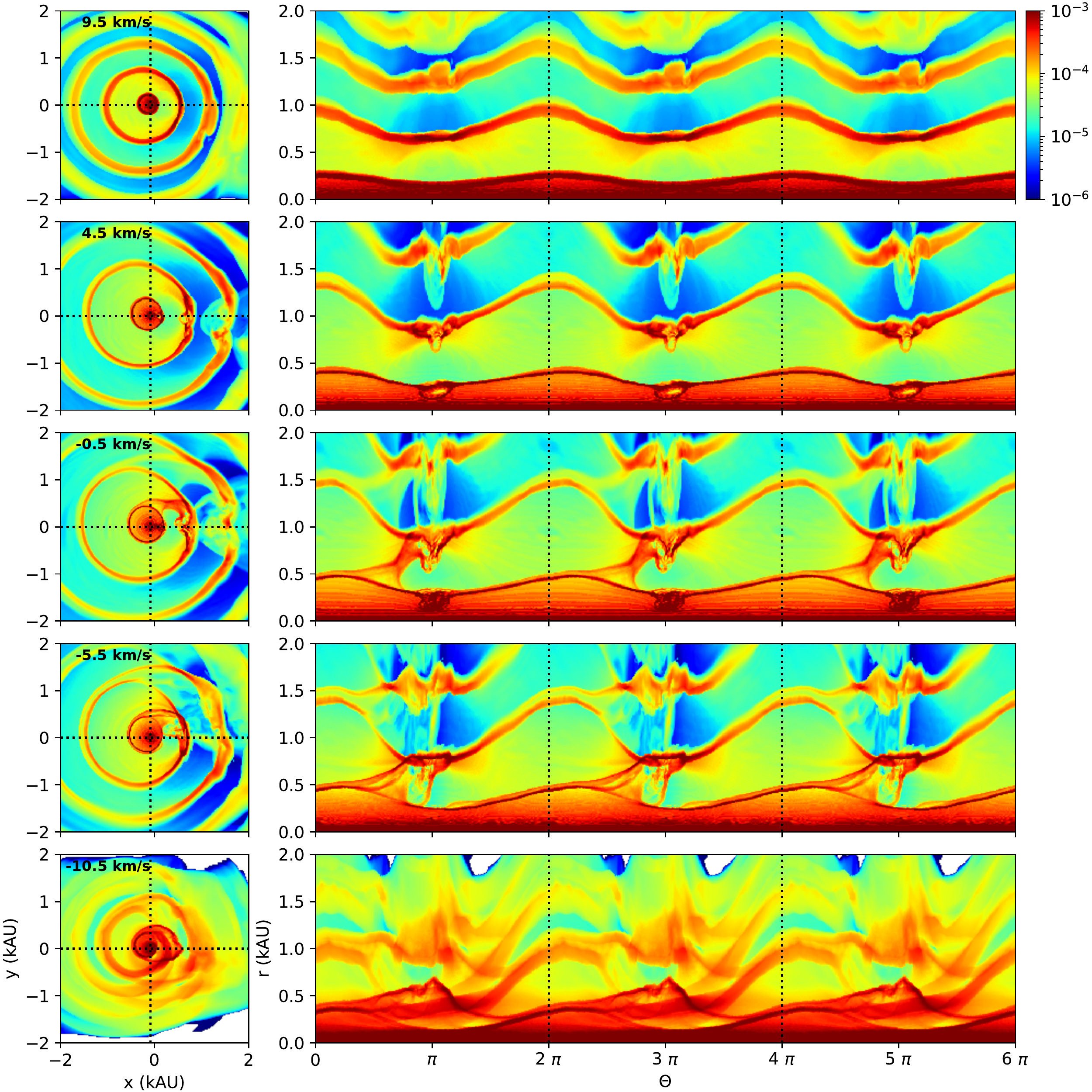}
  \caption{\label{fig:tr8o60}
    Same as Fig.\,\ref{fig:tr8o00} (Model 4) but for $i=60\arcdeg$.
  }
\end{figure*}

\begin{figure*} 
  \plotone{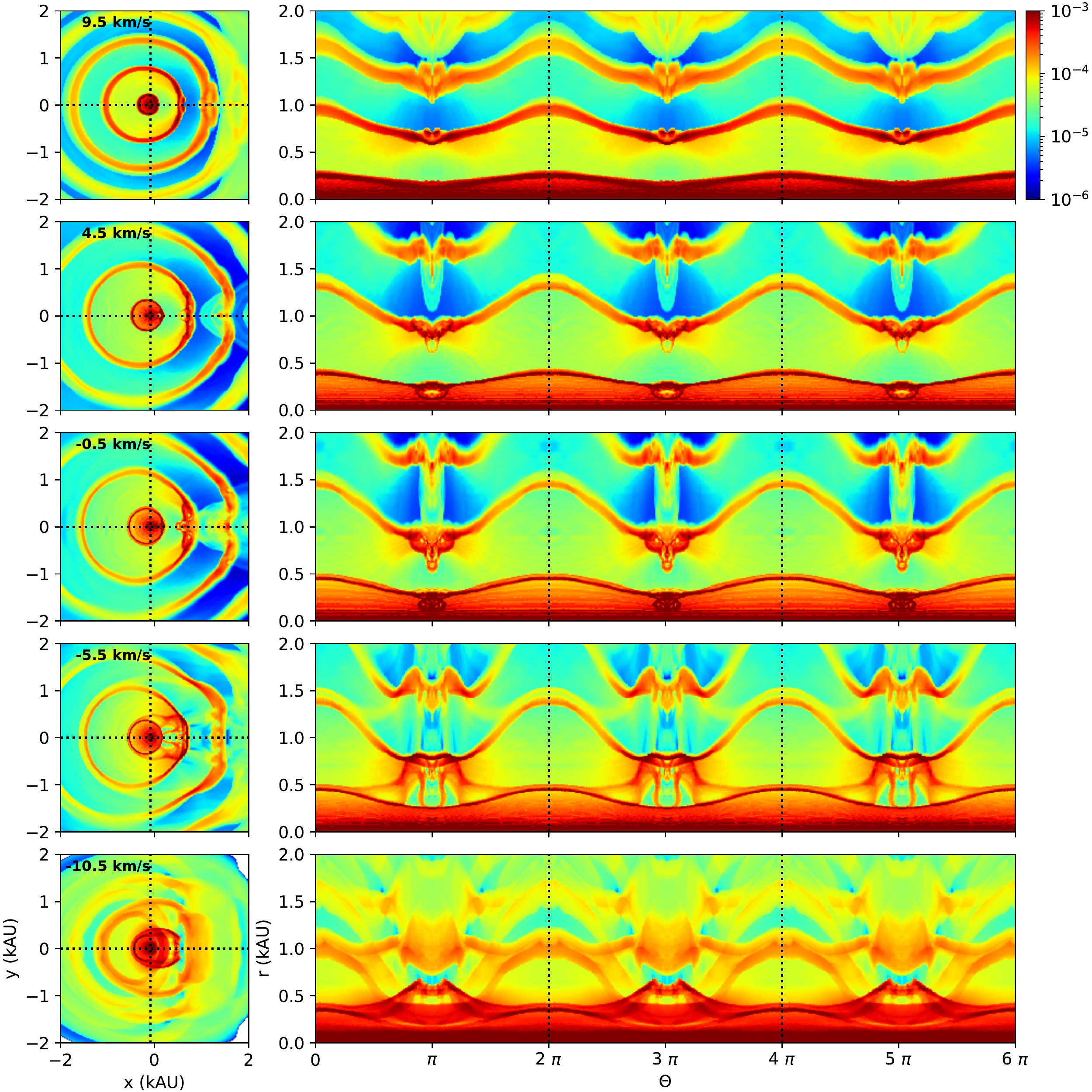}
  \caption{\label{fig:tr8o90}
    Same as Fig.\,\ref{fig:tr8o00} (Model 4) but for $i=90\arcdeg$.
  }
\end{figure*}

\end{document}